\documentclass[a4paper]{article}%
\pdfoutput=1%
\usepackage{libertine}
\RequirePackage{amsthm}
\usepackage{aken}

\usepackage{dirhott}

\RequirePackage{amsmath}
\usepackage{xcolor}
\usepackage{wasysym}
\usepackage{enumitem}

\usepackage{bm}

\usepackage[english]{babel}
\usepackage{a4wide}
\usepackage{graphicx}
\usepackage{wrapfig}
\usepackage{fontawesome}
\usepackage{adjustbox}

\newlist{deriv}{enumerate}{20}
\setlist[deriv]{label={}}

\renewcommand{\todoi}[1]{\textbf{\textcolor{red}{#1 \qed}}\par\noindent}
\newcommand{\todo}[1]{\textbf{\textcolor{red}{\footnote{\textcolor{red}{#1}}}}}
\newcommand{\wip}[1]{}

\renewcommand{\inferenceright}[3]{\dfrac{\begin{array}{l l}#1\end{array}}{\begin{array}{l l}#2\end{array}} #3}
\newcommand{\inference}[3]{\inferenceright{#1}{#2}{\text{#3}}}

\renewcommand{\loch}{{\llcorner\hspace{-.4ex}\lrcorner}}

\newcommand{\Ty}{\mathrm{Ty}}
\newcommand{\Tm}{\mathrm{Tm}}

\newcommand{\seelog}{\textsuperscript{\S \ref{sec:changelog}}}

\newcommand{\pparen}[1]{{\llparenthesis #1 \rrparenthesis}}

\newcommand{\IB}{\mathbb{B}}

\newcommand{\IX}{\mathbb{I}}

\newcommand{\tycode}[1]{\ulcorner #1 \urcorner}

\newcommand{\dra}[2]{\left\langle #1 \,\middle|\, #2 \right\rangle}

\newcommand{\twisted}[1]{\name{Tw}(#1)}
	\newcommand{\twistobj}[3]{(#1 \xrightarrow{#3} #2)}

\newcommand{\yoneda}{\mathbf{y}}

\newcommand{\ftrtm}[2]{{}^{#1} #2}

\newcommand{\splitepislice}[2]{#1 /\hspace{-.5ex}/ #2}

\newcommand{\dimslice}[2]{#1 /\hspace{-.5ex}/ #2}
\newcommand{\orthslice}[2]{#1 /\hspace{-.5ex}/^\partial #2}

\renewcommand{\isvar}{\sharp}
\renewcommand{\invar}{\circ}

\newcommand{\sep}{\mathrel{|}}

\newcommand{\affinecubecat}[1]{\Box^{#1}}

\newcommand{\erasecat}[1]{\name{Erase}_{#1}}

\newcommand{\pointcat}{{\name{Point}}}
\newcommand{\thearrowcat}{{\uparrow}}
\newcommand{\catC}{{\cat C}}
\newcommand{\catD}{{\cat D}}
\newcommand{\catI}{{\cat I}}

\newcommand{\catP}{{\cat P}}
\newcommand{\catW}{{\cat W}}
\newcommand{\catV}{{\cat V}}

\newcommand{\Cod}{\mathrm{Cod}}
\newcommand{\Dom}{\mathrm{Dom}}

\newcommand{\DSub}[2]{#1 \Rightarrow #2}
\newcommand{\Dsez}{\mathrel{\rhd}}

\newcommand{\dsub}[1]{{\left[ #1 \right\rangle}}
\newcommand{\psub}[1]{\angles{#1}}

\newcommand{\textor}{\,\text{or}\,}

\newcommand{\lock}[1]{\text{\faLock}_{#1}}

\newcommand{\ctxmod}[3]{{#1} \shortmid #2 : #3}

\newcommand{\lpsh}[1]{{#1}_!}
\newcommand{\fpsh}[1]{{#1}^*}
\newcommand{\rpsh}[1]{{#1}_*}
\newcommand{\baseslice}[2]{#1^{/#2}}
\newcommand{\lpshpsh}[2]{{#1}_!^{#2 \sep}}
\newcommand{\fpshpsh}[2]{{#1}^{#2 \sep *}}
\newcommand{\rpshpsh}[2]{{#1}_*^{#2 \sep}}

\newcommand{\letin}[2]{\mathsf{let}~#1:=#2~\mathsf{in}~}

\newcommand{\clocksym}{{\text{\clock}}}
\newcommand{\embargo}{\text{\scalebox{1}{\faExclamation}}}

\newcommand{\amaze}{\mathop{\surd}}
\newcommand{\multip}{\ltimes}
\newcommand{\twist}{\mathbin{\bar{\multip}}}

\newcommand{\sumsym}{\exists}
	\newcommand{\dropsym}{\mathsf{drop}}
	\newcommand{\copysym}{\mathsf{copy}}
\newcommand{\freshsym}{\Finv}
	\newcommand{\appsym}{\mathsf{app}}
	\newcommand{\constsym}{\mathsf{const}}
\newcommand{\lollisym}{\forall}
	\newcommand{\unmeridsym}{\mathsf{unmerid}}
	\newcommand{\reindexsym}{\mathsf{reidx}}
\newcommand{\transpsym}{{\between}}  
\newcommand{\pairsym}{\Sigma}
	\newcommand{\hidesym}{\mathsf{hide}}
\newcommand{\wknsym}{\Omega}
	\newcommand{\spoilsym}{\mathsf{spoil}}
\newcommand{\funcsym}{\Pi}
	\newcommand{\cospoilsym}{\mathsf{cospoil}}
\newcommand{\cartranspsym}{\$}

\newcommand{\sumbase}[1]{\sumsym_{#1}}
\newcommand{\pairbase}[1]{\pairsym_{#1}}
	\newcommand{\dropbase}[1]{\dropsym_{#1}}
	\newcommand{\copybase}[1]{\copysym_{#1}}
	\newcommand{\hidebase}[1]{\hidesym_{#1}}
\newcommand{\freshbase}[1]{\freshsym_{#1}}
\newcommand{\wknbase}[1]{\wknsym_{#1}}

	\newcommand{\spoilbase}[1]{\spoilsym_{#1}}
\newcommand{\lollibase}[1]{\lollisym_{#1}}
\newcommand{\funcbase}[1]{\funcsym_{#1}}

\newcommand{\transpbase}[1]{\transpsym_{#1}}
\newcommand{\cartranspbase}[1]{\cartranspsym_{#1}}

\newcommand{\modslice}[3]{#1_{#2}^{/#3}}
\newcommand{\sumslice}[2]{\modslice{\sumsym}{#1}{#2}}
\newcommand{\pairslice}[2]{\modslice{\pairsym}{#1}{#2}}
	\newcommand{\dropslice}[2]{\modslice{\dropsym}{#1}{#2}}
	\newcommand{\copyslice}[2]{\modslice{\copysym}{#1}{#2}}
	\newcommand{\hideslice}[2]{\modslice{\hidesym}{#1}{#2}}
\newcommand{\freshslice}[2]{\modslice{\freshsym}{#1}{#2}}
\newcommand{\wknslice}[2]{\modslice{\wknsym}{#1}{#2}}

	\newcommand{\spoilslice}[2]{\modslice{\spoilsym}{#1}{#2}}
\newcommand{\lollislice}[2]{\modslice{\lollisym}{#1}{#2}}

\newcommand{\freshelem}[2]{\freshsym_{#1}^{\in #2}}

\newcommand{\modpsh}[3]{#1_{#2}^{#3 \sep}}
\newcommand{\sumpsh}[2]{\modpsh{\sumsym}{#1}{#2}}
\newcommand{\pairpsh}[2]{\modpsh{\pairsym}{#1}{#2}}
	\newcommand{\droppsh}[2]{\modpsh{\dropsym}{#1}{#2}}
	\newcommand{\copypsh}[2]{\modpsh{\copysym}{#1}{#2}}
	\newcommand{\hidepsh}[2]{\modpsh{\hidesym}{#1}{#2}}
\newcommand{\freshpsh}[2]{\modpsh{\freshsym}{#1}{#2}}
\newcommand{\wknpsh}[2]{\modpsh{\wknsym}{#1}{#2}}
	\newcommand{\apppsh}[2]{\modpsh{\appsym}{#1}{#2}}
	\newcommand{\constpsh}[2]{\modpsh{\constsym}{#1}{#2}}
	\newcommand{\spoilpsh}[2]{\modpsh{\spoilsym}{#1}{#2}}
\newcommand{\lollipsh}[2]{\modpsh{\lollisym}{#1}{#2}}
\newcommand{\funcpsh}[2]{\modpsh{\funcsym}{#1}{#2}}
	\newcommand{\unmeridpsh}[2]{\modpsh{\unmeridsym}{#1}{#2}}
	\newcommand{\reindexpsh}[2]{\modpsh{\reindexsym}{#1}{#2}}
	\newcommand{\cospoilpsh}[2]{\modpsh{\cospoilsym}{#1}{#2}}
\newcommand{\transppsh}[2]{\modpsh{\transpsym}{#1}{#2}}
\newcommand{\cartransppsh}[2]{\modpsh{\cartranspsym}{#1}{#2}}

\newcommand{\Modbox}[2]{\left\langle #1 \mathrel{\shortmid} #2 \right\rangle}

\newcommand{\modbox}[2]{\mathsf{mod}_{#1}(#2)}

\newcommand{\unmodbox}[2]{\mathsf{mod}\inv_{#1}(#2)}

\theoremstyle{definition}
\newtheorem{theorem}{Theorem}[subsection]
\newtheorem{definition}[theorem]{Definition}
\crefname{definition}{definition}{definitions}
\Crefname{definition}{Definition}{Definitions}
\newtheorem{proposition}[theorem]{Proposition}
\newtheorem{lemma}[theorem]{Lemma}
\newtheorem{example}[theorem]{Example}
\newtheorem{remark}[theorem]{Remark}
\newtheorem{corollary}[theorem]{Corollary}

\newtheorem{notation}[theorem]{Notation}

\newcommand{\thetitle}{The Transpension Type: Technical Report}
\newcommand{\theauthors}{Andreas Nuyts\footnote{Andreas Nuyts holds a Postdoctoral Fellowship from the Research Foundation - Flanders (FWO; 1247922N), and carried out most of this research holding a PhD Fellowship from the Research Foundation - Flanders (FWO; 1110817N). This research was partially conducted at Vrije Universiteit Brussel and funded by the Research Foundation - Flanders (FWO; G0G0519N). This research is partially funded by the Research Fund KU Leuven.}}
\newcommand{\theinstitution}{DistriNet, KU Leuven, Belgium}
\begin{document}
	\addtolength{\voffset}{-.5in}

\title{\thetitle}
\date{\today}
\author{\theauthors{} \\ \theinstitution}
\maketitle

\tableofcontents

\section{Introduction}
The purpose of these notes is to give a categorical semantics for the transpension type \cite{transpension}, which is right adjoint to a potentially substructural dependent function type.
\begin{itemize}
	\item In \cref{sec:prerequisites} we discuss some prerequisites.
	\item In \cref{sec:multipliers}, we define multipliers and discuss their properties.
	\item In \cref{sec:psh}, we study how multipliers lift from base categories to presheaf categories.
	\item In \cref{sec:prior}, we explain how typical presheaf modalities can be used in the presence of the transpension type.
	\item In \cref{sec:commut}, we study commutation properties of prior modalities, substitution modalities and multiplier modalities.
\end{itemize}

\section{Prerequisites} \label{sec:prerequisites}

\subsection{On adjoints}
\subsubsection{Adjoints and natural transformations}
\begin{lemma}\label{thm:cotranspose}
	Let $L \dashv R$.
	\begin{itemize}
		\item Natural transformations $LF \to G$ correspond to natural transformations $F \to RG$, naturally in $F$ and $G$.
		\item Natural transformations $FR \to G$ correspond to natural transformations $F \to GL$, naturally in $F$ and $G$.
	\end{itemize}
\end{lemma}
\begin{proof}
	The first statement is trivial.
	
	To see the second statement, we send $\zeta : FR \to G$ to $\zeta L \circ F \eta : F \to GL$, and conversely $\theta : F \to GL$ to $G \eps \circ \theta R : FR \to G$. Naturality is clear. Mapping $\zeta$ to and fro, we get
	\begin{equation}
		G \eps \circ \zeta LR \circ F \eta R = \zeta \circ FR \eps \circ F \eta R = \zeta.
	\end{equation}
	Mapping $\theta$ to and fro, we get
	\begin{equation*}
		G \eps L \circ \theta RL \circ F \eta = G \eps L \circ GL \eta \circ \theta = \theta. \qedhere
	\end{equation*}
\end{proof}
\begin{lemma} \label{thm:commut-adj}
	Assume 4 triples of adjoint functors: $E \dashv F \dashv G$ and $E' \dashv F' \dashv G'$ and $S_1 \dashv T_1 \dashv U_1$ and $S_2 \dashv T_2 \dashv U_2$ such that the following diagram commutes up to natural isomorphism:
	\begin{equation}
		\xymatrix{
			\cat C_1 \ar[r]^{F} \ar[d]_{T_1}
			& \cat C_2 \ar[d]^{T_2}
			\\
			\cat C_1' \ar[r]_{F'}
			& \cat C_2'
		}
	\end{equation}
	Then we have
	\begin{equation}
		\begin{array}{c c c}
			E S_2 \cong S_1 E'
			& E' T_2 \to T_1 E
			&
			\\
			F S_1 \leftarrow S_2 F'
			& F' T_1 \cong T_2 F
			& F U_1 \to U_2 F'
			\\
			
			& G' T_2 \leftarrow T_1 G
			& G U_2 \cong U_1 G'.
		\end{array}
	\end{equation}
	In fact, any one of these statements holds if only the adjoints used by that statement are given.
\end{lemma}
\begin{proof}
	The central isomorphism is given.
	The other isomorphisms are obtained by taking the left/right adjoints of both hands of the original isomorphism.
	By picking one direction of the central isomorphism, we can step to the left/right/top/bottom by applying \cref{thm:cotranspose}.
\end{proof}

\subsubsection{Adjoints and categories with families}
\begin{proposition}
	If a functor $R : \cat C \to \widehat \catW$ from a CwF $\cat C$ to a presheaf CwF $\widehat \catW$ has a left adjoint $L$, then it is a weak CwF morphism.
\end{proposition}
\begin{proof}
	We use the presheaf notations from \cite{reldtt-techreport} (\cref{sec:prerequisites:psh:notation}).
	
	For $\Gamma \sez_{\cat C} T \type$, define $R \Gamma \sez_{\widehat \catW} RT \type$ by
	\begin{equation}
		(W \Dsez_{\widehat \catW} (RT)\dsub{\delta}) :\cong (L \yoneda W \sez_{\cat C} T[\eps \circ L\delta]).
	\end{equation}
	Naturality of this operation is easy to show, and the action of $R$ on terms is given by $(\ftrtm R t) \dsub \delta = t[\eps \circ L \delta]$.
\end{proof}
\begin{definition}
	Given adjoint functors $L \dashv R$ such that $R$ is a weak CwF morphism, and $A \in \Ty(L\Gamma)$, we write $\dra{R}{A} := (RA)[\eta] \in \Ty(\Gamma)$.
\end{definition}
Note that $\dra{R}{A[\eps]} = (RA)[R\eps][\eta] = RA$.

\subsubsection{Adjoints and slice categories}
\begin{definition}
	For any $U \in \catW$, the \textbf{slice category over $U$}, denoted $\catW/U$, has objects $(W, \psi)$ where $W \in \catW$ and $\psi : W \to U$ and the morphisms $(W, \psi) \to (W', \psi')$ are the morphisms $\chi : W \to W'$ such that $\psi' \circ \chi = \psi$.
\end{definition}
\begin{definition} \label{def:act-slice}
	Given a functor $F : \catV \to \catW$ and $V_0 \in \Obj\catV$, we define the action of $F$ on slice objects over $V_0$ as the functor
	\begin{equation*}
		F^{/V_0} : \catV/V_0 \to \catW/FV_0 : (V, \vfi) \mapsto (FV, F\vfi).
	\end{equation*}
\end{definition}
\begin{proposition}\label{thm:act-slice-adjoint}
	Let $L \dashv R : \catC \to \catD$ with $\alpha : \Hom_\catC(Lc, d) \cong \Hom_\catD(c, Rd)$.\footnote{So $\alpha(\gamma) = R\gamma \circ \eta$ and $\alpha\inv(\delta) = \eps \circ L\delta$.}
	Then $R^{/z} : \catC / c_0 \to \catD / R c_0 : (c, \gamma) \mapsto (Rc, R\gamma)$ has a left adjoint $L_{/z} : \catD / R c_0 \to \catC / c_0 : (d, \delta) \mapsto (Ld, \alpha\inv(\delta))$.
	The transposition operation is simply the restriction of $\alpha$ to morphisms of slice objects.
\end{proposition}
\begin{proof}
	There is a 1-1 correspondence between diagrams
	\begin{equation*}
		\xymatrix{
			Ld
				\ar@{.>}[rr]^{\alpha\inv(\vfi)}
				\ar[rd]_{\alpha\inv(\delta)}
			&&
			c
				\ar[ld]^{\gamma}
			&&&
			d
				\ar@{.>}[rr]^{\vfi}
				\ar[rd]_{\delta}
			&&
			Rc
				\ar[ld]^{R\gamma}
			\\
			&
			c_0
			&&&&&
			Rc_0.
		}\qedhere
	\end{equation*}
\end{proof}

\subsection{Dependent ends and co-ends}
We will use $\forall$ and $\exists$ to denote ends and co-ends as well as their dependent generalizations \cite[\S 2.2.6-7]{nuyts-phd}:
\begin{definition}\label{def:dep-end}
	A \textbf{dependent end} of a functor $F : \twisted \catI \to \catC$, somewhat ambiguously denoted $\forall i . F \twistobj{i}{i}{\id}$, is a limit of $F$.
\end{definition}
\begin{definition}\label{def:dep-coend}
	A \textbf{dependent co-end} of a functor $F : \twisted \catI \op \to \catC$, somewhat ambiguously denoted $\exists i . F \twistobj{i}{i}{\id}$, is a colimit of $F$.
\end{definition}
\begin{example}
	Assume a functor $G : \catC \to \catD$. One way to denote the set of natural transformations $\Id_\catC \to \Id_\catC$ which map to the identity under $G$, is as
	\[
		A := \forall (c \in \catC).\set{\chi : c \to c}{G\chi = \id_{Gc}}.
	\]
	In order to read the above as a dependent end, we must find a functor $G : \twisted \catC \to \Set$ such that $G \twistobj c c \id = \set{\chi : c \to c}{G\chi = \id_{Gc}}$. Clearly every covariant occurrence of $c$ refers to the codomain of $\twistobj c c \id$, whereas every contravariant occurrence refers to the domain. So when we apply $G$ to a general object $\twistobj x y \vfi$ of $\twisted \catC$, we should substitute $x$ for every contravariant $c$ and $y$ for every covariant $c$. We can then throw in $\vfi$ wherever this is necessary to keep things well-typed, as $\vfi$ disappears anyway when $\twistobj x y \vfi = \twistobj c c \id$. Thus, we get
	\[
		G \twistobj x y \vfi = \set{\chi : x \to y}{G\chi = G\vfi}.
	\]
	So we see that using a \emph{dependent} end was necessary in order to mention $\id_c$, as this generalizes to $\vfi : x \to y$ to which we do not have access in a non-dependent end.
	
	An element of $\nu \in A$ is then a function
	\[
		\nu : (c \in \catC) \to \set{\chi : c \to c}{G\chi = \id_{Gc}}
	\] 
	such that, whenever $\vfi : x \to y$, we have $\vfi \circ \nu_x = \nu_y \circ \vfi$.
\end{example}

\subsection{Presheaves}

\subsubsection{Notation} \label{sec:prerequisites:psh:notation}
We use the presheaf notations from \cite{reldtt-techreport}. Concretely:
\begin{itemize}
	\item The application of a presheaf $\Gamma \in \widehat \catW$ to an object $W \in \catW$ is denoted $\DSub W \Gamma$.
	\item The restriction of $\gamma : \DSub W \Gamma$ by $\vfi : V \to W$ is denoted $\gamma \circ \vfi$ or $\gamma \vfi$.
	\item The application of a presheaf morphism $\sigma : \Gamma \to \Delta$ to $\gamma : \DSub W \Gamma$ is denoted $\sigma \circ \gamma$ or $\sigma \gamma$.
	\begin{itemize}
		\item By naturality of $\sigma$, we have $\sigma \circ (\gamma \circ \vfi) = (\sigma \circ \gamma) \circ \vfi$.
	\end{itemize}
	\item If $\Gamma \in \widehat \catW$ and $T \in \Ty(\Gamma)$ (also denoted $\Gamma \sez T \type$), i.e. $T$ is a presheaf over the category of elements $\catW/\Gamma$, then we write the application of $T$ to $(W, \gamma)$ as $(W \Dsez T \dsub \gamma)$ and $t \in (W \Dsez T \dsub \gamma)$ as $W \Dsez t : T \dsub \gamma$.
	\begin{itemize}
		\item By definition of type substitution in a presheaf CwF, we have $(W \Dsez T[\sigma] \dsub \gamma) = (W \Dsez T \dsub{\sigma \gamma})$
	\end{itemize}
	\item The restriction of $W \Dsez t : T \dsub \gamma$ by $\vfi : (V, \gamma \circ \vfi) \to (W, \gamma)$ is denoted as $W \Dsez t \psub \vfi : T \dsub{\gamma \vfi}$.
	\item If $t \in \Tm(\Gamma, T)$ (also denoted $\Gamma \sez t : T$), then the application of $t$ to $(W, \gamma)$ is denoted $V \Dsez t \dsub \gamma : T \dsub \gamma$.
	\begin{itemize}
		\item The naturality condition for terms is then expressed as $t \dsub \gamma \psub \vfi = t \dsub{\gamma \vfi}$.
		\item By definition of term substitution in a presheaf CwF, we have $t[\sigma] \dsub \gamma = t \dsub{\sigma \gamma}$.
	\end{itemize}
	\item We omit applications of the isomorphisms $(\DSub W \Gamma) \cong (\yoneda W \to \Gamma)$ and $(W \Dsez T \dsub \gamma) \cong (\yoneda W \sez T [\gamma])$.
	This is not confusing: e.g. given $W \Dsez t : T \dsub \gamma$, the term $\yoneda W \sez t' : T[\gamma]$ is defined by $t' \dsub \vfi := t \psub \vfi$.
\end{itemize}
One advantage of these notations is that we can put presheaf cells in diagrams; we will use double arrows when doing so.

\subsubsection{On the Yoneda-embedding}
We consider the Yoneda-embedding $\yoneda : \catW \to \widehat \catW$.
\begin{proposition}
	A morphism $\vfi : V \to W$ in $\catW$ is:
	\begin{itemize}
		\item Mono if and only if $\yoneda \vfi$ is mono,
		\item Split epi if and only if $\yoneda \vfi$ is epi.
	\end{itemize}
\end{proposition}
\begin{proof}
	It is well-known that a presheaf morphism $\sigma : \Gamma \to \Delta$ is mono/epi if and only if $\sigma \circ \loch : (\DSub W \Gamma) \to (\DSub W \Delta)$ is injective/surjective for all $W$.
	Now $\yoneda \vfi \circ \loch = \vfi \circ \loch$.
	So $\yoneda \vfi$ is mono if and only if $\vfi \circ \loch$ is injective, which means $\vfi$ is mono.
	On the other hand, $\yoneda \vfi$ is epi if and only if $\vfi \circ \loch$ is surjective, which is the case precisely when $\id$ is in its image, and that exactly means that $\vfi$ is split epi.
\end{proof}

\subsubsection{Lifting functors}
\begin{theorem}\label{thm:adjoint-triple}
	Any functor $F : \catV \to \catW$ gives rise to functors $\lpsh{F} \dashv \fpsh{F} \dashv \rpsh{F}$, with a natural isomorphism $\lpsh F \circ \yoneda \cong \yoneda \circ F : \catV \to \widehat \catW$. We will call $\lpsh{F} : \widehat \catV \to \widehat \catW$ the \textbf{left lifting} of $F$ to presheaves, $\fpsh F : \widehat \catW \to \widehat \catV$ the \textbf{central} and $\rpsh F : \widehat \catV \to \widehat \catW$ the \textbf{right lifting}.%
	\footnote{The central and right liftings are also sometimes called the inverse image and direct image of $F$, but these are actually more general concepts and as such could perhaps cause confusion or unwanted connotations in some circumstances. The left-central-right terminology is very no-nonsense.}%
	\footnote{From the construction, it is evident that $\fpsh F$ is precomposition with $F$ and hence, by definition of Kan extension, $\lpsh F$ and $\rpsh F$ are the left and right Kan extensions of $F$.}
	\cite{stacks-adjoints}
\end{theorem}
\begin{proof}
	Using quantifier symbols for ends and co-ends, we can define:
	\begin{align*}
		\DSub{W}{\lpsh F \Gamma} \quad&:=\quad \exists V . (W \to FV) \times (\DSub V \Gamma), \\
		\DSub{V}{\fpsh F \Delta} \quad&:=\quad \DSub{FV}{\Delta} \\
		\DSub{W}{\rpsh F \Gamma} \quad&:=\quad \forall V . (FV \to W) \to (\DSub V \Gamma) \quad=\quad (\fpsh F \yoneda W \to \Gamma).
	\end{align*}
	By the co-Yoneda lemma, we have:
	\begin{align*}
		\DSub{W}{\lpsh F \yoneda V}
		\quad&=\quad \exists V'.(W \to FV') \times (V' \to V) \quad\cong\quad (W \to FV) \quad=\quad (\DSub W {\yoneda FV}),
	\end{align*}
	i.e. $\lpsh F \yoneda V \cong \yoneda F V$.
	
	Adjointness also follows from applications of the Yoneda and co-Yoneda lemmas.
\end{proof}
\begin{notation} \label{not:adjoint-triple}
	\begin{itemize}
		\item We denote the cell $(V, \vfi, \gamma) : \DSub{W}{\lpsh F \Gamma}$ as $\lpsh F \gamma \circ \vfi$. If we rename $\lpsh F$, then we will also do so in this notation. We will further abbreviate $\lpsh F \gamma \circ \id = \lpsh F \gamma$ and, if $\Gamma = \yoneda V$, also $\lpsh F \id \circ \vfi = \vfi$.
		\item If $\delta : \DSub{FV}{\Delta}$, then we write $\alpha_F(\delta) : \DSub{V}{\fpsh F \Delta}$.
		\item If $\gamma : {\fpsh F \yoneda W} \to {\Gamma}$, then we write $\beta_F(\gamma) : \DSub{W}{\rpsh F \Gamma}$.
	\end{itemize}
\end{notation}

\begin{proposition} \label{thm:left-lifting-ff}
	A functor $F : \catV \to \catW$ is fully faithful if and only if $\lpsh F$ is fully faithful.
\end{proposition}
\begin{proof}
	To see the implication from left to right:
		It is a standard fact of adjoint functors \cite{nlab:adjoint-functor} that the left adjoint $\lpsh F$ is fully faithful if and only if $\eta : \Gamma \to \fpsh F \lpsh F \Gamma$ is a natural isomorphism. If $F$ is fully faithful, then we can apply the co-Yoneda lemma:
		\begin{equation*}
			(\DSub V {\fpsh F \lpsh F \Gamma})
			= (\exists V' . (FV \to FV') \times (\DSub{V'}{\Gamma}))
			\cong (\exists V' . (V \to V') \times (\DSub{V'}{\Gamma}))
			\cong (\DSub{V}{\Gamma})
		\end{equation*}
		i.e.\ $\fpsh F \lpsh F \Gamma \cong \Gamma$ and it is straightforward to see that this isomorphism is indeed the co-unit.
	
	The implication from right to left is straightforward. By full faithfulness of $\yoneda$ and by \cref{thm:adjoint-triple} we have
	\begin{align*}
		(\yoneda U \to \yoneda V) &\cong (U \to V), \\
		(\lpsh F \yoneda U \to \lpsh F \yoneda V)
		&\cong (\yoneda FU \to \yoneda FV)
		\cong (FU \to FV). & \qedhere
	\end{align*}
\end{proof}

\subsubsection{Dependent presheaf categories}
Let $\catW$ be a category. Then $\widehat \catW$ is a category with families (CwF). The following notion is standard:
\begin{definition}
	For any $\Gamma \in \widehat{\catW}$, the \textbf{category of elements} of $\Gamma$, denoted
	\begin{equation}
		\int_\catW \Gamma \quad \textor \quad \catW / \Gamma
	\end{equation}
	has objects $(W, \gamma)$ where $W \in \catW$ and $\gamma : \DSub W \Gamma$, and the morphisms $(W, \gamma) \to (W', \gamma')$ are the morphisms $\chi : W \to W'$ such that $\gamma' \circ \chi = \gamma$.
\end{definition}
Clearly, we have an isomorphism $\catW/U \cong \catW/\yoneda U$ between the slice category over $U$ and the category of elements of $\yoneda U$.\footnote{Depending on pedantic details, we may even have $\catW/U = \catW/\yoneda U$.}

We will use type-theoretic notation to make statements about the CwF $\widehat \catW$, e.g. $\Gamma \ctx$ means $\Gamma \in \widehat \catW$ and $\Gamma \sez T \type$ means $T \in \Ty(\Gamma)$. Now for any context or closed type $\Gamma \in \widehat \catW$, there is another CwF $\widehat{\catW/\Gamma}$. Statements about this category will also be denoted using type-theoretic notation, but prefixed with `$\Gamma \sep$'.

By unfolding the definitions of types and terms in a presheaf CwF, it is trivial to show that there is a correspondence --- which we will treat as though it were the identity --- between both CwFs:
\begin{itemize}
	\item Contexts $\Gamma \sep \Theta \ctx$ correspond to types $\Gamma \sez \Theta \type$ which we will think of as telescopes $\Gamma.\Theta \sez \Ctx$,
	\item Substitutions $\Gamma \sep \sigma : \Theta \to \Theta'$ correspond to functions $\Gamma \sez \sigma : \Theta \to \Theta'$, or equivalently to telescope substitutions $\id_\Gamma.\sigma : \Gamma.\Theta \to \Gamma.\Theta'$,
	\item Types $\Gamma \sep \Theta \sez T \type$ correspond to types $\Gamma.\Theta \sez T \type$,
	\item Terms $\Gamma \sep \Theta \sez t : T$ correspond to terms $\Gamma.\Theta \sez t : T$.
\end{itemize}
In summary, the pipe is equivalent to a dot.

\begin{proposition}\label{thm:preimage}
	We have an equivalence of categories $\widehat{\catW / \Gamma} \simeq \widehat \catW / \Gamma$.
\end{proposition}
\begin{proof}
	\begin{enumerate}
		\item[$\rightarrow$] We map the presheaf $\Gamma \sep \Theta \ctx$ to the slice object $(\Gamma.\Theta, \pi)$.
		\item[$\leftarrow$] We map the slice object $(\Delta, \sigma)$ to the preimage of $\sigma$, i.e. the presheaf $\sigma\inv$ which sends $(W, \gamma)$ to $\set{\delta : \DSub W \Delta}{\sigma \circ \delta = \gamma}$.
		\item[$\widehat{\catW/\Gamma}$] We need a natural isomorphism $\eta : \forall \Theta . (\Gamma \sep \eta : \Theta \cong \pi\inv)$. If $\theta : \DSub{(W, \gamma)}{\Theta}$, then we define $\eta(\theta) = (\gamma, \theta) : \DSub{W}{\Gamma.\Theta}$ and indeed we have $\pi \circ (\gamma, \theta) = \gamma$. This is clearly invertible.
		\item[$\widehat \catW / \Gamma$] We need a natural isomorphism $\eps : \forall (\Delta, \sigma).(\Gamma.\sigma\inv, \pi) \cong (\Delta, \sigma)$. Given $(\gamma, \delta) : \DSub{W}{\Gamma.\sigma\inv}$ (i.e. we know $\sigma \circ \delta = \gamma$), we define $\eps \circ (\gamma, \delta) = \delta : \DSub{W}{\Delta}$. Then
		\begin{equation}
			\sigma \circ \eps \circ (\gamma, \delta) = \sigma \circ \delta = \gamma = \pi \circ (\gamma, \delta),
		\end{equation}
		so indeed we have a morphism in the slice category. It is inverted by sending $\delta : \DSub{W}{\Delta}$ to $(\sigma \circ \delta, \delta) : \DSub{W}{\Gamma.\sigma\inv}$. \qedhere
	\end{enumerate}
\end{proof}
\begin{corollary}
	We have $\widehat{\catW / U} \cong \widehat{\catW / \yoneda U} \simeq \widehat \catW / \yoneda U$. \qed
\end{corollary}

\subsubsection{Substitution and its adjoints} \label{sec:subst-adjoints}
\begin{definition}
	Given $U \in \catW$, we write
	\begin{itemize}
		\item $\pairbase U : \catW / U \to \catW : (W, \psi) \mapsto W$,
		\item $\wknbase U : \catW \to \catW / U : W \to (W \times U, \pi_2)$ (if $\catW$ has cartesian products with $U$).
	\end{itemize}
\end{definition}
\begin{proposition} \label{thm:pair-wkn}
	If $\wknbase U$ exists, then $\pairbase U \dashv \wknbase U$. We denote the unit as $\copybase U : \Id \to \wknbase U \pairbase U$ and the co-unit as $\dropbase U : \pairbase U \wknbase U \to \Id$. \qed
\end{proposition}
\begin{proposition} \label{thm:wknbase-faithful}
	\begin{enumerate}
		\item If $U \to \top$ is split epi, then the functor $\wknbase U$ is faithful.
		\item (Not used). If $U \to \top$ is mono, then $\pairbase U$ is full.\footnote{An earlier version asserted fullness of $\wknbase U$ instead, but proved the current theorem.}
	\end{enumerate}
\end{proposition}
\begin{proof}
	\begin{enumerate}
		\item We have some $\upsilon : \top \to U$, so that the action of $\wknbase U$ on morphisms sending $\vfi \mapsto \vfi \times U$ can be inverted: $\vfi = \pi_1 \circ (\vfi \times U) \circ (\id, \upsilon)$.
		\item Take slice objects $(W_1, \psi_1)$ and $(W_2, \psi_2)$ and a morphism $\vfi : W_1 \to W_2$. The fact that $U \to \top$ is mono just means that morphisms to $U$ are unique if existent. Then $\vfi$ is also a morphism between the slice objects. \qedhere
	\end{enumerate}
\end{proof}
\begin{definition}
	Given $\chi : W_0' \to W_0$ in $\catW$, we write
	\begin{itemize}
		\item $\pairslice{}{\chi} : \catW / W_0' \to \catW / W_0 : (W', \psi') \mapsto (W', \chi \circ \psi')$,
		\item $\wknslice{}{\chi} : \catW / W_0 \to \catW / W_0'$ for the functor that maps $(W, \psi)$ to its pullback along $\chi$ (if $\catW$ has pullbacks along $\chi$).
	\end{itemize}
	If $\chi = \pi_1 : W_0 \times U \to W_0$, we also write $\pairslice{U}{W_0} : \catW / (W_0 \times U) \to \catW / W_0$ and $\wknslice{U}{W_0} : \catW / W_0 \to \catW / (W_0 \times U)$.
\end{definition}
\begin{proposition}
	If $\wknslice{}{\chi}$ exists, then $\pairslice{}{\chi} \dashv \wknslice{}{\chi}$. We denote the unit as $\copyslice{}{\chi} : \Id \to \wknslice{}{\chi} \pairslice{}{\chi}$ and the co-unit as $\dropslice{}{\chi} : \pairslice{}{\chi} \wknslice{}{\chi} \to \Id$. \qed
\end{proposition}
\begin{proposition}[Ultimately not used] \label{thm:subst-ff-slice}
	\begin{enumerate}
		\item If $\chi$ is split epi, then $\wknslice{}{\chi}$ is faithful.
		\item If $\chi$ is mono, then $\pairslice{}{\chi}$ is full.\footnote{An earlier version asserted fullness of $\wknslice{}{\chi}$ instead, but proved the current theorem.}
	\end{enumerate}
\end{proposition}
\begin{proof}
	\begin{enumerate}
		\item We have some $\upsilon : W_0 \to W_0'$ such that $\chi \circ \upsilon = \id$. Then the action of $\wknslice{}{\chi}$ on morphisms sending $\vfi \mapsto \vfi \times_{W_0} W_0'$ can be inverted: given $\vfi : (W_1, \psi_1) \to (W_2, \psi_2) \in \catW/W_0$, we have
		\begin{equation}
			\vfi : W_1 \xrightarrow{(\id, \upsilon \circ \psi_1)} W_1 \times_{W_0} W_0' \xrightarrow{\vfi \times_{W_0} W_0'} W_2 \times_{W_0} W_0' \xrightarrow{\pi_1} W_2.
		\end{equation}
	
		\item Take a morphism $\vfi : (W_1, \chi \circ \psi_1) \to (W_2, \chi \circ \psi_2)$. Then $\chi \circ \psi_2 \circ \vfi = \chi \circ \psi_1$. Because $\chi$ is mono, this implies that $\psi_2 \circ \vfi = \psi_1$, i.e. $\vfi : (W_1, \psi_1) \to (W_2, \psi_2)$. \qedhere
	\end{enumerate}
\end{proof}
\begin{definition}
	Given $\sigma : \Psi' \to \Psi$ in $\widehat \catW$, we write
	\begin{itemize}
		\item $\pairslice{}{\sigma} : \catW / \Psi' \to \catW / \Psi : (W', \psi') \mapsto (W', \sigma \circ \psi')$,
		\item $\wknslice{}{\sigma} : \catW / \Psi \to \catW / \Psi'$ for the functor that maps $(W, \psi)$ to its pullback along $\sigma$ (if $\catW$ has pullbacks along $\sigma$), by which we mean a universal solution $W'$ to the diagram
		\begin{equation}
			\xymatrix{
				W' \ar@{.>}[r] \ar@{:>}[d]
				& W \ar@{=>}[d]^{\psi}
				\\
				\Psi' \ar[r]_{\sigma}
				& \Psi.
			}
		\end{equation}
	\end{itemize}
	If $\sigma = \pi_1 : \Psi \times \Phi \to \Psi$, we also write $\pairslice{\Phi}{\Psi} : \catW / (\Psi \times \Phi) \to \catW / \Psi$ and $\wknslice{\Phi}{\Psi} : \catW / \Psi \to \catW / (\Psi \times \Phi)$.
\end{definition}
\begin{proposition}
	If $\wknslice{}{\sigma}$ exists, then $\pairslice{}{\sigma} \dashv \wknslice{}{\sigma}$. We denote the unit as $\copyslice{}{\sigma} : \Id \to \wknslice{}{\sigma} \pairslice{}{\sigma}$ and the co-unit as $\dropslice{}{\sigma} : \pairslice{}{\sigma} \wknslice{}{\sigma} \to \Id$. \qed
\end{proposition}
\begin{proposition}[Not used] \label{thm:subst-ff-elements}
	\begin{enumerate}
		\item If $\sigma$ is surjective, then $\wknslice{}{\sigma}$ is faithful.
		\item If $\sigma$ is injective, then $\pairslice{}{\sigma}$ is full.\footnote{An earlier version asserted fullness of $\wknslice{}{\sigma}$ instead, but proved the current theorem.}
	\end{enumerate}
\end{proposition}
\begin{proof}
	\begin{enumerate}
		\item If $\sigma$ is surjective, then by the axiom of choice, there is at least a non-natural $f : \Psi \to \Psi'$ such that $\sigma \circ f = \id$. The rest of the proof is as for \cref{thm:subst-ff-slice}.
		\item Same as for \cref{thm:subst-ff-slice}. \qedhere
	\end{enumerate}
\end{proof}
\begin{definition} \label{def:4-cart-functors}
	The functors $\pairslice{}{\sigma} \dashv \wknslice{}{\sigma}$ give rise to four adjoint functors
	\begin{equation}
		\pairpsh{}{\sigma} \dashv \wknpsh{}{\sigma} \dashv \funcpsh{}{\sigma} \dashv \cartransppsh{}{\sigma}
	\end{equation}
	between $\widehat{\catW / \Psi}$ and $\widehat{\catW / \Psi'}$,
	of which the first three exist if only $\pairslice{}{\sigma}$ exists.%
	\footnote{The latter functor is already a cartesian transpension functor; however we have not guaranteed its existence. Later on we will discuss a transpension functor for certain -- not necessarily cartesian -- \emph{shapes}, modelled by \emph{multipliers}, and there we will guarantee existence.}
	
	The units and co-units will be denoted:
	\begin{equation}
		\begin{array}{r c l c r c l}
			\copypsh{}{\sigma} &:& \Id \to \wknpsh{}{\sigma} \pairpsh{}{\sigma}
			&\qquad \qquad&
			\droppsh{}{\sigma} &:& \pairpsh{}{\sigma} \wknpsh{}{\sigma} \to \Id
			\\
			\constpsh{}{\sigma} &:& \Id \to \funcpsh{}{\sigma} \wknpsh{}{\sigma}
			&\qquad \qquad&
			\apppsh{}{\sigma} &:& \wknpsh{}{\sigma} \funcpsh{}{\sigma} \to \Id
			\\
			\reindexpsh{}{\sigma} &:& \Id \to \cartransppsh{}{\sigma} \funcpsh{}{\sigma}
			&\qquad \qquad&
			\unmeridpsh{}{\sigma} &:& \funcpsh{}{\sigma} \cartransppsh{}{\sigma} \to \Id
		\end{array}
	\end{equation}
\end{definition}
We remark that, if we read presheaves over $\catW / \Psi$ as types in context $\Psi$,
then $\wknpsh{}{\sigma} : \widehat{\catW / \Psi} \to \widehat{\catW / \Psi'}$ is the standard interpretation of substitution in a presheaf category. If $\sigma = \pi : \Psi.A \to \Psi$ is a weakening morphism, then $\wknpsh{A}{\Psi} := \wknpsh{}{\pi}$ is the weakening substitution, $\funcpsh{A}{\Psi} := \funcpsh{}{\pi} : \widehat{\catW / \Psi.A} \to \widehat{\catW / \Psi}$ is isomorphic to the standard interpretation of the $\Pi$-type and $\pairpsh{A}{\Psi} := \pairpsh{}{\pi} : \widehat{\catW / \Psi.A} \to \widehat{\catW / \Psi}$ is isomorphic to the standard interpretation of the $\Sigma$-type.

\begin{theorem} \label{thm:commut-subst-subst}
	Given types $\Psi \sez A, B \type$, the projections constitute a pullback diagram:
	\begin{equation}
		\xymatrix{
			\Psi.(A \times B)
				\ar[r]^{\beta'}
				\ar[d]_{\alpha'}
			& \Psi.A \ar[d]^\alpha
			\\
			\Psi.B \ar[r]_\beta
			& \Psi,
		}
	\end{equation}
	and every pullback diagram in a presheaf category is isomorphic to a diagram of this form.
	We have the following commutation properties:
	\begin{equation}
		\begin{array}{c || c | c | c | c}
			& \pairbase B & \wknbase B & \funcbase B & \cartranspbase B
			\\ \hline \hline
			\pairbase A
			& \pairpsh{}{\alpha} \pairpsh{}{\beta'} \cong \pairpsh{}{\beta} \pairpsh{}{\alpha'}
			& \pairpsh{}{\alpha'} \wknpsh{}{\beta'} \cong \wknpsh{}{\beta} \pairpsh{}{\alpha}
			& \pairpsh{}{\alpha} \funcpsh{}{\beta'} \to \funcpsh{}{\beta} \pairpsh{}{\alpha'}
			&
			\\ \hline
			\wknbase A
			& \wknpsh{}{\alpha} \pairpsh{}{\beta} \cong \pairpsh{}{\beta'} \wknpsh{}{\alpha'}
			& \wknpsh{}{\alpha'} \wknpsh{}{\beta} = \wknpsh{}{\beta'} \wknpsh{}{\alpha}
			& \wknpsh{}{\alpha} \funcpsh{}{\beta} \cong \funcpsh{}{\beta'} \wknpsh{}{\alpha'}
			& \wknpsh{}{\alpha'} \cartransppsh{}{\beta} \to \cartransppsh{}{\beta'} \wknpsh{}{\alpha}
			\\ \hline
			\funcbase A
			& \funcpsh{}{\alpha} \pairpsh{}{\beta'} \leftarrow \pairpsh{}{\beta} \funcpsh{}{\alpha'}
			& \funcpsh{}{\alpha'} \wknpsh{}{\beta'} \cong \wknpsh{}{\beta} \funcpsh{}{\alpha}
			& \funcpsh{}{\alpha} \funcpsh{}{\beta'} \cong \funcpsh{}{\beta} \funcpsh{}{\alpha'}
			& \funcpsh{}{\alpha'} \cartransppsh{}{\beta'} \cong \cartransppsh{}{\beta} \funcpsh{}{\alpha}
			\\ \hline
			\cartranspbase A
			&
			& \cartransppsh{}{\alpha'} \wknpsh{}{\beta} \leftarrow \wknpsh{}{\beta'} \cartransppsh{}{\alpha}
			& \cartransppsh{}{\alpha} \funcpsh{}{\beta} \cong \funcpsh{}{\beta'} \cartransppsh{}{\alpha'}
			& \cartransppsh{}{\alpha'} \cartransppsh{}{\beta} \cong \cartransppsh{}{\beta'} \cartransppsh{}{\alpha}
		\end{array}
	\end{equation}
	where every statement holds if the mentioned functors exist.
\end{theorem}
\begin{proof}
	In the base category, it is evident that $\pairslice{}{\alpha} \pairslice{}{\beta'} = \pairslice{}{\beta} \pairslice{}{\alpha'}$. By applying the functor $\fpsh \loch$, we obtain $\wknpsh{}{\alpha'} \wknpsh{}{\beta} = \wknpsh{}{\beta'} \wknpsh{}{\alpha}$, whence by \cref{thm:commut-adj} the entire diagonal of the commutation table.
	
	It is a well-known fact that $\Sigma$- and $\Pi$-types are respected by substitution, which gives us the isomorphisms for swapping $\Omega$ and either $\Sigma$ or $\Pi$. \Cref{thm:commut-adj} then gives the rest.
\end{proof}

\begin{theorem}
	Given $\sigma : \Psi' \to \Psi$, the following operations are invertible:
	\begin{equation}
		\inference{
			\Psi \sep \pairpsh{}{\sigma} \Gamma \sez T \type
		}{
			\Psi' \sep \Gamma \sez (\wknpsh{}{\sigma} T)[\copypsh{}{\sigma}] \type
		}{}
		\qquad
		\inference{
			\Psi \sep \pairpsh{}{\sigma} \Gamma \sez t : T
		}{
			\Psi' \sep \Gamma \sez (\ftrtm{\wknpsh{}{\sigma}}{t})[\copypsh{}{\sigma}] : (\wknpsh{}{\sigma} T)[\copypsh{}{\sigma}]
		}{}
	\end{equation}
\end{theorem}
\begin{proof}
	Note that $T$ is a presheaf over $(\catW/\Psi)/\pairpsh{}{\sigma} \Gamma$, and $(\wknpsh{}{\sigma} T)[\copypsh{}{\sigma}]$ is a presheaf over $(\catW/\Psi')/\Gamma$. We compare the objects of these categories:
	\begin{align*}
		&\Obj((\catW/\Psi)/\pairpsh{}{\sigma} \Gamma) \\
		&= (W \in \catW) \times (\psi : \DSub{W}{\Psi}) \times \exists ((W', \psi') \in \catW/\Psi' . (\chi : (W, \psi) \to \pairslice{}{\sigma}(W', \psi')) \times (\DSub{(W', \psi')}{\Gamma}) \\
		&\cong (W \in \catW) \times (\psi : \DSub{W}{\Psi}) \times \exists W' . (\psi' : \DSub{W'}{\Psi'}) \times (\chi : (W, \psi) \to \pairslice{}{\sigma}(W', \psi')) \times (\DSub{(W', \psi')}{\Gamma}) \\
		&\cong (W \in \catW) \times (\psi : \DSub{W}{\Psi}) \times \exists W' . (\psi' : \DSub{W'}{\Psi'}) \times (\chi : (W, \psi) \to (W', \sigma \circ \psi')) \times (\DSub{(W', \psi')}{\Gamma}) \\
		&\cong (W \in \catW) \times \exists W' . (\psi' : \DSub{W'}{\Psi'}) \times (\chi : W \to W') \times (\DSub{(W', \psi')}{\Gamma}) \\
		& \qquad \text{because $\chi$ is a slice morphism iff $\psi = \sigma \circ \psi' \circ \chi$} \\
		&\cong (W \in \catW) \times (\psi' : \DSub{W}{\Psi'}) \times (\DSub{(W, \psi')}{\Gamma}) \\
		&\cong \Obj((\catW/\Psi')/\Gamma).
	\end{align*}
	A similar consideration of the $\Hom$-sets leads to the conclusion that both categories are isomorphic. Moreover, we remark that the isomorphism sends $((W, \psi'), \gamma)$ on the right to $((W, \sigma \circ \psi'), \pairpsh{}{\sigma} \gamma)$ on the left. When we consider the action of $(\wknpsh{}{\sigma} T)[\copypsh{}{\sigma}]$ on $((W, \psi'), \gamma)$, we find:
	\begin{align*}
		\paren{(W, \psi') \Dsez (\wknpsh{}{\sigma} T)[\copypsh{}{\sigma}] \dsub{\gamma}}
		&= \paren{\pairslice{}{\sigma}(W, \psi') \Dsez T \dsub{\pairpsh{}{\sigma} \gamma}} \\
		&= \paren{(W, \sigma \circ \psi') \Dsez T \dsub{\pairpsh{}{\sigma} \gamma}}
	\end{align*}
	In other words, the types $T$ and $(\wknpsh{}{\sigma} T)[\copypsh{}{\sigma}]$ are equal over an isomorphism of categories. Then certainly $T$ can be retrieved from $(\wknpsh{}{\sigma} T)[\copypsh{}{\sigma}]$. An identical argument works for terms.
\end{proof}

\subsubsection{Reconstructing right adjoints}
\begin{proposition}
	Given a left adjoint functor $L : \widehat \catW \to \catC$, we can construct a right adjoint $R_L : \catC \to \widehat \catW$ without using the axiom of choice.
\end{proposition}
\begin{proof}
	Define $(\DSub{W}{R_L \Gamma}) := (L\yoneda W \to \Gamma)$.
	As a matter of notational hygiene, write $\alpha_L : (L\yoneda W \to \Gamma) \to (\DSub{W}{R_L \Gamma})$ for the identity function.
	Define restriction by $\alpha_L(\gamma) \circ \vfi = \alpha_L(\gamma \circ L\yoneda \vfi)$ and the functorial action by $R_L \sigma \circ \alpha_L(\gamma) = \alpha_L(\sigma \circ \gamma)$.
	This is a well-defined presheaf functor.
	
	Now we show that $L \dashv R_L$. Since $L$ is a left adjoint, it has a right adjoint $R$. We have natural isomorphisms
	\begin{align*}
		(\DSub{W}{R_L \Gamma}) = (L\yoneda W \to \Gamma) \cong (\yoneda W \to R\Gamma) \cong (\DSub W {R\Gamma})
	\end{align*}
	so that $R_L$ is naturally isomorphic to $R$ and indeed right adjoint to $L$.
\end{proof}

\section{Multipliers in the base category} \label{sec:multipliers}

\subsection{Definition}
\begin{definition}\label{def:pointable}
	Let $\catW$ be a category with terminal object $\top$. An object $W$ is \textbf{pointable}\seelog{} if $() : W \to \top$ is split epi. A category is \textbf{objectwise pointable}\seelog{} if every object is pointable.
\end{definition}
We have carefully chosen the above terminology to emphasize (1) that pointability is a property, not structure (the corresponding structure is called \emph{pointed}), and (2) that objectwise pointability does \emph{not} require that the pointings can be chosen naturally.
\begin{definition} \label{def:multiplier}
	Let $\catW$ be a category with terminal object $\top$. A \textbf{multiplier} for an object $U \in \catV$ is a functor $\loch \multip U : \catW \to \catV$ such that $\top \multip U \cong U$.%
	\footnote{$\loch \multip U$ is to be regarded as a single-character symbol, i.e.\  $\multip$ in itself is meaningless.
In most concrete applications, however, the multiplier is defined as some monoidal product $\loch \otimes U$ with a given object $U$.
For this reason, we also refrain from defining $U := \top \multip U$ because we may not have $\top \otimes U = U$ on the nose for the object of interest $U$.}
	This gives us a second projection $\pi_2 : \forall W . W \multip U \to U$.
	We define the \textbf{fresh weakening functor} as $\freshbase{U} : \catW \to \catV/U : W \mapsto (W \multip U, \pi_2)$, which is essentially the action of the multiplier on slice objects over $\top$.
	
	We say that a multiplier is:
	\begin{itemize}
		\item \textbf{Endo} if it is an endofunctor (i.e. $\catV = \catW$), and in that case:
		\begin{itemize}
			\item \textbf{Copointed}\seelog{} if there is also a first projection $\pi_1 : \forall W . W \multip U \to W$,
			\item A \textbf{comonad}\seelog{} if there is additionally a `diagonal' natural transformation $\loch \multip \delta : \forall W . W \multip U \to (W \multip U) \multip U$ such that $\pi_1 \circ (W \multip \delta) = (\pi_1 \multip U) \circ (W \multip \delta) = \id$.
			\item \textbf{Cartesian} if it satisfies the universal property of the cartesian product with $U$,
		\end{itemize}
		\item \textbf{$\top$-slice faithful}\seelog{} if $\freshbase{U}$ is faithful, or equivalently (\cref{thm:tsfaithful-char}) if $\loch \multip U$ is faithful,
		\item \textbf{$\top$-slice full}\seelog{} if $\freshbase{U}$ is full,
		\item \textbf{$\top$-slice objectwise pointable}\seelog{} if $\pi_2 : W \multip U \to U$ is always split epi, and in that case:
		\begin{itemize}
			\item \textbf{$\top$-slice shard-free}\seelog{} if $\freshbase{U}$ is essentially surjective on objects $(V, \psi)$ such that $\psi$ is split epi, i.e. if every such object in $\catV/U$ is isomorphic to some $\freshbase{U} W$.
			
			\item A split epi slice object $(V, \psi)$ that is not in the image of $\freshbase{U}$ even up to isomorphism, will be called a \textbf{shard} of the multiplier.
		\end{itemize}
		\item \textbf{$\top$-slice right adjoint}\seelog{} if $\freshbase{U}$ has a left adjoint $\sumbase U : \catV/U \to \catW$.%
		\footnote{A functor $\loch \multip U$ with this property is usually called a \emph{parametric} or \emph{local right adjoint} \cite{nlab:parametric-right-adjoint}, but the word `local' is overloaded \cite{nlab:locally} and so is `parametric', and we wanted uniform terminology.}
		We denote the unit as $\copybase U : \Id \to \freshbase U \sumbase U$ and the co-unit as $\dropbase U : \sumbase U \freshbase U \to \Id$.
	\end{itemize}
\end{definition}

\subsection{Basic properties}
Some readers may prefer to first consult some examples (\cref{sec:multipliers:ex}).
\begin{proposition}
	For any multiplier, we have $(\loch \multip U) = \pairbase U \circ \freshbase U$. \qed
\end{proposition}
\begin{lemma} \label{thm:tsfaithful-char}
	The functor $\loch \multip U$ is faithful if and only if $\freshbase{U}$ is faithful.
\end{lemma}
\begin{proof}
	We have $(\loch \multip U) = \pairbase U \circ \freshbase U$ and $\pairbase U : \catV/U \to \catV$ is faithful as is obvious from its definition.
\end{proof}

\begin{proposition}
	A multiplier with an objectwise pointable domain is $\top$-slice objectwise pointable.
\end{proposition}
\begin{proof}
	The multiplier, as any functor, preserves split epimorphisms.
\end{proof}

\begin{proposition}
	Cartesian endomultipliers are comonads, and comonads are copointed. \qed
\end{proposition}
\begin{proposition} \label{thm:cartesian-tsra}
	Cartesian endomultipliers are $\top$-slice right adjoint.
\end{proposition}
\begin{proof}
	The left adjoint to $\freshbase U = \wknbase U$ is then given by $\sumbase U (V, \vfi) = \pairbase U (V, \vfi) = V$ (\cref{thm:pair-wkn}).
\end{proof}

\begin{proposition}\label{thm:pointable-tsfaitful}
	Cartesian endomultipliers for pointable objects, are $\top$-slice faithful.
\end{proposition}
\noindent Pointability is not required however: cartesian endomultipliers for unpointable objects may be $\top$-slice faithful (\cref{ex:cartesian-cubes,ex:clocks}).
\begin{proof}
	In this case, $\freshbase U = \wknbase U$ and $U \to \top$ is split epi, so this is part of \cref{thm:wknbase-faithful}.
\end{proof}

Being $\top$-slice full expresses absence of diagonals in the following sense:
\begin{proposition}\label{thm:tsfull-cartesian}
	If an endomultiplier for $U$ is both a comonad and $\top$-slice full, then $U$ is a terminal object. If the endomultiplier is moreover cartesian, then it is naturally isomorphic to the identity functor.
\end{proposition}
\begin{proof}
	Consider the following diagram:
	\begin{equation}
		\xymatrix{
			\top \multip U \ar[rr]^{\top \multip \delta}
				\ar[rd]_{\pi_2}
			&& (\top \multip U) \multip U
				\ar[ld]^{\pi_2}
			\\
			& U
		}
	\end{equation}
	This is a morphism of slice objects $\top \multip \delta : \freshbase{U} \top \to \freshbase{U} (\top \multip U)$ and thus, by fullness of $\freshbase{U}$, of the form $\freshbase{U} \upsilon$ for some $\upsilon : \top \to \top \multip U$. This means in particular that
	\begin{equation}
		\id_{\top \multip U} = \pi_1 \circ (\top \multip \delta) = \pi_1 \circ (\upsilon \multip U) = \upsilon \circ \pi_1 : \top \multip U \to \top \multip U.
	\end{equation}
	Composing on both sides with $\pi_2 : \top \multip U \cong U$, we find that $\id_U = (\pi_2 \circ \upsilon) \circ (\pi_1 \circ \pi_2\inv)$ factors over $\top$, which means exactly that $\pi_2 \circ \upsilon : \top \to U$ and $\pi_1 \circ \pi_2\inv : U \to \top$ constitute an isomorphism, i.e. $U$ is terminal.
	
	If $\loch \multip U$ is cartesian, then it is a cartesian product with a terminal object and therefore naturally isomorphic to the identity functor.
\end{proof}

\subsection{Examples}\label{sec:multipliers:ex}
\begin{example}[Identity] \label{ex:identity-multiplier}
	The identity functor $W \multip \top := W$ is an endomultiplier for $\top$.
	
	It is cartesian, $\top$-slice fully faithful, $\top$-slice objectwise pointable iff $\catW$ is objectwise pointable and in that case $\top$-slice shard-free, and $\top$-slice right adjoint.
	
	The functor $\freshbase \top : \catW \to \catW / \top : W \mapsto (W, ())$ has a left adjoint $\sumbase \top : \catW/\top \to \catW : (W, ()) \mapsto W$.
\end{example}
\begin{example}[Cartesian product] \label{ex:cartesian}
	Let $\catW$ be a category with finite products and $U \in \catW$.
	
	Then $\loch \times U$ is an endomultiplier for $U$.
	
	It is cartesian, $\top$-slice faithful if (but not only if) $U$ is pointable (\cref{thm:pointable-tsfaitful}), $\top$-slice full if and only if $U \cong \top$ (\cref{thm:tsfull-cartesian}) and $\top$-slice right adjoint (\cref{thm:cartesian-tsra}). We do not consider $\top$-slice objectwise pointability for this general case.
	
	The functor $\freshbase{U} = \wknbase U : V \mapsto (V \times U, \pi_2)$ has a left adjoint $\sumbase U = \pairbase U : (W, \psi) \mapsto W$. Hence, we have $\sumbase U \freshbase{U} = \loch \times U$.
\end{example}
\begin{example}[Affine cubes] \label{ex:affine-cubes}
	Let $\affinecubecat k$ be the category of affine non-symmetric $k$-ary cubes $\IX^n$ as used in \cite{model-cubical} (binary) or \cite{moulin-param3} (unary). A morphism $\vfi : \IX^m \to \IX^n$ is a function $\loch\psub{\vfi} : \accol{i_1, \ldots, i_n} \to \accol{i_1 \ldots i_m, 0, \ldots, k-1}$ such that $i\psub{\vfi} = j\psub{\vfi} \not \in \accol{0, \ldots, k-1}$ implies $i = j$.
	We also write $\vfi = (i_1 \psub{\vfi}/i_1, \ldots, i_n \psub \vfi / i_n)$.
	This category is objectwise pointable if and only if $k > 0$.
	
	Consider the functor $\loch * \IX : \Box^k \to \Box^k : \IX^n \mapsto \IX^{n+1}$, which is a multiplier for $\IX$. It acts on morphisms $\vfi : \IX^m \to \IX^n$ by setting $\vfi * \IX = (\vfi, i_{m+1}/i_{n+1})$.
	
	It is straightforwardly seen to be copointed, not a comonad, $\top$-slice fully faithful, $\top$-slice objectwise pointable iff $k \neq 0$ and in that case $\top$-slice shard-free, and $\top$-slice right adjoint.
	
	The functor $\freshbase \IX : \IX^n \mapsto (\IX^{n+1}, (i_{n+1}/i_1))$ has as left adjoint the functor $\sumbase \IX$ which sends $(\IX^{n}, \psi)$ to $\IX^n$ if $i_1 \psub{\psi} \in \accol{0, \ldots, k-1}$ and to $\IX^{n-1}$ (by removing the variable $i_1 \psub{\psi}$ and renaming the next ones) otherwise. The action on morphisms is straightforwardly constructed.
	
	In the case where $k = 2$, we can throw in an involution $\lnot : \IX \to \IX$. This changes none of the above results, except that $i_1 \psub{\psi}$ may be the negation $\lnot j$ of a variable $j$, in which case $\sumbase U$ removes the variable $j$.
\end{example}
\begin{example}[Cartesian cubes] \label{ex:cartesian-cubes}
	Let $\boxslash^k$ be the category of cartesian non-symmetric $k$-ary cubes $\IX^n$. A morphism $\vfi : \IX^m \to \IX^n$ is any function $\loch\psub{\vfi} : \accol{i_1, \ldots, i_n} \to \accol{i_1 \ldots i_m, 0, \ldots, k-1}$. This category is objectwise pointable if and only if $k > 0$.
	
	Consider the functor $\loch \times \IX : \boxslash^k \to \boxslash^k : \IX^n \mapsto \IX^{n+1}$, which is an endomultiplier for $\IX$.
	
	It is cartesian (hence $\top$-slice non-full and right adjoint with $\sumbase \IX(W, \psi) = W$), $\top$-slice full, $\top$-slice objectwise pointable iff $k > 0$ and in that case $\top$-slice shard-free.
	
	Again, involutions change none of the above results.
\end{example}
\begin{example}[CCHM cubes]
	Let $\boxslash_{\vee, \wedge, \lnot}$ be the category of (binary) CCHM cubes \cite{cubical}. What's special here is that we have connections $\vee, \wedge : \IX^2 \to \IX$ (as well as involutions). This category is objectwise pointable.
	
	Again, we consider the functor $\loch \times \IX : \boxslash_{\vee, \wedge, \lnot} \to \boxslash_{\vee, \wedge, \lnot} : \IX^n \mapsto \IX^{n+1}$, which is an endomultiplier for $\IX$.
	
	It is cartesian (hence $\top$-slice non-full and right adjoint with $\sumbase \IX(W, \psi) = W$), $\top$-slice faithful and $\top$-slice objectwise pointable but not shard-free (since $(\IX^2, \vee)$ and $(\IX^2, \wedge)$ are shards).
\end{example}
\begin{example}[Clocks] \label{ex:clocks}
	Let $\clocksym$ be the category of clocks, used as a base category in guarded type theory \cite{clock-cat}.
	Its objects take the form $(i_1 : \clocksym_{k_1}, \ldots, i_n : \clocksym_{k_n})$ where all $k_j \geq 0$.
	We can think of a variable of type $\clocksym_k$ as representing a clock (i.e. a time dimension) paired up with a certificate that we do not care what happens after the time on this clock exceeds $k$.
	Correspondingly, we have a map $\clocksym_k \to \clocksym_\ell$ if $k \leq \ell$.
	These maps, together with weakening, exchange, and contraction, generate the category. The terminal object is $()$ and every other object is unpointable.
	
	Consider in this category the functor $\loch \times (i : \clocksym_k) : \clocksym \to \clocksym : W \mapsto (W, i : \clocksym_k)$, which is an endomultiplier for $(i : \clocksym_k)$.
	
	It is cartesian (hence $\top$-slice non-full and right adjoint with $\sumbase{(i : \clocksym_k)}(W, \psi) = W$), $\top$-slice faithful and not $\top$-slice objectwise pointable.
\end{example}
\begin{example}[Twisting posets] \label{ex:twisting-posets}
	Let $\catP$ be the category of finite non-empty posets and monotonic maps. This category is objectwise pointable.
	
	Let $\IX = \accol{0 < 1}$ and let $W \multip \IX = (W\op \times \accol{0}) \cup (W \times \accol{1})$ with $(x, 0) < (y, 1)$ for all $x, y \in W$. This is an endomultiplier for $\IX$.
	
	It is easily seen to be: not copointed, $\top$-slice faithful but not full, $\top$-slice objectwise pointable but not shard-free, and $\top$-slice right adjoint.
	
	The functor $\freshbase \IX : V \mapsto (V \multip \IX, \pi_2)$ has a left adjoint $\sumbase \IX : (W, \psi) \mapsto \psi\inv(0)\op \uplus \psi\inv(1)$ where elements from different sides of the $\uplus$ are incomparable.
	
	We see this category as a candidate base category for directed type theory. The idea is that a cell over $W$ is a commutative diagram in a category. A problem here is that a cell over a discrete poset such as $\accol{x, y}$ where $x$ and $y$ are incomparable, should then be the same as a pair of cells over $\accol x$ and $\accol y$. This will require that we restrict from presheaves to sheaves, but that makes it notoriously difficult to model the universe \cite{sheaf-universes}. One solution would be to restrict to totally ordered sets, but then we lose the left adjoint $\sumbase \IX$. We address this in \cref{ex:twisting-cubes}.
\end{example}
\begin{example}[Twisted cubes] \label{ex:twisting-cubes}
	Let $\Bowtie$ be the subcategory of $\catP$ whose objects are generated by $\top$ and $\loch \multip \IX$ (note that every object then also has an opposite since $\top\op = \top$ and $(V \multip \IX)\op \cong V \multip \IX$), and whose morphisms are given by
	\begin{itemize}
		\item $(\vfi, 0) : \Bowtie(V, W \multip \IX)$ if $\vfi : \Bowtie(V, W\op)$,
		\item $(\vfi, 1) : \Bowtie(V, W \multip \IX)$ if $\vfi : \Bowtie(V, W)$,
		\item $\vfi \multip \IX : \Bowtie(V \multip \IX, W \multip \IX)$ if $\vfi : \Bowtie(V, W)$,
		\item $() : \Bowtie(V, \top)$.
	\end{itemize}
	Note that this collection automatically contains all identities, composites, and opposites.
	It is isomorphic to Pinyo and Kraus's category of twisted cubes, as can be seen from the ternary representation of said category \cite[def. 34]{pinyo-twisted}.
	This category is objectwise pointable.
	
	Again, we consider the functor $\loch \multip \IX : \Bowtie \to \Bowtie$, which is well-defined by construction of $\Bowtie$ and an endomultiplier for $\IX$. It corresponds to Pinyo and Kraus's twisted prism functor.
	
	It is: not copointed and $\top$-slice fully faithful, objectwise pointable, shard-free and right adjoint.
	
	The left adjoint to $\freshbase{\IX} : W \mapsto (W \multip \IX, \pi_2)$ is now given by
	\begin{equation}
		\sumbase \IX : \left \{
			\begin{array}{l c l}
				(W, ((), 0)) &\mapsto& W\op \\
				(W, ((), 1)) &\mapsto& W \\
				(W \multip \IX, () \multip \IX) &\mapsto& W,
			\end{array}
		\right.
	\end{equation}
	with the obvious action on morphisms.
\end{example}
\begin{example}[Embargoes] \label{ex:embargoes}
	In order to define contextual fibrancy \cite{boulier-tabareau} internally, we need to be able to somehow put a sign in the context $\Gamma.\embargo.\Theta$ in order to be able to say: the type is fibrant over $\Theta$ in context $\Gamma$.
	We call this an embargo and say that $\Theta$ is embargoed whereas $\Gamma$ is not.
	If $\cat C$ is the category of contexts, then $\Gamma.\embargo.\Theta$ can be seen as an object of the arrow category $\cat C^\thearrowcat$, namely the arrow $\Gamma.\Theta \to \Gamma$.
	
	If $\cat C = \widehat \catW$ happens to be a presheaf category, then we have an isomorphism of categories $H : \widehat \catW^\thearrowcat \cong \widehat{\catW \times \thearrowcat}$ where $\thearrowcat = \accol{\bot \to \top}$. Under this isomorphism, we have $\yoneda(W, \top) \cong H(\yoneda W \xrightarrow{\id} \yoneda W)$ which we think of as $\yoneda W.\embargo.\top$ and $\yoneda(W, \bot) \cong H(\bot \xrightarrow{[]} \yoneda W)$ which we think of as $\yoneda W.\embargo.\bot$. Thus, forgetting the second component of $(W, o)$ amounts to forgetting the embargoed part of the context.
	A $(W, \top)$-cell of $\Gamma.\embargo.\Theta$ is a $W$-cell of $\Gamma.\Theta$, i.e.\ a partly embargoed $W$-cell.
	We can extract the unembargoed information by restricting to $(W, \bot)$, as a $(W, \bot)$-cell of $\Gamma.\embargo.\Theta$ is just a $W$-cell of $\Gamma$.
	
	There are 3 adjoint functors $\bot \dashv () \dashv \top$ between $\thearrowcat$ and $\pointcat$ from which we obtain 3 adjoint functors $(\Id, \bot) \dashv \pi_1 \dashv (\Id, \top)$ between $\catW \times \thearrowcat$ and $\catW$. The rightmost functor $(\Id, \top) : \catW \to \catW \times \thearrowcat$ is a multiplier for the terminal object $\embargo := (\top, \top) \in \catW \times \thearrowcat$, denoted $\loch \multip \embargo$.
	
	It is: not endo, $\top$-slice fully faithful, $\top$-slice objectwise pointable iff $\catW$ is and in that case $\top$-slice shard-free, and $\top$-slice right adjoint.
	
	In order to look at the left adjoint, note first that since $\embargo$ is terminal, we have $(\catW \times \thearrowcat) / \embargo \cong \catW \times \thearrowcat$ and clearly $\freshbase{\embargo}$ corresponds to $(\Id, \top)$ under this isomorphism. This functor is part of a chain of \emph{three} adjoint functors $(\Id, \bot) \dashv \pi_1 \dashv (\Id, \top)$ so that the multiplier is not just $\top$-slice right adjoint but $\sumbase{\embargo}$ even has a further left adjoint!
	
	If $\loch \multip U : \catV \to \catW$ is a multiplier, then we can lift it to a multiplier $\loch \multip (U \multip \embargo) : \catV \times \thearrowcat \to \catW \times \thearrowcat$ by applying it to the first component, i.e. $(W, o) \multip (U \multip \embargo) = (W \multip U, o)$. The resulting multiplier inherits all properties in \cref{def:multiplier} from $\loch \multip U$, except that it is never $\top$-slice objectwise pointable.
\end{example}
\begin{example}[Enhanced embargoes] \label{ex:embargoes2}
	If $\loch \multip U$ is a copointed endomultiplier on $\catW$, then we might want to apply it to an arrow $V \xrightarrow{\psi} W$ by sending it to $V \multip U \xrightarrow{\psi \circ \pi_1} W$. This operation is not definable on $\catW \times \thearrowcat$, which only encodes arrows of the forms $W \to W$ (as $(W, \top)$) and $\bot \to W$ (as $(W, \bot)$). For this reason, we move to the comma category $\catW_\embargo := \catW_\bot / \catW$ where $\catW_\bot$ is $\catW$ with a freely added initial object. This comma category has as its objects arrows $V \xrightarrow{\psi} W$ where $V \in \catW_\bot$ and $W \in \catW$. Morphisms are simply commutative squares.
	A $(V \xrightarrow{\psi} W)$-cell is now a non-embargoed $W$-cell $\gamma$ with embargoed information about $\gamma \circ \psi$.
	
	We still have three adjoint functors $(\bot \xrightarrow{[]} \loch) \dashv \Cod \dashv \Delta$ where $\Delta W = (W \xrightarrow{\id} W)$. Further right adjoints would be $\Dom \dashv (\loch \xrightarrow{()} \top)$, but $\Dom$ is not definable as the domain might be $\bot$. We take $\Delta$ as a multiplier for $\embargo := (\top \to \top)$, denoted $\loch \multip \embargo := \Delta$.
	
	The multiplier $\loch \multip \embargo$ is: not endo, $\top$-slice fully faithful, $\top$-slice objectwise pointable iff $\catW$ is objectwise pointable and in that case generally \emph{not} $\top$-slice shard-free (as every non-identity arrow is a shard), and $\top$-slice right adjoint.
	
	Now we can still lift any multiplier $\loch \multip U : \catV \to \catW$ to a multiplier $\loch \multip (U \multip \embargo) : \catV_\embargo \to \catW_\embargo$ for $(U \multip \embargo) = (U \xrightarrow{\id} U)$ by applying it to both domain and codomain, i.e.\ $(V \xrightarrow \psi W) \multip (U \multip \embargo) := (V \multip U \xrightarrow{\psi \multip U} W \multip U)$, where by convention $\bot \multip U = \bot$. It inherits all properties in \cref{def:multiplier} from $\loch \multip U$, except that it is never $\top$-slice objectwise pointable.
	
	For reasons that will become apparent later, we write $\embargo \amaze \loch := (\loch \to \top)$.
	Note that a $(\embargo \amaze U)$-cell is an unembargoed point with embargoed information about the degenerate $U$-cell on that point.
	E.g.\ in a context $\Gamma.\embargo.\Theta$, an $(\embargo \amaze \IX)$-cell is exactly a path in $\Theta$ above a point in $\Gamma$, which is a concept that we need to quantify over when defining internal Kan fibrancy \cite{boulier-tabareau}.
	
	If $\loch \multip U$ is copointed, then we can also lift a multiplier for $U$ to a multiplier for $(\embargo \amaze U)$ by applying the original one only to the domain, i.e.\ $(V \xrightarrow{\psi} W) \multip (\embargo \amaze U) = (V \multip U \xrightarrow{\psi \circ \pi_1} W)$.
	This again inherits all properties in \cref{def:multiplier} from $\loch \multip U$, except that it is never $\top$-slice objectwise pointable, and that $\top$-slice fullness requires that $\pi_1 : \loch \multip U \to \Id$ is objectwise epi (e.g. because $U$ is pointable) and $\loch \multip U$ is slicewise full, and that $\top$-slice right adjointness can only be inherited if $\cat W$ has pushouts. In that case, we have
	\begin{equation}
		\sumbase{(\embargo \amaze U)}(W_1 \xrightarrow{\psi} W_2, (\psi_1, ())) = (\sumbase U (W_1, \psi 1) \to W_2 \uplus_{W_1} \sumbase U (W_1, \psi 1)).
	\end{equation}
	Here, the morphism $W_1 = \pairbase U (W_1, \psi_1) \to \sumbase U (W_1, \psi 1)$ is an instance of the natural transformation $\hidebase U : \pairbase U \to \sumbase U$ obtained by \cref{thm:cotranspose} from $\pi_1 : \loch \multip U = \pairbase U \freshbase U \to \Id$ (\cref{thm:quantification}).
	Indeed, given a morphism of slice objects $(\chi_1, \chi_2) : (W_1 \xrightarrow{\psi} W_2, (\psi_1, ())) \to \freshbase{(\embargo \amaze U)}(V_1 \xrightarrow{\vfi} V_2)$, i.e.
	\begin{equation}
		\xymatrix{
			&&&& V_1 \multip U
				\ar[r]^{\pi_1}
				\ar[lldddd]^{\pi_2}
			& V_1
				\ar[r]^\vfi
			& V_2
				\ar[lldddd]
			\\ \\
			W_1
				\ar[rr]_\psi
				\ar[rrrruu]^{\chi_1}
				\ar[rrdd]_{\psi_1}
			&& W_2
				\ar[rrrruu]^{\chi_2}
				\ar[rrdd]
			\\ \\
			&& U
				\ar[rr]
			&& \top
		}
	\end{equation}
	we get a commutative diagram (the upper right square commutes by construction of $\hidebase U$)
	\begin{equation}
		\xymatrix{
			\sumbase U (W_1, \psi_1)
				\ar[rr]^{\sumbase U \chi_1}
			&& \sumbase U \freshbase U V_1
				\ar[rr]^{\dropbase U}
			&& V_1
			\\
			\pairbase U (W_1, \psi_1)
				\ar[rr]^{\pairbase U \chi_1}
				\ar[u]^{\hidebase U}
			&& \pairbase U \freshbase U V_1
				\ar[u]^{\hidebase U}
			\\
			W_1
				\ar[rr]^{\chi_1}
				\ar@{=}[u]
				\ar[d]_\psi
			&& V_1 \multip U
				\ar[rr]^{\pi_1}
				\ar@{=}[u]
			&& V_1
				\ar@{=}[uu]
				\ar[d]_\vfi
			\\
			W_2
				\ar[rrrr]^{\chi_2}
			&&&& V_2
		}
	\end{equation}
	so the top horizontal line, which is the transpose of $\chi_1$, is a well-typed first component of the transpose of $(\chi_1, \chi_2)$, while the three horizontal lines together constitute an arrow from the pushout to $V_2$ which is a well-typed second component.
	Conversely, given $(\omega_1, \omega_2) : \sumbase{(\embargo \amaze U)}(W_1 \xrightarrow{\psi} W_2, (\psi_1, ())) \to (V_1 \xrightarrow \vfi V_2)$, i.e.\ (unwrapping the pushout)
	\begin{equation}
		\xymatrix{
			W_1
				\ar[rr]^{\hidebase U}
				\ar[d]_{\psi}
			&& \sumbase U (W_1, \psi_1)
				\ar[rr]^{\omega_1}
			&& V_1
				\ar[d]^\vfi
			\\
			W_2
				\ar[rrrr]_{\chi_2}
			&&&& V_2
		}
	\end{equation}
	we can take the transpose of $\omega_1$ as a first component and $\chi_2$ as a second component of the transpose of $(\omega_1, \omega_2)$. It remains to show that these form a commutative diagram with $\psi: W_1 \to W_2$ and $\vfi \circ \pi_1 : V_1 \multip U \to V_2$. But we have a commutative diagram
	\begin{equation*}
		\xymatrix{
			W_1
				\ar@{=}[r]
				\ar@{=}[dd]
			& \pairbase U (W_1, \psi_1)
				\ar[rr]^{\pairbase U \copybase U}
				\ar[d]^{\hidebase U}
			&& \pairbase U \freshbase U \sumbase U (W_1, \psi_1)
				\ar[rr]^{\pairbase U \freshbase U \omega_1}
				\ar[d]^{\hidebase U}
			&& \pairbase U \freshbase U V_1
				\ar@{=}[r]
				\ar[d]^{\hidebase U}
			& V_1 \multip U
				\ar[d]^{\pi_1}
			\\
			& \sumbase U (W_1, \psi_1)
				\ar@{=}[d]
				\ar[rr]_{\sumbase U \copybase U}
			&& \sumbase U \freshbase U \sumbase U (W_1, \psi_1)
				\ar[rr]_{\sumbase U \freshbase U \omega_1}
			&& \sumbase U \freshbase U V_1
				\ar[r]_{\dropbase U}
			& V_1
				\ar@{=}[d]
			\\
			W_1
				\ar[r]_(0.35){\hidebase U}
			& \sumbase U (W_1, \psi_1)
				\ar[rrrrr]_{\omega_1}
			&&&&& V_1
		}
	\end{equation*}
	which can be pasted on top of the previous one to settle the matter. Finally, it is surprisingly easy to verify that the transposition operations just defined are mutually inverse.
\end{example}

\begin{example}[Depth $d$ cubes] \label{ex:depth}
	Let $\boxslash_d$ with $d \geq -1$ be the category of depth $d$ cubes, used as a base category in degrees of relatedness \cite{reldtt,reldtt-techreport}.%
	\footnote{For $d = -1$, we get the point category. For $d = 0$, we get the category of binary cartesian cubes $\boxslash^2$. For $d = 1$, we get the category of bridge/path cubes \cite{paramdtt,reldtt-techreport}.}
	Its objects take the form $(i_1 : \pparen{k_1}, \ldots, i_n : \pparen{k_n})$ where all $k_j \in \accol{0, \ldots, d}$.
	Conceptually, we have a map $\pparen k \to \pparen \ell$ if $k \geq \ell$.
	Thus, morphisms $\vfi : (i_1 : \pparen{k_1}, \ldots, i_n : \pparen{k_n}) \to (j_1 : \pparen{\ell_1}, \ldots, j_m : \pparen{\ell_m})$ send every variable $j : \pparen \ell$ of the codomain to a value $j \psub \vfi$, which is either 0, 1 or a variable $i : \pparen k$ of the domain such that $k \geq \ell$.
	The terminal object is $()$ and the category is objectwise pointable.
	
	Consider in this category the functor $\loch \times (i : \pparen k) : \boxslash_d \to \boxslash_d : W \mapsto (W, i : \pparen k)$, which is an endomultiplier for $(i : \pparen k)$.
	
	It is cartesian (hence $\top$-slice non-full and right adjoint with $\sumbase{(i : \pparen k)}(W, \psi) = W$), $\top$-slice faithful, objectwise pointable and shard-free.
\end{example}

\begin{example}[Erasure] \label{ex:erasure}
	Let $\erasecat{d} = \accol{\top \leftarrow 0 \leftarrow 1 \leftarrow \ldots \leftarrow d}$ with $d \geq -1$. This category has cartesian products $m \times n = \max(m, n)$ and only the terminal object is pointable. We remark that $\widehat{\erasecat{0}}$ is the Sierpi\'nski topos.
	
	We consider the endomultiplier $\loch \times i : \erasecat d \to \erasecat d$.
	
	It is cartesian (hence $\top$-slice non-full and right adjoint with $\sumbase i(j, \psi) = j$), $\top$-slice faithful and not $\top$-slice objectwise pointable.
	
	We believe that this base category is a good foundation for studying the semantics of erasure of irrelevant subterms in Degrees of Relatedness \cite{reldtt}. The idea is that, for a presheaf $\Gamma$, the set $\DSub{\top}{\Gamma}$ is the set of elements, whereas the set $\DSub i \Gamma$ is the set of elements considered up to $i$-relatedness, but also whose existence is only guaranteed by a derivation up to $i$-relatedness.
\end{example}
\wip{\begin{example}[Simplices] \label{ex:simplices}
	\todoi{The monoidal sum is not functorial!}
	The simplex category $\Delta$ has as objects non-empty finite linear orders $\Delta^n = \accol{0 \leq \ldots \leq n}$ for $n \geq 0$ and as morphisms all monotone functions. We have a monoidal structure $(\Delta, \Delta^0, \oplus)$, where $\Delta^m \oplus \Delta^n = \Delta^{m + n}$ by taking the disjoint union and identifying $m \in \Delta^m$ with $0 \in \Delta^n$.
	
	We consider the endomultiplier $\loch \oplus \Delta^1 : \Delta \to \Delta$, which is a multiplier for $\Delta^1$.
	
	It is 3/4-cartesian but not cartesian, cancellative, not affine, non-spooky, not connection-free and quantifiable. Connection-freedom is violated e.g. by $\id \oplus () : \Delta^1 \oplus \Delta^1 \to \Delta^1$.
	
	The left adjoint to $\freshbase{\Delta^0} : \Delta \to \Delta/\Delta^1$ is given by $\sumbase{\Delta^0} : \Delta/\Delta^1 \to \Delta : (W, \vfi) \mapsto \vfi\inv(0)$.
\end{example}}
\begin{example}[Counterexample for $\top$-slice faithful] \label{ex:multip:initial}
	Let $\boxslash^2_\bot$ be the category of binary cartesian cubes extended with an initial object. We consider the cartesian product $\loch \times \bot$ which sends everything to $\bot$. This is not $\top$-slice faithful, as $\freshbase \bot$ sends both $(0/i)$ and $(1/i) : () \to (i : \IX)$ to $[] : (\bot, []) \to (\bot, [])$.
	It is not $\top$-slice full, as there is no $\psi : () \to \bot$ such that $\psi \times \bot = [] : \freshbase \bot () \to \freshbase \bot \bot$.
\end{example}

\subsection{Properties}
\subsubsection{Functoriality}
\begin{definition} \label{def:multip-hom}
	A \textbf{multiplier morphism} or \textbf{morphism multiplier} for $\upsilon : U \to U'$ is a natural transformation $\loch \multip \upsilon : \loch \multip U \to \loch \multip U'$ such that $\pi_2 \circ (\top \multip \upsilon) \circ \pi_2\inv = \upsilon : U \to U'$ (or equivalently $\pi_2 \circ (W \multip \upsilon) = \upsilon \circ \pi_2 : W \multip U \to U'$ for all $W$).
	\begin{itemize}
		\item If both multipliers are copointed, then $\upsilon$ is said to be a \textbf{morphism of copointed multipliers}\seelog{} if it is a morphism of copointed endofunctors, i.e. if $\pi_1 \circ (W \multip \upsilon) = \pi_1$,
		\item If both multipliers are comonads, then $\upsilon$ is said to be a \textbf{comonad morphism of multipliers}\seelog{} if it is a comonad morphism, i.e. if additionally $(W \multip \delta) \circ (W \multip \upsilon) = ((W \multip \upsilon) \multip \upsilon) \circ (W \multip \delta)$,
		\item A morphism of cartesian multipliers is \textbf{cartesian} if it is the cartesian product with $\upsilon$.
	\end{itemize}
\end{definition}
\begin{proposition}
	A morphism of copointed multipliers, whose domain and codomain happen to be cartesian multipliers, is cartesian.
\end{proposition}
\begin{proof}
	We have $\pi_2 \circ (W \multip \upsilon) = \upsilon \circ \pi_2$ and $\pi_1 \circ (W \multip \upsilon) = \pi_1$. Hence, $(W \multip \upsilon) = (\pi_1, \upsilon \circ \pi_2) = W \times \upsilon$.
\end{proof}
\begin{proposition}[Functoriality] \label{thm:functoriality}
	A multiplier morphism $\loch \multip \upsilon : \loch \multip U \to \loch \multip U'$ gives rise to a natural transformation $\pairslice{}{\upsilon} \circ \freshbase U \to \freshbase {U'}$. Hence, for $\top$-slice right adjoint multipliers, we also have $\sumbase{U'} \circ \pairslice{}{\upsilon} \to \sumbase U$.
\end{proposition}
\begin{proof}
	We have to show that for every $W \in \catW$, we get $(W \multip U, \upsilon \circ \pi_2) \to (W \multip U', \pi_2)$. The morphism $W \multip \upsilon : W \multip U \to W \multip U'$ does the job. The second statement follows from \cref{thm:cotranspose}.
\end{proof}

\subsubsection{Quantification and quotient theorem}
\begin{theorem}[$\top$-slice quantification theorem] \label{thm:quantification}
	If $\loch \multip U$ is
	\begin{enumerate}
		\item $\top$-slice fully faithful and right adjoint, then we have a natural isomorphism $\dropbase U : \sumbase U \freshbase{U} \cong \Id$.
		\item copointed, then we have:
		\begin{enumerate}
			\item $\hidebase U : \pairbase U \to \sumbase U$ (if $\top$-slice right adjoint),
			\item $\spoilbase U : \freshbase U \to \wknbase U$ (if $\wknbase U$ exists),
			\item in any case $\pairbase U \freshbase U \to \Id$.
		\end{enumerate}
		\item a comonad, then there is a natural transformation $\pairslice{}{\delta} \circ \freshbase U \to \freshbase{U \multip U}$, where we compose multipliers as in \cref{thm:compose-multip}.
		\item cartesian, then we have:
		\begin{enumerate}
			\item $\sumbase U \cong \pairbase U$,
			\item $\freshbase U \cong \wknbase U$,
			\item $\sumbase U \freshbase U \cong \pairbase U \wknbase U = (\loch \times U) \cong (\loch \multip U)$.
		\end{enumerate}
		Moreover, these isomorphisms become equalities by choosing $\sumbase U$ and $\wknbase U$ wisely (both are defined only up to isomorphism).
	\end{enumerate}
\end{theorem}
\begin{proof}
	\begin{enumerate}
		\item This is a standard fact of fully faithful right adjoints such as $\freshbase{U}$.
		\item By \cref{thm:cotranspose}, it is sufficient to prove $\pairbase U \freshbase U \to \Id$. But $\pairbase U \freshbase U = (\loch \multip U)$, so this is exactly the statement that the multiplier is copointed.
		\item This is a special case of \cref{thm:functoriality}.
		\item By uniqueness of the cartesian product, we have $\freshbase U \cong \wknbase U$. Then the multiplier is $\top$-slice right adjoint with $\sumbase U \cong \pairbase U$. The last point is now trivial. \qedhere
	\end{enumerate}
\end{proof}
\begin{theorem}[$\top$-slice quotient theorem\seelog{} for $\top$-slice objectwise pointable multipliers] \label{thm:quotient}
	If $\loch \multip U : \catW \to \catV$ is $\top$-slice objectwise pointable, fully faithful and shard-free, then $\freshbase{U} : \catW \simeq \splitepislice{\catV}{U}$ is an equivalence of categories, where $\splitepislice{\catV}{U}$ is the full subcategory of $\catV / U$ whose objects are the split epimorphic slice objects.
\end{theorem}
\begin{proof}
	By $\top$-slice objectwise pointability, $\freshbase{U}$ lands in $\splitepislice{\catV}{U}$. The other properties assert that $\freshbase{U}$ is fully faithful and essentially surjective as a functor $\catW \to \splitepislice{\catV}{U}$.
\end{proof}
This quotient theorem applies to \cref{ex:affine-cubes,ex:twisting-cubes,ex:embargoes,ex:identity-multiplier}.
However, we can extend the quotient theorem to also consider multipliers that are not $\top$-slice objectwise pointable \cref{thm:quotient-exorcised}, and then it will apply to more examples.

We will use the quotient theorem in \cref{thm:transp-elim} on transpension elimination, a dependent eliminator for the transpension type from which we can build a dependent eliminator for BCM's $\Psi$-type and prove BCM's $\Phi$-rule \cite{moulin,moulin-param3}.

\subsubsection{Dealing with unpointability}
Since multipliers that are not $\top$-slice objectwise pointable, do not guarantee that $\freshbase U$ produces split epi slice objects, we need to come up with a larger class of suitable epi-like morphisms to $U$ before we can proceed.
\begin{definition} \label{def:dim-split}
	Given a multiplier $\loch \multip U : \catW \to \catV$, we say that a morphism $\vfi : V \to U$ is \textbf{dimensionally split} if there is some $W \in \catW$ such that $\pi_2 : W \multip U \to U$ factors over $\vfi$. The other factor $\chi$ such that $\pi_2 = \vfi \circ \chi$ will be called a \textbf{dimensional section} of $\vfi$. We write $\dimslice{\catV}{U}$ for the full subcategory of $\catV / U$ of dimensionally split slice objects.
\end{definition}
The $\top$-slice objectwise pointability condition for multipliers is automatically satisfied if we replace `split epi' with `dimensionally split':
\begin{corollary}
	For any multiplier $\loch \multip U$, any projection $\pi_2 : W \multip U \to U$ is dimensionally split. \qed
\end{corollary}
\begin{proposition} \label{thm:dim-split}
	Take a multiplier $\loch \multip U : \catW \to \catV$.
	\begin{enumerate}
		\item If $\vfi \circ \chi$ is dimensionally split, then so is $\vfi$.
		\item The identity morphism $\id_U : U \to U$ is dimensionally split.
		\item If $\vfi : V \to U$ is dimensionally split and $\chi : V' \to V$ is split epi, then $\vfi \circ \chi : V' \to U$ is dimensionally split.
		\item Every split epimorphism to $U$ is dimensionally split.
		\item If $\loch \multip U$ is $\top$-slice objectwise pointable, then every dimensionally split morphism is split epi.
	\end{enumerate}
\end{proposition}
\begin{proof}
	\begin{enumerate}
		\item If $\pi_2 : W \multip U \to U$ factors over $\vfi \circ \chi$, then it certainly factors over $\vfi$.
		\item Since $\pi_2 : \top \multip U \to U$ factors over $\id_U$.
		\item Let $\vfi'$ be a dimensional section of $\vfi$ and $\chi'$ a section of $\chi$. Then $\chi' \circ \vfi'$ is a dimensional section of $\vfi \circ \chi$.
		\item From the previous two points, or (essentially by composition of the above reasoning) because if $\chi : U \to V$ is a section of $\vfi : V \to U$, then $\chi \circ \pi_2 : \top \multip U \to V$ is a dimensional section of $\vfi$.
		\item If $\vfi : V \to U$ is dimensionally split, then some $\pi_2 : W \multip U \to U$ factors over $\vfi$. Since $\pi_2$ is split epi, $\id_U$ factors over $\pi_2$ and hence over $\vfi$, i.e. $\vfi$ is split epi. \qedhere
	\end{enumerate}
\end{proof}
We can now extend the notions of shard and shard-freedom to multipliers that are not $\top$-slice objectwise pointable without changing their meaning for those that are:
\begin{definition} \label{def:tssf}
	We say that a multiplier $\loch \multip U : \catW \to \catV$ is \textbf{$\top$-slice shard-free} if $\freshbase U$ is essentially surjective on $\dimslice{\catV}{U}$, the full subcategory of $\catV/U$ of dimensionally split slice objects.
	A dimensionally split slice object $(V, \psi)$ that is not in the image of $\freshbase{U}$ even up to isomorphism, will be called a \textbf{shard} of the multiplier.
\end{definition}
Note that a multiplier is $\top$-slice shard-free if every dimensionally split slice object has an \emph{invertible} dimensional section.
\begin{theorem}[$\top$-slice quotient theorem\seelog{}] \label{thm:quotient-exorcised}
	If a multiplier $\loch \multip U : \catW \to \catV$ is $\top$-slice fully faithful and shard-free, then $\freshbase{U} : \catW \simeq \dimslice{\catV}{U}$ is an equivalence of categories. \qed
\end{theorem}
\begin{example}[Identity] \label{ex:identity-multiplier-dim}
	In the category $\catW$ with the identity multiplier $W \multip \top = W$, every morphism $W \to \top$ is dimensionally split with $\id_W$ as an invertible dimensional section. The multiplier is $\top$-slice fully faithful and shard-free, so the quotient theorem applies.
\end{example}
\begin{example}[Nullary cubes] \label{ex:cubes-dim}
	In the categories of $k$-affine cubes $\Box^k$ (\cref{ex:affine-cubes}) and $k$-ary cartesian cubes $\boxslash^k$ (\cref{ex:cartesian-cubes}) ($k \geq 0$), a morphism $\vfi : \IX^n \to \IX$ is dimensionally split if $i_1 \psub \vfi$ is a variable.
	The multipliers $\loch * \IX : \Box^k \to \Box^k$ and $\loch \times \IX : \boxslash^k \to \boxslash^k$ are $\top$-slice shard-free.
	The multiplier for affine cubes is also $\top$-slice fully faithful so the quotient theorem applies.
\end{example}
\begin{example}[Clocks] \label{ex:clocks-dim}
	In the category of clocks $\clocksym$ (\cref{ex:clocks}), a morphism $\vfi : V \to (i : \clocksym_k)$ is dimensionally split if $i \psub \vfi$ has clock type $\clocksym_k$.
	The multiplier $\loch \times (i : \clocksym_k)$ is $\top$-slice fully faithful and shard-free, so the quotient theorem applies.
\end{example}
\wip{\begin{example}[Charged cubes] \label{ex:charged-dim}
	In the category $\Box_{\twist}$ of charged twisted cubes, we considered two multipliers.
	
	For $\loch \twist \IX$, a morphism to $\top \twist \IX$ is dimensionally split if it is of the form $() \twist \IX$. The multiplier is connection-free.
	
	For $\IX \twist \loch$, a morphism to $\IX \twist \top$ is dimensionally split if it is of the form $\IX \twist ()$. The multiplier is connection-free.
\end{example}}
\begin{example}[Embargoes] \label{ex:embargoes-dim}
	For the embargo multiplier $\loch \multip \embargo := (\Id, \top) : \catW \to \catW \times \uparrow$ (\cref{ex:embargoes}) for $\embargo := (\top, \top)$, a morphism $((), ()) : (W, o) \to \embargo$ is dimensionally split if $o = \top$, with the identity as an invertible dimensional section. The multiplier $\loch \multip \embargo$ is $\top$-slice shard-free.
	
	For $\loch \multip (\embargo \multip U) : (W, o) \mapsto (W \multip U, o)$, a morphism $(\vfi, ()) : (W, o) \to (U, \top) = (\embargo \multip U)$ is dimensionally split if $\vfi : W \to U$ is dimensionally split for $\loch \multip U$. If $\chi : W' \multip U \to W$ is a dimensional section for $\vfi$, then $(\chi, \id_o) : (W' \multip U, o) \to (W, o)$ is a dimensional section for $(\vfi, ())$. $\top$-slice shard-freedom is then inherited from $\loch \multip U$.
\end{example}
\begin{example}[Enhanced embargoes] \label{ex:embargoes2-dim}
	For the enhanced embargo multiplier $\loch \multip \embargo : \catW \to \catW_\embargo = \catW_\bot / \catW : W \mapsto (W \xrightarrow{\id} W)$ (\cref{ex:embargoes2}), a morphism $(V \xrightarrow \vfi W) \to (\top \to \top) = \embargo$ is dimensionally split if $V \neq \bot$, with dimensional section $(\id_V, \vfi) : (V \to V) \to (V \xrightarrow \vfi W)$. This multiplier is generally not $\top$-slice shard-free: since it only produces identity arrows, any dimensionally split non-identity arrow is a shard.
	
	For $\loch \multip (U \multip \embargo) : (V \to W) \mapsto (V \multip U \to W \multip U)$, a morphism $(V \to W) \to (U \to U) = (U \multip \embargo)$ is dimensionally split (with section $([], \chi) : (\bot \to W' \multip U) \to (V \to W)$) if the morphism $W \to U$ is dimensionally split for $\loch \multip U$ with section $\chi : W' \multip U \to W$. The multiplier $\loch \multip (U \multip \embargo)$ is generally not $\top$-slice shard-free, as the domain part of a dimensionally split morphism could be anything.
	
	For $\loch \multip (\embargo \amaze U) : (V \to W) \mapsto (V \multip U \to W)$, any morphism $(V \to W) \to (U \to \top) = (\embargo \amaze U)$ is dimensionally split by
	\begin{equation}
		([], \id) \quad:\quad (\bot \to W) \multip (\embargo \amaze U) \quad=\quad (\bot \to W) \quad\to\quad (V \to W).
	\end{equation}
	This multiplier is therefore generally not $\top$-slice shard-free.
	
	To conclude, we have made the base category more complicated in order to be able to define the latter multiplier, but as a trade-off we now have shards to deal with.
\end{example}
\begin{example}[Erasure] \label{ex:erasure-dim}
	In the category $\erasecat d$ (\cref{ex:erasure}) with multiplier $\loch \times i$, all morphisms to $i$ are dimensionally split with the identity as an invertible dimensional section. The multiplier is shard-free.
\end{example}

\subsubsection{Boundaries}
\begin{definition}
	The boundary $\partial U$ of a multiplier $\loch \multip U : \catW \to \catV$ is a presheaf over $\catV$ such that the cells $\DSub{V}{\partial U}$ are precisely the morphisms $V \to U$ that are \emph{not} dimensionally split.
\end{definition}
This is a valid presheaf by \cref{thm:dim-split}.
\begin{proposition}
	If $\loch \multip U$ is $\top$-slice objectwise pointable, then $\partial U$ is the largest strict subobject of $\yoneda U$.
\end{proposition}
\begin{proof}
	Recall that if the multiplier is $\top$-slice objectwise pointable, then dimensionally split and split epi are synonymous.

	Clearly, $\partial U \subseteq \yoneda U$. Since $\id : U \to U$ is split epi, we have $\partial U \subsetneq \yoneda U$. Now take another strict subobject $\Upsilon \subsetneq \yoneda U$. We show that $\Upsilon \subseteq \partial U$.
	
	We start by showing that $\id \not\in \DSub{U}{\Upsilon}$. Otherwise, every $\vfi \in \DSub{V}{\yoneda U}$ would have to be a cell of $\Upsilon$ as it is a restriction of $\id$, which would imply $\Upsilon = \yoneda U$.
	
	Now $\id$ is a restriction of any split epimorphism, so $\Upsilon$ contains no split epimorphisms, i.e. $\Upsilon \subseteq \partial U$.
\end{proof}
\begin{remark} \label{rem:cosieve-base}
	$\top$-slice shard-freedom can also be formulated using (co)sieves \cite{nlab:sieve}.
	A \textbf{sieve in $\catW$} is a full subcategory $\cat S$ such that if $W \in \cat S$ and $\vfi : V \to W$, then $V \in \cat S$.
	The dual (where $\vfi$ points the other way) is called a \textbf{cosieve in $\catW$}.
	Being full subcategories, (co)sieves can be regarded as subsets of $\Obj{\catW}$.
	A \textbf{sieve on $U \in \catW$} is a sieve in $\catW/U$ or, equivalently, a subpresheaf of $\yoneda U$.
	
	A multiplier is $\top$-slice shard-free if either of the following equivalent criteria is satisfied:
	\begin{itemize}
		\item The objects in the essential image of $\freshbase U$ constitute a cosieve in $\catW/U$ \cite{mathoverflow:image-cosieve}.
		\item The objects \emph{outside} the essential image of $\freshbase U$ constitute a sieve in $\catW/U$, i.e.\ a sieve on $U$.
	\end{itemize}
	The slice objects of the cosieve generated by objects of the essential image of $\freshbase U$, are called dimensionally split.
	The boundary $\partial U$ is the largest sieve on $U$ that is disjoint with the objects of the essential image of $\freshbase U$.
	
	If $\loch \multip U$ is $\top$-slice fully faithful, then the above conditions are furthermore equivalent to $\freshbase U$ being a Street opfibration.
\end{remark}
\begin{example}
	In all the binary cube categories mentioned in \cref{sec:multipliers:ex}, $\partial\IX$ is isomorphic to the constant presheaf of booleans.
	
	For affine cubes, if we define a multiplier $\loch * \IX^2$ in the obvious way, then $\partial \IX^2$ is isomorphic to a colimit of four times $\yoneda \IX$ and four times $\yoneda \top$, i.e. a square without filler. For cartesian asymmetric cubes, the square also gains a diagonal. For symmetric cubes (with an involution $\lnot : \IX \to \IX$), the other diagonal also appears.
\end{example}

\subsection{Acting on slice objects} \label{sec:act-slices}
\begin{definition} \label{def:act-slices}
	Given a multiplier $\loch \multip U : \catW \to \catV$, we define
	\begin{equation}
		\freshslice{U}{W_0} : \catW / W_0 \to \catV / (W_0 \multip U) : (W, \psi) \mapsto (W \multip U, \psi \multip U),
	\end{equation}
	which is an instance of \cref{def:act-slice}.
	We say that $\loch \multip U$ is:
	\begin{itemize}
		\item \textbf{Slicewise faithful}\seelog{} if for all $W_0$, the functor $\freshslice{U}{W_0}$ is faithful,
		\item \textbf{Slicewise full}\seelog{} if for all $W_0$, the functor $\freshslice{U}{W_0}$ is full,
		\item \textbf{Indirectly slicewise shard-free}\seelog{} (obsolete%
		\footnote{This was the original notion of shard-freedom\seelog{}, as referred to in \cite{nuyts-phd}, but it leads to a boundary predicate (\cref{def:boundary-predicate}) that is respected by the substitution functor $\wknpsh{\yoneda U} \Psi$.
		This is in contrast to the transpension type, which in general is not respected by the substitution functor (it is if the multiplier is $\top$-slice fully faithful, see \cref{thm:commut-multip-subst}).
		As a result, with this notion of indirect slicewise shard-freedom, the boundary \cref{thm:boundary} can only be stated over $\top \multip \yoneda U$ as opposed to general $\Psi \multip \yoneda U$.
		For this reason, we prefer the notion of \emph{direct slicewise shard-freedom} below.}%
		) if for all $W_0$, the functor $\freshslice{U}{W_0}$ is essentially surjective on slice objects $(V, \vfi) \in {\catV}/{(W_0 \multip U)}$ such that $\vfi$ is indirectly dimensionally split,
		\begin{itemize}
			\item We say that $\vfi : V \to W_0 \multip U$ is \textbf{indirectly dimensionally split} if $\pi_2 \circ \vfi : V \to U$ is dimensionally split.
			\item We point out that the full subcategory of such slice objects is isomorphic to $(\dimslice{\catV}{U}) / (W_0 \multip U, \pi_2)$.
			\item An indirectly dimensionally split slice object $(V, \psi) \in \catV / W_0 \multip U$ that is not in the image of $\freshslice{U}{W_0}$ even up to isomorphism, will be called an \textbf{indirect shard}\seelog{} of the multiplier.
		\end{itemize}
		\item \textbf{Directly slicewise shard-free}\seelog{} if for all $W_0$, the functor $\freshslice{U}{W_0}$ is essentially surjective on slice objects $(V, \vfi) \in {\catV}/{(W_0 \multip U)}$ such that $\vfi : V \to W_0 \multip U$ is directly dimensionally split:
		\begin{itemize}
			\item We say that $\vfi : V \to W_0 \multip U$ is \textbf{directly dimensionally split} with direct dimensional section $\chi : W \multip U \to V$ if $\vfi \circ \chi$ is of the form $\psi \multip U$. The section can alternatively be presented as a morphism of slice objects $\chi : \freshslice{U}{W_0} (W, \psi) \to (V, \vfi)$.
			\item We denote the full subcategory of directly dimensionally split slice objects as $\dimslice{\catV}{(W_0 \multip U)}$.
			\item A directly dimensionally split slice object $(V, \psi) \in \catV / W_0 \multip U$ that is not in the image of $\freshslice{U}{W_0}$ even up to isomorphism, will be called a \textbf{direct shard}\seelog{} of the multiplier.
		\end{itemize}
		\item \textbf{Slicewise right adjoint}\seelog{} if for all $W_0$, the functor $\freshslice{U}{W_0}$ has a left adjoint $\sumslice{U}{W_0} : \catV / (W_0 \multip U) \to \catW / W_0$.
		We denote the unit as $\copyslice U{W_0} : \Id \to \freshslice{U}{W_0} \sumslice{U}{W_0}$ and the co-unit as $\dropslice U{W_0} : \sumslice{U}{W_0} \freshslice{U}{W_0} \to \Id$.
	\end{itemize}
\end{definition}
The above definition generalizes the functor $\freshbase U$ that we already had:
\begin{proposition}
	The functor $\freshslice{U}{\top} : \catW / \top \to \catV / (\top \multip U)$ is equal to $\freshbase U : \catW \to \catV / U$ over the obvious isomorphisms between their domains and codomains. Hence, each of the slicewise properties implies the $\top$-slice property. (Both notions of slicewise shard-freedom imply basic shard-freedom.) \qed
\end{proposition}
Note that both notions of slicewise shard-freedom are well-defined:
\begin{proposition} \label{thm:swsf-well-defined}
	\begin{enumerate}
		\item (Obsolete.) The functor $\freshslice{U}{W_0}$ factors over $(\dimslice{\catV}{U}) / (W_0 \multip U, \pi_2)$.
		\item The functor $\freshslice{U}{W_0}$ factors over $\dimslice{\catV}{(W_0 \multip U)}$.
		\item Directly dimensionally split morphisms are indirectly dimensionally split with the same section. As such, there is a functor $\dimslice{\catV}{(W_0 \multip U)} \to (\dimslice{\catV}{U}) / (W_0 \multip U, \pi_2)$. Hence, direct shards are indirect shards and indirect slicewise shard-freedom implies direct slicewise shard-freedom.
	\end{enumerate}
\end{proposition}
\begin{proof}
	\begin{enumerate}
		\item The functor $\freshslice{U}{W_0}$ sends $(W, \psi)$ to $(W \multip U, \psi \multip U)$. Since $\pi_2 \circ (\psi \multip U) = \pi_2$, it is dimensionally split with the identity as a section.
		\item The identity is a direct dimensional section.
		\item Let $\vfi : V \to W_0 \multip U$ be directly dimensionally split with section $\chi$, i.e.\ $\vfi \circ \chi = \psi \multip U$. Then $\pi_2 \circ \vfi \circ \chi = \pi_2 \circ (\psi \multip U) = \pi_2$, so $\pi_2 \circ \vfi$ is dimensionally split with section $\chi$. \qedhere
	\end{enumerate}
\end{proof}
\begin{proposition} \label{thm:swfaithful}
	If $\loch \multip U : \catW \to \catV$ is $\top$-slice faithful, then it is slicewise faithful.
\end{proposition}
\begin{proof}
	Pick morphisms $\vfi, \chi : (W, \psi) \to (W', \psi')$ in $\catW / W_0$ such that $\freshslice{U}{W_0} \vfi = \freshslice{U}{V} \chi$. Expanding the definition of $\freshslice{U}{W_0}$, we see that this means that $\vfi \multip U = \chi \multip U$, and hence $\vfi = \chi$ by faithfulness of $\loch \multip U$ (\cref{thm:tsfaithful-char}).
\end{proof}
\begin{proposition} \label{thm:swfull}
	If $\loch \multip U : \catW \to \catV$ is $\top$-slice fully faithful, then it is slicewise full.
\end{proposition}
\begin{proof}
	Pick $(W, \psi)$ and $(W', \psi')$ in $\catW / W_0$, and a morphism $\chi : \freshslice{U}{W_0} (W, \psi) \to \freshslice{U}{W_0} (W', \psi')$. This amounts to a diagram:
	\begin{equation}
		\xymatrix{
			W \multip U \ar[rr]^{\chi} \ar[rd]_{\psi \multip U}
				\ar@/_{2em}/[rdd]_{\pi_2}
			&& W' \multip U \ar[ld]^{\psi' \multip U}
				\ar@/^{2em}/[ldd]^{\pi_2}
			\\
			& W_0 \multip U \ar[d]^{\pi_2}
			\\
			& U,
		}
	\end{equation}
	i.e. a triangle in $\catW / U$, the objects of which are in the image of $\freshbase{U} : \catW \to \catW / U$. Then, by fullness of $\freshbase U$ we get $\chi_0 : W \to W'$ such that $\freshbase U \chi_0 = \chi$, which by faithfulness of $\freshbase U$ makes the following diagram commute:
	\begin{equation}
		\xymatrix{
			W \ar[rr]^{\chi_0} \ar[rd]_{\psi}
			&& W' \ar[ld]^{\psi'}
			\\
			& W_0
		}
	\end{equation}
	Then $\chi_0$ is a morphism $\chi_0 : (W, \psi) \to (W', \psi')$ in $\catW/W_0$ and $\freshslice{U}{W_0} \chi_0 = \chi$.
\end{proof}
\begin{proposition} \label{thm:swsf}
	\begin{enumerate}
		\item If $\loch \multip U : \catW \to \catV$ is $\top$-slice full, then direct and indirect dimensional splitness are equivalent, with the same dimensional sections.
		\item (Obsolete.) If $\loch \multip U : \catW \to \catV$ is $\top$-slice full and shard-free, then it is indirectly slicewise shard-free.
		\item If $\loch \multip U : \catW \to \catV$ is $\top$-slice full and shard-free, then it is directly slicewise shard-free.
	\end{enumerate}
\end{proposition}
\begin{proof}
	\begin{enumerate}
		\item We already know that direct dimensional splitness implies indirect dimensional splitness with the same section (\cref{thm:swsf-well-defined}). We prove the other implication.
		
		Pick some $(V, \vfi) \in \catV/(W_0 \multip U)$ such that $\pi_2 \circ \vfi : V \to U$ is dimensionally split with section $\chi : W \multip U \to V$.
		Because $\freshbase{U}$ is full, there is a morphism $\psi : W \to W_0$ such that $\psi \multip U = \vfi \circ \chi : W \multip U \to W_0 \multip U$.
		Thus, $\vfi$ is directly dimensionally split.
		\begin{equation}
			\xymatrix{
				V \ar@{<-}[rr]^{\chi} \ar[rd]_{\vfi}
				&& W \multip U \ar@{.>}[ld]^{\psi \multip U} \ar@/^{2em}/[ldd]^{\pi_2}
				\\
				& W_0 \multip U \ar[d]^{\pi_2}
				\\
				& U
			}
		\end{equation}
		
		\item Pick some $(V, \vfi) \in \catV/(W_0 \multip U)$ such that $\pi_2 \circ \vfi : V \to U$ is dimensionally split. Because $\freshbase{U}$ is essentially surjective on $\dimslice \catV U$, there must be some $W \in \catW$ such that $\iota : \freshbase{U} W = (W \multip U, \pi_2) \cong (V, \pi_2 \circ \vfi)$ as slice objects over $U$. Because $\freshbase{U}$ is full, there is a morphism $\psi : W \to W_0$ such that $\psi \multip U = \vfi \circ \iota : W \multip U \to W_0 \multip U$. Thus, $\iota\inv : (V, \vfi) \cong (W \multip U, \psi \multip U) = \freshslice{U}{W_0}(W, \psi)$ as slice objects over $W_0 \multip U$.
		\begin{equation}
			\xymatrix{
				V \ar@{<-}[rr]^{\iota}_{\cong} \ar[rd]_{\vfi}
				&& W \multip U \ar@{.>}[ld]^{\psi \multip U} \ar@/^{2em}/[ldd]^{\pi_2}
				\\
				& W_0 \multip U \ar[d]^{\pi_2}
				\\
				& U
			}
		\end{equation}
		\item Since indirect slicewise shard-freedom implies direct slicewise shard-freedom (\cref{thm:swsf-well-defined}). \qedhere
	\end{enumerate}
\end{proof}
\begin{example}[Obsolete]
	In the category $\boxslash^k$ of $k$-ary cartesian cubes (\cref{ex:cartesian-cubes}), the diagonal $\delta : \IX \to \IX \times \IX$ has the property that $\pi_2 \circ \delta$ is split epi, but $(\IX, \delta)$ is not in the image of $\freshslice{\IX}{\IX}$. Thus, $\loch \multip \IX$ is not \emph{indirectly} slicewise shard-free, despite being $\top$-slice shard-free.
\end{example}
\begin{proposition} \label{thm:swra}
	If $\loch \multip U : \catW \to \catV$ is $\top$-slice right adjoint, then it is slicewise right adjoint, with
	\begin{align*}
		\sumslice U{W_0}(V, \vfi) &= (\sumbase U(V, \pi_2 \circ \vfi), \dropbase{U} \circ \sumbase U \vfi), \\
		\dropslice{U}{W_0}(W, \psi) &= \dropbase U W : \sumslice{U}{W_0} \freshslice{U}{W_0} (W, \psi) \to (W, \psi), \\
		\copyslice{U}{W_0}(V, \vfi) &= \copybase U (V, \pi_2 \circ \vfi) : (V, \vfi) \to \freshslice{U}{W_0} \sumslice{U}{W_0} (V, \vfi).
	\end{align*}
\end{proposition}
\begin{proof}
	Note that a slice category over a slice category is just a slice category, i.e.\ $(\catC/y)/(x, \vfi) \cong \catC/x$.
	In this light, the functor $\freshslice{U}{W_0}$ is not just the action of $\loch \multip U$ on slice objects over $W_0$, but also the action of $\freshbase U$ on slice objects over $W_0$.
	Now since $\freshbase U$ has left adjoint $\sumbase U$, we get a left adjoint to $\freshslice{U}{W_0}$ by \cref{thm:act-slice-adjoint}.
\end{proof}
%
\begin{proposition}[Functoriality of the slice category] \label{thm:sw-functoriality}
	A morphism of multipliers $\loch \multip \upsilon : \loch \multip U \to \loch \multip U'$ gives rise to a natural transformation $\pairslice{}{W_0 \multip \upsilon} \circ \freshslice{U}{W_0} \to \freshslice{U'}{W_0}$.
	Hence, if both multipliers are $\top$-slice (or equivalently slicewise) right adjoint, we also get $\sumslice{U'}{W_0} \circ \pairslice{}{W_0 \multip \upsilon} \to \sumslice{U}{W_0}$.
\end{proposition}
\begin{proof}
	For any $(W, \psi) \in \catW / W_0$, we have to prove $(W \multip U, (W_0 \multip \upsilon) \circ (\psi \multip U)) \to (W \multip U', \psi \multip U')$. The morphism $W \multip \upsilon : W \multip U \to W \multip U'$ does the job.
	The second statement follows from \cref{thm:cotranspose}.
\end{proof}
\begin{theorem}[Slicewise quantification theorem] \label{thm:sw-quantification}
	If $\loch \multip U$ is
	\begin{enumerate}
		\item $\top$-slice (or equivalently slicewise) fully faithful and right adjoint, then we have a natural isomorphism $\dropslice U{W_0} : \sumslice U{W_0} \freshslice U{W_0} \cong \Id$.
		\item copointed, then we have
		\begin{enumerate}
			\item $\hideslice U{W_0} : \pairslice U{W_0} \to \sumslice U{W_0}$ (if $\top$-slice, or equivalently presheafwise, right adjoint),
			\item $\spoilslice U{W_0} : \freshslice U{W_0} \to \wknslice U{W_0}$ (if $\wknslice U{W_0}$ exists),
			\item in any case $\pairslice U{W_0} \freshslice U{W_0} \to \Id$.
		\end{enumerate}
		\item a comonad, then there is a natural transformation $\pairslice{}{W_0 \multip \delta} \circ \freshslice U{W_0} \to \freshslice{U \multip U}{W_0}$, where we compose multipliers as in \cref{thm:compose-multip}.
		\item cartesian, then we have natural isomorphisms:
		\begin{enumerate}
			\item $\sumslice{U}{W_0}(V, \vfi) \cong \pairslice{U}{W_0}(V, \vfi) = (V, \pi_1 \circ \vfi)$,
			\item $\freshslice{U}{W_0}(W, \psi) \cong \wknslice{U}{W_0}(W, \psi)$,
			\item $\sumslice{U}{W_0} \freshslice{U}{W_0}(W, \psi) \cong \pairslice{U}{W_0} \wknslice{U}{W_0}(W, \psi) \cong (W \times U, \psi \circ \pi_1)$.
		\end{enumerate}
		Moreover, these isomorphisms become equality if $\sumslice{U}{W_0}$ is constructed from $\sumbase U = \pairbase U$ as in the proof of \cref{thm:swra}, and $\wknslice{U}{W_0}(W, \psi)$ is chosen wisely. (Both functors are defined only up to isomorphism.)
	\end{enumerate}
\end{theorem}
\begin{proof}
	\begin{enumerate}
		\item This is a standard fact about fully faithful right adjoints such as $\freshslice U{W_0}$.
		\item By \cref{thm:cotranspose}, it is sufficient to prove $\pairslice U{W_0} \freshslice U{W_0} \to \Id$, and indeed we have
		\[
			\pi_1 : \pairslice U{W_0} \freshslice U{W_0}(W, \psi) = (W \multip U, \pi_1 \circ (\psi \multip U)) = (W \multip U, \psi \circ \pi_1) \to (W, \psi).
		\]
		\item This is a special case of \cref{thm:sw-functoriality}.
		\item \begin{enumerate}
			\item The isomorphism is obtained from the next point by uniqueness of adjoints. We prove the equality if $\sumbase U = \pairbase U$.
			The co-unit is then given by $\dropbase U = \pi_1 : W \times U \to W$.
			The construction of $\sumslice{U}{W_0}$ then reveals that $\sumslice{U}{W_0}(V, \vfi) = (V, \pi_1 \circ \vfi)$, which is the definition of $\pairslice{U}{W_0}(V, \vfi)$.
			\item This follows from the definitions.
			\item We have
			\begin{equation*}
				\sumslice{U}{W_0} \freshslice{U}{W_0}(W, \psi) = \sumslice{U}{W_0} (W \times U, \psi \times U) \cong (W \times U, \pi_1 \circ (\psi \times U))
				= (W \times U, \psi \circ \pi_1). \qedhere
			\end{equation*}
		\end{enumerate}
	\end{enumerate}
\end{proof}
\begin{theorem}[Slicewise quotient theorem\seelog{}]
	If $\loch \multip U : \catW \to \catV$ is $\top$-slice (or equivalently slicewise, for either notion of shard-freedom) fully faithful and shard-free, then
	\begin{enumerate}
		\item (Obsolete.) $\freshslice{U}{W_0} : \catW/W_0 \simeq (\dimslice{\catV}{U}) / (W_0 \multip U, \pi_2)$ is an equivalence of categories,\footnote{We use a slight abuse of notation by using $(\dimslice{\catV}{U}) / (W_0 \multip U, \pi_2)$ as a subcategory of $\catV/(W_0 \multip U)$.}
		\item $\freshslice{U}{W_0} : \catW/W_0 \simeq \dimslice{\catV}{(W_0 \multip U)}$ is an equivalence of categories. \qed
	\end{enumerate}
\end{theorem}

\subsection{Composing multipliers}
\begin{theorem} \label{thm:compose-multip}
	If $\loch \multip U : \catW \to \catV$ is a multiplier for $U$ and $\loch \multip U' : \catV \to \catV'$ is a multiplier for $U'$, then their composite $\loch \multip (U \multip U') := (\loch \multip U) \multip U'$ is a multiplier for $U \multip U'$.
	\begin{enumerate}
		\item The functor $\freshbase{U \multip U'} : \catW \to \catV' /(U \multip U')$ equals $\freshslice{U'}{U} \circ \freshbase U$.
		\item The functor $\freshslice{U \multip U'}{W_0} : \catW \to \catV' /(U \multip U')$ equals $\freshslice{U'}{W_0 \multip U} \circ \freshslice U{W_0}$.
		\item Assume both multipliers are endo. Then:
		\begin{enumerate}
			\item The composite $\loch \multip (U \multip U')$ is copointed if $\loch \multip U$ and $\loch \multip U'$ are copointed,
			\item \sout{The composite $\loch \multip (U \multip U')$ is a comonad if $\loch \multip U$ and $\loch \multip U'$ are comonads,}
			\item The composite $\loch \multip (U \multip U')$ is cartesian if $\loch \multip U$ and $\loch \multip U'$ are cartesian.
		\end{enumerate}
		\item The composite $\loch \multip (U \multip U')$ is $\top$-slice faithful if $\loch \multip U$ and $\loch \multip U'$ are $\top$-slice faithful.
		\item The composite $\loch \multip (U \multip U')$ is $\top$-slice full if $\loch \multip U$ is $\top$-slice full and $\loch \multip U'$ is slicewise full.
		\item The composite $\loch \multip (U \multip U')$ is slicewise full if $\loch \multip U$ and $\loch \multip U'$ are slicewise full.
		\item The composite $\loch \multip (U \multip U')$ is $\top$-slice shard-free if $\loch \multip U$ is $\top$-slice shard-free and $\loch \multip U'$ is slicewise full and shard-free.
		\item \begin{enumerate}
			\item (Obsolete). The composite $\loch \multip (U \multip U')$ is indirectly slicewise shard-free if $\loch \multip U$ is indirectly slicewise shard-free and $\loch \multip U'$ is slicewise full and indirectly slicewise shard-free.
			\item The composite $\loch \multip (U \multip U')$ is directly slicewise shard-free if $\loch \multip U$ is directly slicewise shard-free and $\loch \multip U'$ is slicewise full and directly slicewise shard-free.
		\end{enumerate}
		\item The composite $\loch \multip (U \multip U')$ is $\top$-slice right adjoint if $\loch \multip U$ and $\loch \multip U'$ are $\top$-slice right adjoint, and in that case we have:
		\begin{enumerate}
			\item $\sumbase{U \multip U'} = \sumbase U \circ \sumslice{U'}{U}$,
			\item $\sumslice{U \multip U'}{W_0} = \sumslice{U}{W_0} \circ \sumslice{U'}{W_0 \multip U}$.
		\end{enumerate}
	\end{enumerate}
\end{theorem}
\begin{proof}
	Since $\top \multip U \cong U$, we see that $(\top \multip U) \multip U' \cong U \multip U'$, so the composite is indeed a multiplier for $U \multip U'$.
	\begin{enumerate}
		\item[1-2.] Follows from expanding the definitions.
		\item[3.] \begin{enumerate}
			\item Copointed endofunctors compose.
			\item \sout{Comonads compose.} They most certainly do not!
			\item By associativity of the cartesian product.
		\end{enumerate}
		\item[4.] $\top$-slice faithful multipliers are slicewise faithful (\cref{thm:swfaithful}), and the composite $\freshbase{U \multip U'} = \freshslice{U'}{U} \freshbase{U}$ of faithful functors is faithful.
		\item[5-6.] Follows from the first two properties, since the composite of full functors is full.
		\item[7.] Analogous to the next point, with $W_0 = \top$.
		\item[8.] Recall that the assumptions imply that $\loch \multip U'$ is slicewise fully faithful and slicewise indirectly and directly shard-free.
		
		\begin{enumerate}
			\item Pick a slice object $(V', \vfi') \in {\catV'}/{(W_0 \multip U \multip U')}$ such that $\pi_2^{U \multip U'} \circ \vfi' : V' \to U \multip U'$ is dimensionally split with section $\chi' : W_1 \multip U \multip U' \to V'$. Then $\pi_2^{U'} \circ \pi_2^{U \multip U'} \circ \vfi' = \pi_2^{U'} \circ \vfi' : V' \to U'$ is also dimensionally split with the same section.
			
			Because $\loch \multip U'$ is indirectly slicewise shard-free, we find some $(V, \vfi) \in \catV / (W_0 \multip U)$ such that $\iota' : (V', \vfi') \cong \freshslice{U'}{W_0 \multip U}(V, \vfi) \in \catV' / (W_0 \multip U \multip U')$.
			
			\begin{equation*}
				\xymatrix{
					& W_1 \multip U
						\ar[d]_{\chi}
						\ar@/^{3em}/[ddd]^{\pi_2^U}
					&&
					& W_1 \multip U \multip U'
						\ar[ld]_{\chi'}
						\ar[d]|{\chi \multip U'}
						\ar@/^{3em}/[ddd]|{\pi_2^{U \multip U'}}
						\ar@/^{6em}/[dddd]^{\pi_2^{U'}}
					\\
					& V
						\ar[ld]_\vfi
						\ar[d]^{\iota}_\cong
					&&
					V'
						\ar[d]_{\vfi'}
						\ar[r]^{\iota'}_\cong
					& V \multip U'
						\ar[ld]^{\vfi \multip U'}
					\\
					W_0 \multip U
						\ar[rd]_{\pi_2^U}
					&
					W \multip U
						\ar[l]^{\psi \multip U}
					&&
					W_0 \multip U \multip U'
						\ar[rd]|{\pi_2^{U \multip U'}}
						\ar@/_{2em}/[rdd]_{\pi_2^{U'}}
					\\
					& U
					&&
					& U \multip U' \ar[d]_{\pi_2^{U'}}
					\\
					&&&
					& U'
				}
			\end{equation*}
			
			Note that $\pi_2^{U \multip U'} = \pi_2^U \multip U'$. Because $\freshslice{U'}{U}$ is full, the morphism $\iota' \circ \chi' : \freshslice{U'}{U}(W_1 \multip U, \pi_2^U) \to \freshslice{U'}{U} (V, \pi_2^U \circ \vfi)$ has a preimage $\chi : (W_1 \multip U, \pi_2^U) \to (V, \pi_2^U \circ \vfi)$ under $\freshslice{U'}{U}$.
			Thus, we see that $\pi_2 \circ \vfi : V \to U$ is dimensionally split. Because $\loch \multip U$ is indirectly slicewise shard-free, we find some slice object $(W, \psi) \in \catW / W_0$ so that $\iota : (V, \vfi) \cong \freshslice{U}{W_0}(W, \psi) \in \catV / (W_0 \multip U)$. We conclude that
			\begin{equation}
				(V', \vfi') \cong \freshslice{U'}{W_0 \multip U} (V, \vfi) \cong \freshslice{U'}{W_0 \multip U} \freshslice{U}{W_0} (W, \psi) = \freshslice{U \multip U'}{W_0}(W, \psi).
			\end{equation}
		
			\item Pick a slice object $(V', \vfi') \in {\catV'}/{(W_0 \multip U \multip U')}$ that is directly dimensionally split for the composite multiplier with section $\chi' : W_1 \multip U \multip U' \to V'$, composing to $\vfi' \circ \chi' = \psi_1 \multip U \multip U'$. Then $\vfi'$ is also directly dimensionally split for $\loch \multip U'$.
			
			Because $\loch \multip U'$ is directly slicewise shard-free, we find some $(V, \vfi) \in \catV / (W_0 \multip U)$ such that $\iota' : (V', \vfi') \cong \freshslice{U'}{W_0 \multip U}(V, \vfi) \in \catV' / (W_0 \multip U \multip U')$.
			
			\begin{equation*}
				\xymatrix{
					& W_1 \multip U
						\ar[d]_{\chi}
						\ar `r[ddd] `[ddd] `[dddl]_{\psi_1 \multip U} [ddl]
					&&
					& W_1 \multip U \multip U'
						\ar[ld]_{\chi'}
						\ar[d]|{\chi \multip U'}
						\ar `r[ddd] `[ddd] `[dddl]_{\psi_1 \multip U \multip U'} [ddl]
					\\
					& V
						\ar[ld]_\vfi
						\ar[d]^{\iota}_\cong
					&&
					V'
						\ar[d]_{\vfi'}
						\ar[r]^{\iota'}_\cong
					& V \multip U'
						\ar[ld]^{\vfi \multip U'}
					\\
					W_0 \multip U
					&
					W \multip U
						\ar[l]^{\psi \multip U}
					&&
					W_0 \multip U \multip U'
					&
					\\
					&&&&
				}
			\end{equation*}
			
			Because $\freshslice{U'}{W_0 \multip U}$ is full, the morphism $\iota' \circ \chi' : \freshslice{U'}{W_0 \multip U}(W_1 \multip U, \psi_1 \multip U) \to \freshslice{U'}{W_0 \multip U} (V, \vfi)$ has a preimage $\chi : (W_1 \multip U, \psi_1 \multip U) \to (V, \vfi)$ under $\freshslice{U'}{W_0 \multip U}$.
			Thus, we see that $\vfi : V \to U$ is directly dimensionally split with section $\chi$. Because $\loch \multip U$ is directly slicewise shard-free, we find some slice object $(W, \psi) \in \catW / W_0$ so that $\iota : (V, \vfi) \cong \freshslice{U}{W_0}(W, \psi) \in \catV / (W_0 \multip U)$. We conclude that
			\begin{equation}
				(V', \vfi') \cong \freshslice{U'}{W_0 \multip U} (V, \vfi) \cong \freshslice{U'}{W_0 \multip U} \freshslice{U}{W_0} (W, \psi) = \freshslice{U \multip U'}{W_0}(W, \psi).
			\end{equation}
		\end{enumerate}
		
		\item[9.] $\top$-slice right adjoint multipliers are slicewise right adjoint (\cref{thm:swra}), and the composite of the left adjoints is a left adjoint to the composite. \qedhere
	\end{enumerate}
\end{proof}

\section{Multipliers and presheaves} \label{sec:psh}
\begin{definition} \label{def:3-multip-functors}
	Every multiplier $\loch \multip U : \catW \to \catV$ gives rise to three adjoint endofunctors between $\widehat \catW$ and $\widehat \catV$ via \cref{thm:adjoint-triple}, which we will denote
	\begin{equation}
		(\loch \multip \yoneda U) \dashv (\yoneda U \multimap \loch) \dashv (\yoneda U \amaze \loch).
	\end{equation}
	Correspondingly, a morphism of multipliers $\loch \multip \upsilon$ gives rise to natural transformations $\loch \multip \yoneda \upsilon$, $\yoneda \upsilon \multimap \loch$ and $\yoneda \upsilon \amaze \loch$.
\end{definition}
We will not actually be using the latter two of these functors, although they can be retrieved at least up to isomorphism from the functors in \cref{def:4-multip-functors,def:4-cart-functors} via the equation $\loch \multip U = \pairbase U \freshbase U$.

Note that the functor $\loch \multip \yoneda U : \widehat \catW \to \widehat \catV$ is quite reminiscent of the Day-convolution with $\yoneda U$, which is the reason for our choice of notation.
However, each of the notations is to be regarded as a single symbol, i.e.\ $\multip$, $\multimap$ and $\amaze$ by themselves have no meaning.

\subsection{Acting on elements}
In \cref{sec:act-slices}, we generalized $\freshbase U : \catW \to \catV/U$ to act on slice objects as $\freshslice U{W_0} : \catW/W_0 \to \catV/(W_0 \multip U)$. Here, we further generalize to a functor whose domain is the category of elements:
\begin{definition} \label{def:act-elements}
	We define (using \cref{not:adjoint-triple}):
	\begin{itemize}
		\item $\freshslice U \Psi : \catW/\Psi \to \catV/(\Psi \multip \yoneda U) : (W, \psi) \mapsto (W \multip U, \psi \multip \yoneda U)$,
		\item $\freshelem{U} \Psi : (\DSub W \Psi) \to \set{\vfi : \DSub{W \multip U}{\Psi \multip \yoneda U}}{\pi_2 \circ \vfi = \pi_2 : W \multip U \to U} : \psi \mapsto \psi \multip \yoneda U$.
	\end{itemize}
	We say that $\loch \multip U$ is:
	\begin{itemize}
		\item \textbf{Presheafwise faithful}\seelog{} if for all $\Psi$, the functor $\freshslice{U}{\Psi}$ is faithful,
		\item \textbf{$\top$-slice elementally faithful}\seelog{} if for all $\Psi$, the natural transformation $\freshelem{U}{\Psi}$ is componentwise injective,
		\item \textbf{Presheafwise full}\seelog{} if for all $\Psi$, the functor $\freshslice{U}{\Psi}$ is full,
		\item \textbf{$\top$-slice elementally full}\seelog{} if for all $\Psi$, the natural transformation $\freshelem{U}{\Psi}$ is componentwise surjective,
		\item \textbf{Indirectly presheafwise shard-free}\seelog{} (obsolete\footnote{see \cref{def:act-slices}}) if for all $\Psi$, the functor $\freshslice{U}{\Psi}$ is essentially surjective on elements $(V, \vfi) \in {\catV}/{(\Psi \multip \yoneda U)}$ such that $\vfi$ is indirectly dimensionally split:
		\begin{itemize}
			\item We say that $\vfi : \DSub{V}{\Psi \multip \yoneda U}$ is \textbf{indirectly dimensionally split} if $\pi_2 \circ \vfi : V \to U$ is dimensionally split.
			\item An indirectly dimensionally split element $(V, \psi) \in \catV / (\Psi \multip \yoneda U)$ that is not in the image of $\freshslice{U}{\Psi}$ even up to isomorphism, will be called an \textbf{indirect shard}\seelog{} of the multiplier.
		\end{itemize}
		\item \textbf{Directly presheafwise shard-free}\seelog{} if for all $\Psi$, the functor $\freshslice{U}{\Psi}$ is essentially surjective on elements $(V, \vfi) \in {\catV}/{(\Psi \multip \yoneda U)}$ such that $\vfi : V \to \Psi \multip \yoneda U$ is directly dimensionally split:
		\begin{itemize}
			\item We say that $\vfi : \DSub{V}{\Psi \multip \yoneda U}$ is \textbf{directly dimensionally split} with direct dimensional section $\chi : W \multip U \to V$ if $\vfi \circ \chi$ is of the form $\psi \multip \yoneda U$. The section can alternatively be presented as a morphism of elements $\chi : \freshslice{U}{\Psi} (W, \psi) \to (V, \vfi)$.
			\item We denote the full subcategory of directly dimensionally split elements as $\dimslice{\catV}{(\Psi \multip \yoneda U)}$.
			\item A directly dimensionally split element $(V, \psi) \in \catV / (\Psi \multip \yoneda U)$ that is not in the image of $\freshslice{U}{\Psi}$ even up to isomorphism, will be called a \textbf{direct shard}\seelog{} of the multiplier.
		\end{itemize}
		\item \textbf{Presheafwise right adjoint}\seelog{} if for all $\Psi$, the functor $\freshslice{U}{\Psi}$ has a left adjoint $\sumslice{U}{\Psi} : \catV / (\Psi \multip \yoneda U) \to \catW / \Psi$.
		We denote the unit as $\copyslice U{\Psi} : \Id \to \freshslice{U}{\Psi} \sumslice{U}{\Psi}$ and the co-unit as $\dropslice U{\Psi} : \sumslice{U}{\Psi} \freshslice{U}{\Psi} \to \Id$.
	\end{itemize}
\end{definition}
This is indeed a generalization:
\begin{proposition}
	The functor $\freshslice{U}{\yoneda W_0} : \catW / \yoneda W_0 \to \catV/(\yoneda W_0 \multip \yoneda U)$ is equal to $\freshslice{U}{W_0} : \catW / W_0 \to \catV / (W_0 \multip U)$ over the obvious isomorphisms between their domains and codomains.
	Hence, each of the presheafwise notions implies the slicewise notion (\cref{def:act-slices}). Moreover, each of the $\top$-slice elemental notions implies the basic $\top$-slice notion.
\end{proposition}
\begin{proof}
	Most of this is straightforward after extracting the construction of the isomorphism $\yoneda W_0 \multip \yoneda U \cong \yoneda(W_0 \multip U)$ from the proof of \cref{thm:adjoint-triple}. To see the last claim, note that
	\begin{align*}
		\set{\vfi : \DSub{W \multip U}{\yoneda W_0 \multip \yoneda U}}{\pi_2 \circ \vfi = \pi_2} \cong ((W \multip U, \pi_2) \to (W_0 \multip U, \pi_2)) = (\freshbase U W \to \freshbase U W_0).
	\end{align*}
	So if injectivity/surjectivity holds for all $W$ and $W_0$, then we can conclude that $\freshbase U$ is faithful/full.
\end{proof}
Note that both notions of presheafwise shard-freedom are well-defined:
\begin{proposition} \label{thm:pwsf-well-defined}
	\begin{enumerate}
		\item (Obsolete.) The functor $\freshslice{U}{\Psi}$ produces indirectly dimensionally split elements.
		\item The functor $\freshslice{U}{\Psi}$ produces directly dimensionally split elements.
		\item Directly dimensionally split elements are indirectly dimensionally split with the same section. Hence, direct shards are indirect shards and indirect presheafwise shard-freedom implies direct presheafwise shard-freedom.
	\end{enumerate}
\end{proposition}
\begin{proof}
	See \cref{thm:swsf-well-defined}.
\end{proof}
\begin{proposition}\label{thm:pwfaithful}
	If $\loch \multip U$ is $\top$-slice faithful, then it is presheafwise faithful.
\end{proposition}
\begin{proof}
	Analogous to \cref{thm:swfaithful}.
\end{proof}
\begin{proposition}\label{thm:elementally-faithful}
	If $\loch \multip U$ is $\top$-slice fully faithful, then it is $\top$-slice elementally faithful.
\end{proposition}
\begin{proof}
	We have
	\begin{align}
		&\set{\vfi : \DSub{W \multip U}{\Psi \multip \yoneda U}}{\pi_2 \circ \vfi = \pi_2} \nn \\
		&\cong \exists W_0 . (\vfi' : W \multip U \to W_0 \multip U) \times (\psi : \DSub{W_0}{\Psi}) \times (\pi_2 \circ (\psi \multip \yoneda U) \circ \vfi' = \pi_2) \nn \\
		&\cong \exists W_0 . (\vfi' : W \multip U \to W_0 \multip U) \times (\psi : \DSub{W_0}{\Psi}) \times (\pi_2 \circ \vfi' = \pi_2) \nn \\
		&\cong \exists W_0 . (\vfi' : \freshbase U W \to \freshbase U W_0) \times (\psi : \DSub{W_0}{\Psi}) \label{eq:coend-elemental-cod}
	\end{align}
	and
	\begin{align}
		(\DSub W \Psi) \cong \exists W_0 . (W \to W_0) \times (\DSub{W_0}{\Psi}).
		 \label{eq:coend-elemental-dom}
	\end{align}
	Moreover, the action of $\freshelem{U}{\Psi}$ sends $(W_0, \chi, \psi)$ in \cref{eq:coend-elemental-dom} to $(W_0, \freshbase U \chi, \psi)$ in \cref{eq:coend-elemental-cod}. Naively, one would say that this proves injectivity, but some care is required with the equality relation for co-ends. It might be that $(W_0, \chi, \psi)$ and $(W_0, \chi', \psi)$ are sent to the same object.
	This would mean that there exists a zigzag $\zeta$ from $W_0$ to itself and jagwise morphisms $\freshbase U W \to \freshbase U \zeta$ (a priori not necessarily in the image of $\freshbase U$ which is why we need $\top$-slice fullness) and jagwise cells $\DSub{\zeta}{\Psi}$ such that the following diagrams commute:
	\begin{equation}
		\xymatrix{
			& \freshbase U W_0
				\ar@{~}[dd]^{\freshbase U \zeta}
			& W_0
				\ar@{~}[dd]_{\zeta}
				\ar@{=>}[rd]^{\psi}
			\\
			\freshbase U W
				\ar[ru]^{\freshbase U \chi}
				\ar[rd]_{\freshbase U \chi'}
				\ar@<1ex>[r]
				\ar@{}@<0ex>[r]|{\cdots}
				\ar@<-1ex>[r]
			&&
				\ar@<1ex>@{=>}[r]
				\ar@{}@<0ex>[r]|{\cdots}
				\ar@<-1ex>@{=>}[r]
			& \Psi.
			\\
			& \freshbase U W_0
			& W_0
				\ar@{=>}[ru]_{\psi}
		}
	\end{equation}
	Then by full faithfulness of $\freshbase U$, we see that the unique preimage of the left triangle exists and also commutes and hence $\psi \circ \chi = \psi \circ \chi'$, so that $(W_0, \chi, \psi) = (W, \id, \psi \circ \chi) = (W, \id, \psi \circ \chi') = (W_0, \chi', \psi)$.
\end{proof}
\begin{proposition}\label{thm:pwfull}
	If $\loch \multip U$ is $\top$-slice fully faithful, then it is presheafwise full.
\end{proposition}
\begin{proof}
	Pick $(W, \psi)$ and $(W', \psi')$ in $\catW / \Psi$ and a morphism $\chi : \freshslice{U}{\Psi}(W, \psi) \to \freshslice{U}{\Psi}(W', \psi')$. Then we also have $\chi : \freshbase U W \to \freshbase U W'$ and by fullness, we find a preimage $\chi_0 : W \to W'$ under $\freshbase U$.
	We have $(\psi' \multip \yoneda U) \circ \chi = \psi \multip \yoneda U$, so by $\top$-slice elemental faithfulness, we see that $\psi' \circ \chi_0 = \psi$, so that $\chi_0$ is a morphism of slice objects $\chi_0 : (W, \psi) \to (W', \psi') \in \catW / \Psi$ and $\freshslice U \Psi \chi_0 = \chi$.
\end{proof}
\begin{proposition}\label{thm:elementally-full}
	If $\loch \multip U$ is $\top$-slice full, then it is $\top$-slice elementally full.
\end{proposition}
\begin{proof}
	In the proof of \cref{thm:elementally-faithful}, we saw that $\freshelem{U}{\Psi}$ essentially sends $(W_0, \chi, \psi_0)$ to $(W_0, \freshbase U \chi, \psi_0)$. Then if $\freshbase U \chi$ is full, it is immediate that this operation is surjective.
\end{proof}
\begin{proposition}\label{thm:pwsf}
	\begin{enumerate}
		\item If $\loch \multip U : \catW \to \catV$ is $\top$-slice full, then direct and indirect dimensional splitness are equivalent, with the same dimensional sections.
		\item (Obsolete.) If $\loch \multip U$ is indirectly slicewise shard-free, then it is indirectly presheafwise shard-free.
		\item If $\loch \multip U$ is $\top$-slice full and shard-free, then it is directly presheafwise shard-free.
	\end{enumerate}
\end{proposition}
\begin{proof}
	\begin{enumerate}
		\item We already know that direct dimensional splitness implies indirect dimensional splitness with the same section (\cref{thm:swsf-well-defined}). We prove the other implication.
		
		Pick some $(V, \vfi) \in \catV/(\Psi \multip \yoneda U)$ that is indirectly dimensionally split with section $\chi$.
		By $\top$-slice elemental fullness, there is a cell $\psi : \DSub W \Psi$ such that $\psi \multip \yoneda U = \vfi \circ \chi : \DSub{W \multip U}{\Psi \multip \yoneda U}$. Then $\vfi$ is directly dimensionally split with section $\chi$.
		\begin{equation}
			\xymatrix{
				V \ar@{<-}[rr]^{\chi} \ar@{=>}[rd]_{\vfi}
				&& W \multip U \ar@{:>}[ld]^{\psi \multip U} \ar@/^{2em}/@{=>}[ldd]^{\pi_2}
				\\
				& \Psi \multip \yoneda U \ar[d]^{\pi_2}
				\\
				& \yoneda U
			}
		\end{equation}
		
		\item Pick a slice object $(V, \vfi) \in \catV/(\Psi \multip \yoneda U)$ such that $\pi_2 \circ \vfi$ is dimensionally split.
		By definition of $\loch \multip \yoneda U$, there is some $W_0$ such that $\vfi$ factors as $\vfi = (\psi^{\DSub{W_0}{\Psi}} \multip \yoneda U) \circ \chi$. Clearly, $\pi_2 \circ \vfi = \pi_2 \circ \chi$ is dimensionally split. Hence, by indirect slicewise shard-freedom, $(V, \chi) \cong \freshslice{U}{W_0}(W, \chi') \in \catV/(W_0 \multip U)$ for some $(W, \chi') \in \catW / W_0$. Then we also have $(V, \vfi) = (V, (\psi \multip \yoneda U) \circ \chi) \cong \freshslice{U}{\Psi}(W, \psi \circ \chi')$.
		\item Pick some $(V, \vfi) \in \catV/(\Psi \multip \yoneda U)$ that is directly dimensionally split. Then $\pi_2 \circ \vfi$ is dimensionally split.
		Because $\freshbase{U}$ is essentially surjective on $\dimslice \catV U$, there must be some $W \in \catW$ such that $\iota : \freshbase{U} W = (W \multip U, \pi_2) \cong (V, \pi_2 \circ \vfi)$ as slice objects over $U$.
		By $\top$-slice elemental fullness, there is a cell $\psi : \DSub W \Psi$ such that $\psi \multip \yoneda U = \vfi \circ \iota : \DSub{W \multip U}{\Psi \multip \yoneda U}$. Thus, $\iota\inv : (V, \vfi) \cong (W \multip U, \psi \multip \yoneda U) = \freshslice{U}{\Psi}(W, \psi)$ as slice objects over $\Psi \multip \yoneda U$.
		\begin{equation}
			\xymatrix{
				V \ar@{<-}[rr]^{\iota}_{\cong} \ar@{=>}[rd]_{\vfi}
				&& W \multip U \ar@{:>}[ld]^{\psi \multip U} \ar@/^{2em}/@{=>}[ldd]^{\pi_2}
				\\
				& \Psi \multip \yoneda U \ar[d]^{\pi_2}
				\\
				& \yoneda U
			}
		\end{equation}
	\end{enumerate}
\end{proof}
\begin{proposition} \label{thm:pwra}
	If $\loch \multip U$ is $\top$-slice right adjoint, then it is presheafwise right adjoint, with
	\begin{align*}
		\sumslice{U}{\Psi}(V, (\psi \multip \yoneda U) \circ \vfi_0) &= \pairslice{}{\psi} \sumslice{U}{W_0} (V, \vfi_0), \\
		\dropslice{U}{\Psi}(W, \psi) &= \dropbase U W, \\
		\copyslice{U}{\Psi}(V, \vfi) &= \copybase U (V, \pi_2 \circ \vfi).
	\end{align*}
\end{proposition}
\begin{proof}
	Pick $(V, \vfi) \in \catV / (\Psi \multip \yoneda U)$. Then $\vfi$ factors as $(\psi^{\DSub{W_0}{\Psi}} \multip \yoneda U) \circ \vfi_0^{V \to W_0 \multip U}$. Then $(V, \vfi_0) \in \catV / (W_0 \multip U)$ and hence $\sumslice{U}{W_0} (V, \vfi_0) \in \catW/W_0$. We define
	\begin{align*}
		\sumslice{U}{\Psi}(V, \vfi)
		&:= \pairslice{}{\psi} \sumslice{U}{W_0} (V, \vfi_0) \\
		&= \pairslice{}{\psi} (\sumbase U (V, \pi_2 \circ \vfi_0), \dropbase U \circ \sumbase U \vfi_0) \\
		&= (\sumbase U (V, \pi_2^{W_0 \multip U \to U} \circ \vfi_0), \psi \circ \dropbase U \circ \sumbase U \vfi_0) \\
		&= (\sumbase U (V, \pi_2^{\Psi \multip \yoneda U \to \yoneda U} \circ \vfi), \psi \circ \dropbase U \circ \sumbase U \vfi_0).
	\end{align*}
	We need to prove that this is well-defined, i.e. respects equality on the co-end that defines $\DSub{V}{\Psi \multip \yoneda U}$. To this end, assume that $\vfi = (\psi_0^{\DSub{W_0}{\Psi}} \multip \yoneda U) \circ \vfi_0^{V \to W_0 \multip U} = (\psi_1^{\DSub{W_1}{\Psi}} \multip \yoneda U) \circ \vfi_1^{V \to W_1 \multip U}$. This means there are a zigzag $\zeta$ from $W_0$ to $W_1$, jagwise morphisms $V \to \zeta \multip U$ and jagwise cells $\DSub \zeta \Psi$ such that the following triangles commute:
	\begin{equation}
		\xymatrix{
			& W_0 \multip U
				\ar@{~}[dd]^{\zeta \multip U}
			& W_0
				\ar@{~}[dd]_{\zeta}
				\ar@{=>}[rd]^{\psi_0}
			\\
			V
				\ar[ru]^{\vfi_0}
				\ar[rd]^{\vfi_1}
				\ar@<-1ex>[r]
				\ar@{}@<0ex>[r]|{\cdots}
				\ar@<1ex>[r]
			&&
				\ar@{=>}@<-1ex>[r]
				\ar@{}@<0ex>[r]|{\cdots}
				\ar@{=>}@<1ex>[r]
			& \Psi.
			\\
			& W_1 \multip U
			& W_1
				\ar@{=>}[ru]_{\psi_1}
		}
	\end{equation}
	By naturality of $\pi_2$, we find that $(V, \pi_2 \circ \vfi_0) = (V, \pi_2 \circ \vfi_1) \in \catV/U$. By naturality of $\dropbase U$, we find that $\psi_0 \circ \dropbase U \circ \sumbase U \vfi_0 = \psi_1 \circ \dropbase U \circ \sumbase U \vfi_1 : \DSub{(V, \pi_2 \circ \vfi_0) = (V, \pi_2 \circ \vfi_1)}{\Psi}$. We conclude that $\sumslice{U}{\Psi}(V, \vfi)$ is well-defined.
	
	To prove adjointness, we first show how $\freshslice U \Psi$ on the right can be turned into $\sumslice U \Psi$ on the left. Pick a morphism $\chi : (V, \vfi) \to \freshslice U \Psi (W, \psi) = (W \multip U, \psi \multip \yoneda U)$ in $\catV / (\Psi \multip \yoneda U)$. Then one representation of $\vfi$ is $\vfi = (\psi \multip \yoneda U) \circ \chi$ so by definition, $\sumslice U \Psi (V, \vfi) = (\sumbase U (V, \pi_2 \circ \vfi), \psi \circ \dropbase U \circ \sumbase U \chi)$ which clearly factors over $\psi$, i.e. has a morphism $\dropbase U \circ \sumbase U \chi : \sumslice U \Psi (V, \vfi) \to (W, \psi)$. If $\chi = \id$, then we obtain the co-unit $\dropslice U \Psi = \dropbase U \circ \sumbase U \id = \dropbase U$.
	
	Next, we construct the unit $\copyslice U \Psi : (V, \vfi) \to \freshslice U \Psi \sumslice U \Psi (V, \vfi)$. If $\vfi = (\psi \multip \yoneda U) \circ \vfi_0$, then we have
	\begin{align*}
		\freshslice U \Psi \sumslice U \Psi (V, \vfi)
		&= \freshslice U \Psi \pairslice{}{\psi} \sumslice{U}{W_0} (V, \vfi_0) \\
		&= \pairslice{}{\psi \multip \yoneda U} \freshslice U {W_0} \sumslice{U}{W_0} (V, \vfi_0).
	\end{align*}
	On the other hand, $(V, \vfi) = \pairslice{}{\psi \multip \yoneda U}(V, \vfi_0)$, so as the unit we can take $\copyslice{U}{\Psi} = \pairslice{}{\psi \multip \yoneda U} \copyslice{U}{W_0} = \copyslice{U}{W_0} = \copybase U$.
	
	The adjunction laws are then inherited from $\sumbase U \dashv \freshbase U$.
\end{proof}
\begin{proposition}[Functoriality of the category of elements] \label{thm:pw-functoriality}
	A morphism of multipliers $\loch \multip \upsilon : \loch \multip U \to \loch \multip U'$ gives rise to a natural transformation $\pairslice{}{\Psi \multip \yoneda \upsilon} \circ \freshslice{U}{\Psi} \to \freshslice{U'}{\Psi}$.
	Hence, if both multipliers are $\top$-slice right adjoint, we also get $\sumslice{U'}{\Psi} \circ \pairslice{}{\Psi \multip \yoneda \upsilon} \to \sumslice{U}{\Psi}$.
\end{proposition}
\begin{proof}
	For any $(W, \psi) \in \catW / \Psi$, we have to prove $(W \multip U, (\Psi \multip \yoneda \upsilon) \circ (\psi \multip \yoneda U)) \to (W \multip U', \psi \multip \yoneda U')$. The morphism $W \multip \upsilon : W \multip U \to W \multip U'$ does the job.
	The second statement follows from \cref{thm:cotranspose}.
\end{proof}
\begin{theorem}[Presheafwise quantification theorem] \label{thm:pw-quantification}
	If $\loch \multip U$ is
	\begin{enumerate}
		\item $\top$-slice (or equivalently presheafwise) fully faithful and right adjoint, then we have a natural isomorphism $\dropslice U \Psi : \sumslice U{\Psi} \freshslice U{\Psi} \cong \Id$.
		\item copointed, then we have
		\begin{enumerate}
			\item $\hideslice U{\Psi} : \pairslice U{\Psi} \to \sumslice U{\Psi}$ (if $\top$-slice, or equivalently presheafwise, right adjoint),
			\item $\spoilslice U{\Psi} : \freshslice U{\Psi} \to \wknslice U{\Psi}$ (if $\wknslice U{\Psi}$ exists),
			\item in any case $\pairslice U{\Psi} \freshslice U{\Psi} \to \Id$.
		\end{enumerate}
		\item a comonad, then there is a natural transformation $\pairslice{}{\Psi \multip \yoneda \delta} \circ \freshslice U{\Psi} \to \freshslice{U \multip U}{\Psi}$.
		\item cartesian, then we have natural isomorphisms:
		\begin{enumerate}
			\item $\sumslice{U}{\Psi}(V, \vfi) \cong \pairslice{U}{\Psi}(V, \vfi) = (V, \pi_1 \circ \vfi)$,
			\item $\freshslice{U}{\Psi}(W, \psi) \cong \wknslice{U}{\Psi}(W, \psi)$,
			\item $\sumslice{U}{\Psi} \freshslice{U}{\Psi}(W, \psi) \cong \pairslice{U}{\Psi} \wknslice{U}{\Psi}(W, \psi) \cong (W \times \yoneda U, \psi \circ \pi_1)$.
		\end{enumerate}
		Moreover, these isomorphisms become equality if $\sumslice{U}{\Psi}$ is constructed as above from $\sumslice{U}{W_0} = \pairslice{U}{W_0}$, and $\wknslice{U}{\Psi}(W, \psi)$ is chosen wisely. (Both functors are defined only up to isomorphism.)
	\end{enumerate}
\end{theorem}
\begin{proof}
	\begin{enumerate}
		\item This is a standard fact about fully faithful right adjoints such as $\freshslice U{\Psi}$.
		\item By \cref{thm:cotranspose}, it is sufficient to prove $\pairslice U \Psi \freshslice U \Psi \to \Id$, and indeed we have $\pi_1 : \pairslice U \Psi \freshslice U \Psi(W, \psi) = (W \multip U, \pi_1 \circ (\psi \multip \yoneda U)) = (W \multip U, \psi \circ \pi_1) \to (W, \psi)$.
		\item This is a special case of \cref{thm:pw-functoriality}.
		\item \begin{enumerate}
			\item Let $\vfi = (\psi \multip \yoneda U) \circ \vfi_0$. Then we have
			\begin{align*}
				\sumslice{U}{\Psi}(V, \vfi)
				&= \pairslice{}{\psi} \sumslice{U}{W_0} (V, \vfi_0) \\
				&\cong \pairslice{}{\psi} \pairslice{U}{W_0} (V, \vfi_0) \\
				&= \pairslice{}{\psi} (V, \pi_1 \circ \vfi_0) \\
				&= (V, \psi \circ \pi_1 \circ \vfi_0) = (V, \pi_1 \circ (\psi \multip \yoneda U) \circ \vfi_0) = (V, \pi_1 \circ \vfi).
			\end{align*}
			\item This follows from the definitions.
			\item We have
			\begin{equation}
				\sumslice{U}{\Psi} \freshslice{U}{\Psi}(W, \psi) = \sumslice{U}{\Psi} (W \times U, \psi \times \yoneda U) \cong (W \times U, \pi_1 \circ (\psi \times \yoneda U))
			\end{equation}
			and of course $\pi_1 \circ (\psi \times \yoneda U) = \psi \circ \pi_1 : W \times U \to \Psi$. \qedhere
		\end{enumerate}
	\end{enumerate}
\end{proof}
\begin{theorem}[Presheafwise quotient theorem\seelog{}] \label{thm:pw-quotient}
	If $\loch \multip U : \catW \to \catV$ is $\top$-slice (or equivalently presheafwise, for either notion of shard-freedom) fully faithful and shard-free, then
	\begin{enumerate}
		\item (Obsolete.) $\freshslice{U}{\Psi} : \catW/\Psi \simeq (\dimslice{\catV}{U}) / (\Psi \multip \yoneda U, \pi_2)$ is an equivalence of categories.\footnote{We use a slight abuse of notation as $(\dimslice{\catV}{U}) / (\Psi \multip \yoneda U, \pi_2)$ is in fact neither a slice category nor a category of elements.}
		\item $\freshslice{U}{\Psi} : \catW/\Psi \simeq \dimslice{\catV}{(\Psi \multip \yoneda U)}$ is an equivalence of categories. \qed
	\end{enumerate}
\end{theorem}

\subsection{Acting on presheaves} \label{sec:act-psh}
\begin{proposition} \label{thm:psh-multiplier-properties}
	The functor $\loch \multip \yoneda U : \widehat \catW \to \widehat \catV$:
	\begin{enumerate}
		\item is a multiplier for $\yoneda U$,
		\item has the property that $\freshbase{\yoneda U} : \widehat \catW \to \widehat \catV / \yoneda U$ is naturally isomorphic to $\lpsh{(\freshbase U)} : \widehat \catW \to \widehat{\catV / U}$ over the equivalence between their codomains,
		\item has the property that the slice functor $\freshslice{\yoneda U}{\Psi} : \widehat \catW / \Psi \to \widehat \catV / (\Psi \multip \yoneda U)$ is naturally isomorphic to the left lifting of the elements functor $\lpsh{(\freshslice{U}{\Psi})} : \widehat{\catW / \Psi} \to \widehat{\overbrace{\catV / (\Psi \multip \yoneda U)}}$ over the equivalences between their domains and codomains,
		\item is copointed if and only if $\loch \multip U$ is,
		\item is a comonad if and only if $\loch \multip U$ is,
		\item is cartesian if and only if $\loch \multip U$ is,
		\item is $\top$-slice fully faithful if and only if $\loch \multip U$ is $\top$-slice fully faithful,
		\item is slicewise fully faithful if and only if $\loch \multip U$ is presheafwise fully faithful,
		\item is $\top$-slice right adjoint if $\loch \multip U$ is $\top$-slice right adjoint, and
		\begin{itemize}
			\item $\sumbase{\yoneda U}$ is naturally isomorphic to $\lpsh{(\sumbase U)}$ over the equivalence $\widehat{\catV / U} \simeq \widehat \catV / \yoneda U$,
			\item $\sumslice{\yoneda U}{\Psi}$ is naturally isomorphic to $\lpsh{(\sumslice U \Psi)}$ over the equivalences between their domain and codomain.
		\end{itemize}
	\end{enumerate}
\end{proposition}
\begin{proof}
	\begin{enumerate}
		\item Since $\top \multip \yoneda U \cong \yoneda \top \multip \yoneda U \cong \yoneda (\top \multip U) \cong \yoneda U$. We use, in order, that $\yoneda$ preserves the terminal object, that $\lpsh F \circ \yoneda \cong \yoneda \circ F$ (\cref{thm:adjoint-triple}) and that $\loch \multip U$ is a multiplier for $U$.
		
		\item The functor $\lpsh{(\freshbase{U})}$ sends a presheaf $\Gamma \in \widehat \catW$ to the presheaf in $\widehat{\catV/U}$ determined by
		\begin{equation}
			\DSub{(V, \vfi)}{\lpsh{(\freshbase U)} \Gamma} \quad = \quad \exists W . ((V, \vfi) \to \freshbase U W) \times (\DSub W \Gamma).
		\end{equation}
		On the other hand, $\freshbase{\yoneda U} \Gamma$ is the slice object $(\Gamma \multip \yoneda U, \pi_2) \in \widehat \catV / \yoneda U$. Taking the preimage of $\pi_2$ (\cref{thm:preimage}), we get a presheaf $\Delta \in \widehat{\catV / U}$ determined by
		\begin{align*}
			\DSub{(V, \vfi)}{\Delta} \quad
			&= \quad \set{(\gamma \multip \yoneda U) \circ \chi : \DSub{V}{\Gamma \multip \yoneda U}}{\pi_2 \circ (\gamma \multip \yoneda U) \circ \chi = \vfi} \\
			&= \quad \set{(\gamma \multip \yoneda U) \circ \chi : \DSub{V}{\Gamma \multip \yoneda U}}{\pi_2 \circ \chi = \vfi} \\
			&\cong \quad \exists W . (\chi : V \to W \multip U) \times (\gamma : \DSub W \Gamma) \times (\pi_2 \circ \chi = \vfi) \\
			&\cong \quad \exists W . (\chi : (V, \vfi) \to \freshbase U W) \times (\DSub W \Gamma).
		\end{align*}
		Indeed, we see that these functors are isomorphic.
		
		\item The functor $\lpsh{(\freshslice{U}{\Psi})}$ sends a presheaf $\Psi \sep \Gamma \ctx$ over $\catW/\Psi$ to the presheaf $\Psi \multip \yoneda U \sep \lpsh{(\freshbase{U}{\Psi})} \Gamma \ctx$ over $\catV/(\Psi \multip \yoneda U)$ determined by:
		\begin{equation}
			\DSub{(V, \vfi^{\DSub{V}{\Psi \multip \yoneda U}})}{\lpsh{\paren{\freshslice{U}{\Psi}}} \Gamma} \quad = \quad \exists (W, \psi^{\DSub W \Psi}) . ((V, \vfi) \to \freshslice U \Psi (W, \psi)) \times (\DSub{(W, \psi)}{\Gamma}).
		\end{equation}
		On the other hand, $\freshslice{\yoneda U}{\Psi} (\Psi.\Gamma, \pi)$ is the slice $(\Psi.\Gamma \multip \yoneda U, \pi \multip \yoneda U) \in \widehat \catV / (\Psi \multip \yoneda U)$. Taking the preimage of $\pi \multip \yoneda U$ (\cref{thm:preimage}), we get a presheaf $\Psi \multip \yoneda U \sep \Delta \ctx$ over $\catV / (\Psi \multip \yoneda U)$ determined by
		\begin{align*}
			&\DSub{(V, \vfi^{\DSub{V}{\Psi \multip \yoneda U}})}{\Delta} \\
			&= \set{(\psi.\gamma \multip \yoneda U) \circ \chi : \DSub{V}{\Psi.\Gamma \multip \yoneda U}}{(\pi \multip \yoneda U) \circ (\psi.\gamma \multip \yoneda U) \circ \chi = \vfi} \\
			&= \set{(\psi.\gamma \multip \yoneda U) \circ \chi : \DSub{V}{\Psi.\Gamma \multip \yoneda U}}{(\psi \multip \yoneda U) \circ \chi = \vfi} \\
			&\cong \exists W . (\chi : V \to W \multip U) \times (\psi : \DSub W \Psi) \times (\gamma : \DSub{(W, \psi)}{\Gamma}) \times ((\psi \multip \yoneda U) \circ \chi = \vfi) \\
			&\cong \exists (W, \psi^{\DSub{W}{\Psi}}).(\chi : (V, \vfi) \to \freshslice U \Psi (W, \psi)) \times (\gamma : \DSub{(W, \psi)}{\Gamma}).
		\end{align*}
		Indeed, we see that these functors are isomorphic.
		
		\item Assume that $\loch \multip U$ is copointed. It is immediate from the construction of $\lpsh \loch$ that $\lpsh \loch$ preserves natural transformations. Moreover, we have $\lpsh \Id \cong \Id$, so we get $\pi_1 : (\loch \multip \yoneda U) \to \Id$.
		
		Conversely, assume that $\loch \multip \yoneda U$ is copointed. Then we have $\yoneda(\loch \multip U) \cong (\yoneda \loch \multip \yoneda U) \to \yoneda$. Since $\yoneda$ is fully faithful, we have proven $(\loch \multip U) \to \Id$.
		
		\item Analogous to the previous point.
		
		\item Assume that $\loch \multip U$ is cartesian. We apply the universal property of the cartesian product, and the co-Yoneda lemma:
		\begin{align*}
			\DSub{V}{(\Gamma \multip \yoneda U)}
			&= \exists W.(V \to W \multip U) \times (\DSub W \Gamma) \\
			&\cong \exists W.(V \to W) \times (V \to U) \times (\DSub W \Gamma) \\
			&\cong (V \to U) \times (\DSub V \Gamma) \\
			&\cong (\DSub{V}{\yoneda U}) \times (\DSub V \Gamma).
		\end{align*}
		
		Conversely, if $\loch \multip \yoneda U$ is cartesian, we have
		\begin{align*}
			{V} \to {W \multip U}
			&= \DSub{V}{\yoneda(W \multip U)} \\
			&\cong \DSub{V}{\yoneda W \multip \yoneda U} \\
			&\cong (\DSub{V}{\yoneda W}) \times (\DSub{V}{\yoneda U}) \\
			&\cong (V \to W) \times (V \to U).
		\end{align*}
		
		
		
		\item This follows from point 2 and \cref{thm:left-lifting-ff}.
		
		\item This follows from point 3 and \cref{thm:left-lifting-ff}.
		\item
		\begin{itemize}
			\item We know that $\lpsh{(\sumbase U)} \dashv \lpsh{(\freshbase U)}$ so moving it through the natural isomorphism yields a left adjoint to $\freshbase{\yoneda U}$.
			\item By \cref{thm:pwra}, $\sumslice U \Psi$ exists. We know that $\lpsh{(\sumslice U \Psi)} \dashv \lpsh{(\freshslice U \Psi)}$ so moving it through the natural isomorphism yields a left adjoint to $\freshslice{\yoneda U}{\Psi}$. \qedhere
		\end{itemize}
	\end{enumerate}
\end{proof}

\subsection{Four adjoint functors}
Unlike the slice category $\widehat \catW / \Psi$, the equivalent category $\widehat{\catW / \Psi}$ is a presheaf category and therefore immediately a model of dependent type theory. Therefore, we prefer to work with that category, and to use the corresponding functors:
\begin{definition} \label{def:4-multip-functors}
	The adjoint functors $\sumslice U \Psi \dashv \freshslice U \Psi$ give rise to four adjoint functors between presheaf categories over slice categories, which we denote
	\begin{equation}
		\sumpsh{\yoneda U}{\Psi} \dashv
		\freshpsh{\yoneda U}{\Psi} \dashv
		\lollipsh{\yoneda U}{\Psi} \dashv
		\transppsh{\yoneda U}{\Psi}.
	\end{equation}
	We call the fourth functor \textbf{transpension}.
	
	The units and co-units will be denoted:
	\begin{equation}
		\begin{array}{r c l c r c l}
			\copypsh{\yoneda U}{\Psi} &:& \Id \to \freshpsh{\yoneda U}{\Psi} \sumpsh{\yoneda U}{\Psi}
			&\qquad \qquad&
			\droppsh{\yoneda U}{\Psi} &:& \sumpsh{\yoneda U}{\Psi} \freshpsh{\yoneda U}{\Psi} \to \Id
			\\
			\constpsh{\yoneda U}{\Psi} &:& \Id \to \lollipsh{\yoneda U}{\Psi} \freshpsh{\yoneda U}{\Psi}
			&\qquad \qquad&
			\apppsh{\yoneda U}{\Psi} &:& \freshpsh{\yoneda U}{\Psi} \lollipsh{\yoneda U}{\Psi} \to \Id
			\\
			\reindexpsh{\yoneda U}{\Psi} &:& \Id \to \transppsh{\yoneda U}{\Psi} \lollipsh{\yoneda U}{\Psi}
			&\qquad \qquad&
			\unmeridpsh{\yoneda U}{\Psi} &:& \lollipsh{\yoneda U}{\Psi} \transppsh{\yoneda U}{\Psi} \to \Id
		\end{array}
	\end{equation}
	For now, we define all of these functors only up to isomorphism, i.e. for the middle two we do not specify whether they arise as a left, central or right lifting.
\end{definition}
Note that, if in a judgement $\Psi \sep \Gamma \vdash J$, we view the part before the pipe ($\sep$) as part of the context, then $\sumpsh{\yoneda U}{\Gamma}$ and $\lollipsh{\yoneda U}{\Gamma}$ bind a (substructural) variable of type $\yoneda U$, whereas $\freshpsh{\yoneda U}{\Gamma}$ and $\transppsh{\yoneda U}{\Gamma}$ depend on one.

It is worth mentioning that, since $\loch \multip U = \pairbase U \freshbase U$, the functors in \cref{def:3-multip-functors} can be (essentially) retrieved as
\begin{equation}
	\loch \multip \yoneda U = \pairpsh{\yoneda U}{\top} \freshpsh{\yoneda U}{\top}
	\quad \dashv \quad
	\yoneda U \multimap \loch = \lollipsh{\yoneda U}{\top} \wknpsh{\yoneda U}{\top}
	\quad \dashv \quad
	\yoneda U \amaze \loch = \funcpsh{\yoneda U}{\top} \transppsh{\yoneda U}{\top}.
\end{equation}
\begin{corollary}
	The properties asserted by \cref{thm:psh-multiplier-properties} for $\freshslice{\yoneda U}{\Psi}$ also hold for $\freshpsh{\yoneda U}{\Psi}$.
\end{corollary}
\begin{proof}
	Follows from the fact that $\freshpsh{\yoneda U}{\Psi} \cong \lpsh{(\freshslice U \Psi)}$, and the observation in \cref{thm:psh-multiplier-properties} that this functor in turn corresponds to $\freshslice{\yoneda U}{\Psi}$.
\end{proof}
\begin{proposition}[Presheaf functoriality] \label{thm:psh-functoriality}
	A morphism of multipliers $\loch \multip \upsilon : \loch \multip U \to \loch \multip U'$ gives rise to natural transformations
	\begin{itemize}
		\item $\sumpsh{\yoneda U'}{\Psi} \circ \pairpsh{}{\Psi \multip \yoneda \upsilon} \to \sumpsh{\yoneda U}{\Psi}$ (if $\top$-slice (hence presheafwise) right-adjoint),
		\item $\pairpsh{}{\Psi \multip \yoneda \upsilon} \circ \freshpsh{\yoneda U}{\Psi} \to \freshpsh{\yoneda U'}{\Psi}$ and $\freshpsh{\yoneda U}{\Psi} \to \wknpsh{}{\Psi \multip \yoneda \upsilon} \circ \freshpsh{\yoneda U'}{\Psi}$,
		\item $\lollipsh{\yoneda U'}{\Psi} \to \lollipsh{\yoneda U}{\Psi} \circ \wknpsh{}{\Psi \multip \yoneda \upsilon}$ and $\lollipsh{\yoneda U'}{\Psi} \circ \funcpsh{}{\Psi \multip \yoneda \upsilon} \to \lollipsh{\yoneda U}{\Psi}$,
		\item $\funcpsh{}{\Psi \multip \yoneda \upsilon} \circ \transppsh{\yoneda U}{\Psi} \to \transppsh{\yoneda U'}{\Psi}$,
	\end{itemize}
\end{proposition}
\begin{proof}
	Follows directly from \cref{thm:pw-functoriality}.
\end{proof}
\begin{proposition}[Contextual quantification theorem]
	If $\loch \multip U$ is
	\begin{enumerate}
		\item $\top$-slice (or equivalently presheafwise) fully faithful, then $\droppsh{\yoneda U}{\Psi}$ (if $\top$-slice right adjoint), $\constpsh{\yoneda U}{\Psi}$ and $\unmeridpsh{\yoneda U}{\Psi}$ are natural isomorphisms.
		\item copointed, then we have
		\begin{enumerate}
			\item $\hidepsh{\yoneda U}{\Psi} : \pairpsh{\yoneda U}{\Psi} \to \sumpsh{\yoneda U}{\Psi}$ (if $\top$-slice, or equivalently presheafwise, right adjoint),
			\item $\spoilpsh{\yoneda U}{\Psi} : \freshpsh{\yoneda U}{\Psi} \to \wknpsh{\yoneda U}{\Psi}$,
			\item $\cospoilpsh{\yoneda U}{\Psi} : \funcpsh{\yoneda U}{\Psi} \to \lollipsh{\yoneda U}{\Psi}$.
		\end{enumerate}
		\item a comonad, then we can apply \cref{thm:psh-functoriality} to $\loch \multip \delta : \loch \multip U \to \loch \multip (U \multip U)$.
		\item cartesian, then we have natural isomorphisms:
		\begin{enumerate}
			\item $\sumpsh{\yoneda U}{\Psi} \cong \pairpsh{\yoneda U}{\Psi}$,
			\item $\freshpsh{\yoneda U}{\Psi} \cong \wknpsh{\yoneda U}{\Psi}$,
			\item $\lollipsh{\yoneda U}{\Psi} \cong \funcpsh{\yoneda U}{\Psi}$,
			\item $\transppsh{\yoneda U}{\Psi} \cong \cartransppsh{\yoneda U}{\Psi}$ (if $\wknslice U \Psi$ exists).
		\end{enumerate}
		Equality is achieved for any pair of functors if they are lifted in the same way from functors that were equal in \cref{thm:pw-quantification}.
	\end{enumerate}
\end{proposition}
\begin{proof}
	\begin{enumerate}
		\item This is a standard fact about fully faithful left/right adjoints.
		\item By \cref{thm:cotranspose}, it is sufficient to prove $\pairpsh{\yoneda U}{\Psi} \freshpsh{\yoneda U}{\Psi} \to \Id$, which follows immediately from $\pi_1 : \pairslice{U}{\Psi} \freshslice{U}{\Psi} \to \Id$.
		\item Of course we can.
		\item This is an immediate corollary of \cref{thm:pw-quantification}. \qedhere
	\end{enumerate}
\end{proof}
\begin{proposition}[Fresh exchange]
	If $\Psi \sep \Gamma \ctx$, i.e. $\Gamma \in \widehat{\catW / \Psi}$, then we have an isomorphism of slice objects (natural in $\Gamma$):
	\begin{equation}
		\xymatrix{
			(\Psi \multip \yoneda U) . \freshpsh{\yoneda U}{\Psi} \Gamma
				\ar[rr]^\cong
				\ar[rd]_{\pi}
			&& \Psi.\Gamma \multip \yoneda U
				\ar[ld]^{\pi \multip \yoneda U}
			\\
			& \Psi \multip \yoneda U.
		}
	\end{equation}
\end{proposition}
This proposition explains the meaning of $\freshslice{\yoneda U}{\Gamma}$: it is the type depending on a variable of type $\yoneda U$ whose elements are required to be fresh for that variable, where the meaning of `fresh' depends on the nature of the multiplier. If the multiplier is cartesian, then $\freshslice{\yoneda U}{\Gamma}$ is clearly just weakening over $\yoneda U$.
\begin{proof}
	The slice object on the right is $\freshslice{\yoneda U}{\Psi} (\Psi.\Gamma, \pi)$. By \cref{thm:psh-multiplier-properties}, this is isomorphic to $\freshpsh{\yoneda U}{\Psi} \Gamma$ over the equivalence from \cref{thm:preimage} which sends $\Delta$ to $((\Psi \multip \yoneda U).\Delta, \pi)$.
\end{proof}

\subsection{Investigating the transpension functor}
\begin{definition} \label{def:boundary-multip}
	\begin{enumerate}
		\item We define the \textbf{indirect boundary} $\Psi \multip \partial U$ as the pullback
		\begin{equation}
			\xymatrix{
				\Psi \multip \partial U \ar[r]^{\subseteq} \ar[d]^{\pi_2}
				& \Psi \multip \yoneda U \ar[d]^{\pi_2}
				\\
				\partial U \ar[r]^{\subseteq}
				& \yoneda U,
			}
		\end{equation}
		i.e.\ the subpresheaf of $\Psi \multip \yoneda U$ consisting of all cells $\vfi$ such that $\pi_2 \circ \vfi$ is \emph{not} dimensionally split.
		
		\item We define the \textbf{direct boundary}, also denoted $\Psi \multip \partial U$, as the subpresheaf of $\Psi \multip \yoneda U$ consisting of all cells $\vfi$ that are \emph{not} directly dimensionally split.
	\end{enumerate}
\end{definition}
By \cref{thm:swsf-well-defined}, the indirect boundary is a subpresheaf of the direct boundary.
\begin{remark} \label{rem:cosieve-elements}
	Just like $\top$-slice shard-freedom (\cref{rem:cosieve-base}),
	direct presheafwise shard-freedom can be formulated using (co)sieves.
	A multiplier is directly presheafwise shard-free if either of the following equivalent criteria is satisfied:
	\begin{itemize}
		\item The objects in the essential image of $\freshslice U \Xi$ constitute a cosieve in $\catW/\Xi \multip \yoneda U$.
		\item The objects \emph{outside} the essential image of $\freshslice U \Xi$ constitute a sieve in $\catW/\Xi \multip \yoneda U$.
	\end{itemize}
	The objects of the cosieve generated by objects of the essential image of $\freshslice U \Xi$, are called directly dimensionally split.
	The boundary $\Xi \multip \partial U$ is the largest sieve in $\catW/\Xi \multip \yoneda U$ (largest subpresheaf of $\Xi \multip \yoneda U$) that is disjoint with the objects of the essential image of $\freshslice U \Xi$.
	
	If $\loch \multip U$ is presheafwise fully faithful, then the above conditions are furthermore equivalent to $\freshslice U \Xi$ being a Street opfibration.
\end{remark}
\begin{definition} \label{def:boundary-predicate}
	For either notion of boundary, write $(\in \partial U)$ for the inverse image of $\Psi \multip \partial U \subseteq \Psi \multip \yoneda U$, which is a presheaf over $\catW / (\Psi \multip \yoneda U)$ such that $(\Psi \multip \yoneda U).(\in \partial U) \cong \Psi \multip \partial U$.
	We also write $(\in \partial U)$ for the inverse image of $\partial U \subseteq \yoneda U$.
	Finally, we write $\pairslice{(\in \partial U)}{\Psi \multip \yoneda U} \dashv \ldots$ for the functors arising from $\Psi \multip \partial U \subseteq \Psi \multip \yoneda U$.
\end{definition}
\begin{theorem}[Poles of the transpension] \label{thm:poles}
	For either notion of boundary and any multiplier $\loch \multip U : \catW \to \catV$,
	the functor $\wknpsh{(\in \partial U)}{\Psi \multip \yoneda U} \circ \transppsh{\yoneda U}{\Psi} : \widehat{\catW / \Psi} \to \widehat{\overbrace{\catV / (\Psi \multip \partial U)}}$ sends any presheaf to the terminal presheaf, i.e. $\wknpsh{(\in \partial U)}{\Psi \multip \yoneda U} \circ \transppsh{\yoneda U}{\Psi} = \top$.
\end{theorem}
\begin{proof}
	We show that there is always a unique cell $\DSub{(V, \vfi^{\DSub{V}{\Psi \multip \partial U}})}{\wknpsh{(\in \partial U)}{\Psi \multip \yoneda U} \transppsh{\yoneda U}{\Psi}} \Gamma$. We have
	\begin{align*}
		& \DSub{(V, \vfi^{\DSub{V}{\Psi \multip \partial U}})}{\wknpsh{(\in \partial U)}{\Psi \multip \yoneda U} \transppsh{\yoneda U}{\Psi}} \Gamma \\
		&= \DSub{\pairslice{(\in \partial U)}{\Psi \multip \yoneda U} (V, \vfi^{\DSub{V}{\Psi \multip \partial U}})}{\transppsh{\yoneda U}{\Psi} \Gamma} \\
		&= \DSub{(V, \vfi^{\DSub{V}{\Psi \multip \yoneda U}})}{\transppsh{\yoneda U}{\Psi} \Gamma} \\
		&= {\lollipsh{\yoneda U}{\Psi} \yoneda (V, \vfi^{\DSub{V}{\Psi \multip \yoneda U}})} \to {\Gamma} \\
		&= \forall (W, \psi^{\DSub W \Psi}) . \paren{\DSub{(W, \psi)}{\lollipsh{\yoneda U}{\Psi} \yoneda (V, \vfi^{\DSub{V}{\Psi \multip \yoneda U}})}} \to \paren{\DSub{(W, \psi)}{\Gamma}} \\
		&= \forall (W, \psi^{\DSub W \Psi}) . \paren{\DSub{\freshslice{U}{\Psi}(W, \psi)}{\yoneda (V, \vfi^{\DSub{V}{\Psi \multip \yoneda U}})}} \to \paren{\DSub{(W, \psi)}{\Gamma}} \\
		&= \forall (W, \psi^{\DSub W \Psi}) . \paren{(W \multip U, \psi \multip \yoneda U) \to (V, \vfi^{\DSub{V}{\Psi \multip \yoneda U}})} \to \paren{\DSub{(W, \psi)}{\Gamma}},
	\end{align*}
	and then we see that the last argument $\chi$ cannot exist. Indeed, suppose we have a commuting diagram (where the dotted part only applies in the indirect setting)
	\begin{equation}
		\xymatrix{
			W \multip U \ar[rr]^{\chi}
				\ar@{=>}[rd]_{\psi \multip \yoneda U}
				\ar@{:>}@/_{2em}/[rdd]_{\pi}
			&& V \ar@{=>}[d]^\vfi
			\\
			& \Psi \multip \yoneda U \ar@{.>}[d]^{\pi_2}
			& \Psi \multip \partial U \ar[l]^{\supseteq} \ar@{.>}[d]^{\pi_2}
			\\
			& \yoneda U
			& \partial U. \ar@{.>}[l]^{\supseteq}
		}
	\end{equation}
	\begin{description}
		\item[indirect boundary] Then we see that $\pi_2 \circ \vfi : V \to U$ is dimensionally split with section $\chi$ but is also a cell of $\partial U$ which means exactly that it is not dimensionally split.
		\item[direct boundary] Then we see that $\vfi : \DSub{V}{\Psi \multip \yoneda U}$ is directly dimensionally split with section $\chi$ but it is also a cell of $\Psi \multip \partial U$ which means exactly that it is not directly dimensionally split. \qedhere
	\end{description}
\end{proof}

\wip{\begin{lemma} \label{thm:inverse-image-sum}
	Given a presheaf morphism $\sigma : \Phi \to \Psi$, the inverse image $\Psi \sep \sigma \inv \ctx$ is isomorphic to $\Psi \sep \sigma \inv \cong \pairpsh{}{\sigma} \top \ctx$.
	
	\todoi{We're not really using this at the moment.}
\end{lemma}
\begin{proof}
	We have
	\begin{align*}
		\DSub{(W, \psi)}{\pairpsh{}{\sigma} \top}
		&= \exists (V, \vfi^{\DSub V \Phi}) . (\chi : (W, \psi) \to \pairslice{}{\sigma} (V, \vfi)) \times (\DSub{(V, \vfi)}{\top}) \\
		&\cong \exists (V, \vfi^{\DSub V \Phi}) . (\chi : (W, \psi) \to \pairslice{}{\sigma} (V, \vfi)) \\
		&= \exists (V, \vfi^{\DSub V \Phi}) . (\chi : (W, \psi) \to (V, \sigma \circ \vfi)) \\
		&\cong \exists V . (\vfi : {\DSub V \Phi}) \times (\chi : W \to V) \times (\sigma \circ \vfi \circ \chi = \psi) \\
		&\cong (\vfi' : \DSub{W}{\Phi}) \times (\sigma \circ \vfi' = \psi) \\
		&\cong \set{\vfi' : \DSub{W}{\Phi}}{\sigma \circ \vfi' = \psi} \\
		&= \DSub{(W, \psi)}{\sigma\inv}. \qedhere
	\end{align*}
\end{proof}}

The following theorem shows that dimensionally split morphisms are an interesting concept:
\begin{theorem}[Boundary theorem] \label{thm:boundary}
	\begin{enumerate}
		\item (Obsolete.) Using the indirect boundary, we have
		\[
			\top \multip \yoneda U \sep (\in \partial U) \cong \transppsh{\yoneda U}{\top} \bot \ctx
		\]		
		and more generally 
		\[
			\Psi \multip \yoneda U \sep (\in \partial U) \cong \wknpsh{}{() \multip \yoneda U} \transppsh{\yoneda U}{\top} \bot \ctx.
		\]
		\item Using the direct boundary, we have 
		\[
			\Psi \multip \yoneda U \sep (\in \partial U) \cong \transppsh{\yoneda U}{\Psi} \bot \ctx.
		\]
	\end{enumerate}
\end{theorem}
\begin{proof}
	\begin{enumerate}
		\item We prove the first statement by characterizing the right hand side of the isomorphism. We have
		\begin{align*}
			&\DSub{(V, \vfi^{\DSub{V}{\top \multip \yoneda U}})}{\transppsh{\yoneda U}{\top} \bot} \\
			&= \lollipsh{\yoneda U}{\top} \yoneda (V, \vfi^{\DSub{V}{\top \multip \yoneda U}}) \to \bot \\
			&= \forall (W, ()^{\DSub W \top}) . (\DSub{(W, ())}{\lollipsh{\yoneda U}{\top} \yoneda (V, \vfi^{\DSub{V}{\top \multip \yoneda U}})}) \to (\DSub{(W, ())}{\bot}) \\
			&= \forall (W, ()^{\DSub W \top}) . (\DSub{(W, ())}{\lollipsh{\yoneda U}{\top} \yoneda (V, \vfi^{\DSub{V}{\top \multip \yoneda U}})}) \to \eset \\
			&= \forall (W, ()^{\DSub W \top}) . ({\freshslice{U}{\top}(W, ())} \to {(V, \vfi^{\DSub{V}{\top \multip \yoneda U}})}) \to \eset \\
			&= \forall (W, ()^{\DSub W \top}) . ((W \multip U, () \multip U) \to {(V, \vfi^{\DSub{V}{\top \multip \yoneda U}})}) \to \eset \\
			&\cong \forall W . ((W \multip U, \pi_2) \to (V, \pi_2 \circ \vfi)) \to \eset \\
			&\cong (\exists W . (W \multip U, \pi_2) \to (V, \pi_2 \circ \vfi)) \to \eset.
		\end{align*}
		Clearly, the left hand side of the last line is inhabited if and only if $\pi_2 \circ \vfi$ is dimensionally split. Hence, there is a unique cell $\DSub{(V, \vfi^{\DSub{V}{\top \multip \yoneda U}})}{\transppsh{\yoneda U}{\top} \bot}$ if and only if $\pi_2 \circ \vfi$ is \emph{not} dimensionally split, showing that $\transppsh{\yoneda U}{\top} \bot$ is indeed isomorphic to $(\in \partial U)$.
		
		The second statement follows from applying $\wknpsh{}{() \multip \yoneda U}$ to both sides of the first statement and observing that, being defined by pullback, the indirect boundary predicate is preserved by the substitution functor.
		
		\item We prove this by characterizing the right hand side of the isomorphism. We have
		\begin{align*}
			&\DSub{(V, \vfi^{\DSub{V}{\Psi \multip \yoneda U}})}{\transppsh{\yoneda U}{\Psi} \bot} \\
			&= \lollipsh{\yoneda U}{\Psi} \yoneda (V, \vfi^{\DSub{V}{\Psi \multip \yoneda U}}) \to \bot \\
			&= \forall (W, \psi^{\DSub W \Psi}) . (\DSub{(W, \psi)}{\lollipsh{\yoneda U}{\Psi} \yoneda (V, \vfi^{\DSub{V}{\Psi \multip \yoneda U}})}) \to (\DSub{(W, \psi)}{\bot}) \\
			&= \forall (W, \psi^{\DSub W \Psi}) . (\DSub{(W, \psi)}{\lollipsh{\yoneda U}{\Psi} \yoneda (V, \vfi^{\DSub{V}{\Psi \multip \yoneda U}})}) \to \eset \\
			&= \forall (W, \psi^{\DSub W \Psi}) . ({\freshslice{U}{\Psi}(W, \psi)} \to {(V, \vfi^{\DSub{V}{\Psi \multip \yoneda U}})}) \to \eset \\
			&= \forall (W, \psi^{\DSub W \Psi}) . ((W \multip U, \psi \multip U) \to {(V, \vfi^{\DSub{V}{\Psi \multip \yoneda U}})}) \to \eset \\
			&\cong (\exists (W, \psi^{\DSub W \Psi}) . (W \multip U, \psi \multip U) \to {(V, \vfi^{\DSub{V}{\Psi \multip \yoneda U}})}) \to \eset.
		\end{align*}
		Clearly, the left hand side of the last line is inhabited if and only if $\vfi$ is directly dimensionally split. Hence, there is a unique cell $\DSub{(V, \vfi^{\DSub{V}{\Psi \multip \yoneda U}})}{\transppsh{\yoneda U}{\Psi} \bot}$ if and only if $\vfi$ is \emph{not} directly dimensionally split, showing that $\transppsh{\yoneda U}{\Psi} \bot$ is indeed isomorphic to $(\in \partial U)$. \qedhere
	\end{enumerate}
\end{proof}
\begin{remark} \label{rem:boundary}
	In \cref{sec:commut-multip-subst} (\cref{thm:commut-multip-subst}), we will see that unless the multiplier is $\top$-slice (or equivalently presheafwise) fully faithful, the transpension type may not be stable under substitution. Instead, for $\sigma : \Psi_1 \to \Psi_2$, we only have $\wknpsh{}{\sigma \multip \yoneda U} \circ \transppsh{\yoneda U}{\Psi_2} \to \transppsh{\yoneda U}{\Psi_1} \circ \wknpsh{}{\sigma}$.
	
	Instantiating this with $\sigma = () : \Psi \to \top$ and applying both hands to $\bot$, which is preserved by the substitution functor, we find $\wknpsh{}{() \multip \yoneda U} \transppsh{\yoneda U}{\top} \bot \to \transppsh{\yoneda U}{\Psi} \bot$, i.e.\ the indirect boundary predicate implies the direct boundary predicate.
	
	Since the transpension type \emph{is} stable under substitution for $\top$-slice (or equivalently presheafwise) fully faithful multipliers, we can conclude that for those multipliers, both notions of boundary coincide.
	In fact, we already proved this for $\top$-slice full multipliers (\cref{thm:pwsf}).
\end{remark}

\begin{theorem}[Transpension elimination] \label{thm:transp-elim}
	Let $\loch \multip U : \catW \to \catV$ be $\top$-slice (or equivalently presheafwise) fully faithful and shard-free. Then we have\footnote{regardless of the notion of boundary, as these coincide for $\top$-slice full multipliers (\cref{thm:pwsf}); we do not even have to distinguish cases in the proof as we will simply apply the appropriate version of the quotient \cref{thm:pw-quotient}.}
	\begin{equation}
		\inference{
			\Psi \multip \yoneda U \sep \Gamma \ctx \\
			\Psi \sep \lollipsh{\yoneda U}{\Psi} \Gamma \sez A \type \\
			\Psi \multip \yoneda U \sep \Gamma . \dra{ \transppsh{\yoneda U}{\Psi} }{ A } \sez B \type \\
			\Psi \multip \partial U \sep \wknpsh{(\in \partial U)}{\Psi \multip \partial U} \Gamma \sez b_\partial : \paren{ \wknpsh{(\in \partial U)}{\Psi \multip \partial U} B }[(\id, \_)] \\
			\Psi \sep \paren{ \lollipsh{\yoneda U}{\Psi} \Gamma } . A \sez \mathring b : \paren{\lollipsh{\yoneda U}{\Psi} B} \brac{  \paren{  \pi, \paren{\unmeridpsh{\yoneda U}{\Psi}}\inv(\xi)  }  } \\
			\Psi \multip \partial U \sep \wknpsh{(\in \partial U)}{\Psi \multip \partial U} \freshpsh{\yoneda U}{\Psi} \paren{  \paren{ \lollipsh{\yoneda U}{\Psi} \Gamma } . A  } \sez \ftrtm{\wknpsh{(\in \partial U)}{\Psi \multip \partial U}}{  \paren{  \apppsh{\yoneda U}{\Psi} \paren{  \ftrtm{\freshpsh{\yoneda U}{\Psi}}{\mathring b}  }  }  } = b_\partial \brac{  \wknpsh{(\in \partial U)}{\Psi \multip \partial U} \paren{  \apppsh{\yoneda U}{\Psi} \circ \pi  }  } \\
			\qquad : \paren{ \wknpsh{(\in \partial U)}{\Psi \multip \partial U} B } [(\id, \_)] \brac{  \wknpsh{(\in \partial U)}{\Psi \multip \partial U} \paren{  \apppsh{\yoneda U}{\Psi} \circ \pi  }  }
		}{
			\Psi \multip \yoneda U \sep \Gamma . \dra{ \transppsh{\yoneda U}{\Psi} }{ A } \sez b : B
		}{}
	\end{equation}
	and $b$ reduces to $b_\partial$ and $\mathring b$ if we apply to it the same functors and substitutions that have been applied to $B$ in the types of $b_\partial$ and $\mathring b$.
	
	(If the multiplier is not $\top$-slice (or equivalently presheafwise) right adjoint, then $\freshpsh{\yoneda U}{\Psi}$ may not be a CwF morphism, but the term $\apppsh{\yoneda U}{\Psi} \paren{  \ftrtm{\freshpsh{\yoneda U}{\Psi}}{\mathring b}  }$ is essentially a dependent transposition for the adjunction $\freshpsh{\yoneda U}{\Psi} \dashv \lollipsh{\yoneda U}{\Psi}$ which even exists if only the right adjoint is a CwF morphism \cite{reldtt-techreport}).
	
	In words: if we want to eliminate an element of the transpension type, then we can do so by induction. We distinguish two cases and a coherence condition:
	\begin{itemize}
		\item In the first case ($b_\partial$), we are on the boundary of $U$ and the transpension type trivializes.
		\item In the second case, we are defining an action on cells that live over all of $\yoneda U$. In the transpension type, such cells are in 1-1 correspondence with cells of type $A$ under the isomorphism $\unmeridpsh{\yoneda U}{\Psi} : \lollipsh{\yoneda U}{\Psi}\transppsh{\yoneda U}{\Psi} \cong \Id$.
		\item The boundary of the image of cells in the second case, must always be $b_\partial$.
	\end{itemize}
\end{theorem}
Note that right adjoint weak CwF morphisms such as $\transppsh{\yoneda U}{\Psi}$ give rise to a DRA by applying the CwF morphism and then substituting with the unit of the adjunction. As such, the transpension \emph{type} is modelled by the DRA sending $A$ to $\dra{ \transppsh{\yoneda U}{\Psi} }{ A } = \paren{\transppsh{\yoneda U}{\Psi} A} \brac{\reindexpsh{\yoneda U}{\Psi}}$.
\begin{proof}
	\textbf{Well-formedness.} We first show that the theorem is well-formed.
	\begin{itemize}
		\item The rule for $\Gamma$ just assumes that $\Gamma$ is a presheaf over $\catV / (\Psi \multip \yoneda U)$.
		
		\item Then $\lollipsh{\yoneda U}{\Psi} \Gamma$ is a presheaf over $\catW / \Psi$ and we assume that $A$ is a type in that context, i.e. a presheaf over the category of elements of $\lollipsh{\yoneda U}{\Psi} \Gamma$.
		
		\item Then the DRA of $\transppsh{\yoneda U}{\Psi}$ applied to $A$ is a type in context $\Gamma$.
		We assume that $B$ is a type over the extended context.
		
		\item Being a central lifting, $\wknpsh{(\in \partial U)}{\Psi \multip \partial U}$ is a CwF morphism and can be applied to $B$, yielding a type in context
		\begin{align*}
			\wknpsh{(\in \partial U)}{\Psi \multip \partial U} \paren{  \Gamma . \paren{ \transppsh{\yoneda U}{\Psi} A }\brac{ \reindexpsh{\yoneda U}{\Psi} }  }
			&=
			\wknpsh{(\in \partial U)}{\Psi \multip \partial U}\Gamma . \paren{ \wknpsh{(\in \partial U)}{\Psi \multip \partial U} \transppsh{\yoneda U}{\Psi} A }\brac{ \wknpsh{(\in \partial U)}{\Psi \multip \partial U} \reindexpsh{\yoneda U}{\Psi} } \\
			&\cong
			\wknpsh{(\in \partial U)}{\Psi \multip \partial U}\Gamma . \top,
		\end{align*}
		where the isomorphism is an application of \cref{thm:poles}. The substitution $(\id, \_) = \pi\inv$ yields a type in context $\wknpsh{(\in \partial U)}{\Psi \multip \partial U}\Gamma$. We assume that $b_\partial$ has this type.
		
		\item Being a central lifting, $\lollipsh{\yoneda U}{\Psi}$ is a CwF morphism and can be applied to $B$, yielding a type in context
		\begin{align*}
			\lollipsh{\yoneda U}{\Psi} \paren{
				\Gamma . \paren{ \transppsh{\yoneda U}{\Psi} A }\brac{ \reindexpsh{\yoneda U}{\Psi} }
			}
			=
			\lollipsh{\yoneda U}{\Psi} \Gamma . 
			\paren{ \lollipsh{\yoneda U}{\Psi} \transppsh{\yoneda U}{\Psi} A }\brac{ \lollipsh{\yoneda U}{\Psi} \reindexpsh{\yoneda U}{\Psi} }.
		\end{align*}
		The natural transformation $(\unmeridpsh{\yoneda U}{\Psi})\inv$ gives rise \cite{reldtt-techreport} to a function
		\begin{equation}
			(\unmeridpsh{\yoneda U}{\Psi})\inv : A \to \paren{  \lollipsh{\yoneda U}{\Psi} \transppsh{\yoneda U}{\Psi} A  }[(\unmeridpsh{\yoneda U}{\Psi})\inv].
		\end{equation}
		Now, by the adjunction laws, $\lollipsh{\yoneda U}{\Psi} \reindexpsh{\yoneda U}{\Psi} \circ \unmeridpsh{\yoneda U}{\Psi} = \id$, so
		\begin{equation}
			\lollipsh{\yoneda U}{\Psi} \reindexpsh{\yoneda U}{\Psi}
			= \lollipsh{\yoneda U}{\Psi} \reindexpsh{\yoneda U}{\Psi} \circ \unmeridpsh{\yoneda U}{\Psi} \circ (\unmeridpsh{\yoneda U}{\Psi})\inv
			= (\unmeridpsh{\yoneda U}{\Psi})\inv.
		\end{equation}
		Then we have
		\begin{equation}
			(\unmeridpsh{\yoneda U}{\Psi})\inv : A \to \paren{  \lollipsh{\yoneda U}{\Psi} \transppsh{\yoneda U}{\Psi} A  }\brac{ \lollipsh{\yoneda U}{\Psi} \reindexpsh{\yoneda U}{\Psi} }.
		\end{equation}
		Thus, we can substitute $\lollipsh{\yoneda U}{\Psi} B$ with $(\pi, (\unmeridpsh{\yoneda U}{\Psi})\inv(\xi))$, yielding a type in the desired context. We assume that $\mathring b$ has this type.
		
		\item In the coherence criterion, we have applied operations to $b_\partial$ and $\mathring b$ before equating them. We have to ensure that the resulting terms are well-typed in the given context and type.
		\begin{itemize}
			\item If we apply $\freshpsh{\yoneda U}{\Psi}$ to the term $\mathring b$, we get
			\begin{align*}
				\Psi \multip \yoneda U \sep \paren{ \freshpsh{\yoneda U}{\Psi} \lollipsh{\yoneda U}{\Psi} \Gamma } . \freshpsh{\yoneda U}{\Psi} A \sez \ftrtm{\freshpsh{\yoneda U}{\Psi}}{\mathring b} : \paren{\freshpsh{\yoneda U}{\Psi}\lollipsh{\yoneda U}{\Psi} B} \brac{  \freshpsh{\yoneda U}{\Psi} \paren{  \pi, \paren{\unmeridpsh{\yoneda U}{\Psi}}\inv(\xi)  }  }.
			\end{align*}
			If we subsequently apply $\apppsh{\yoneda U}{\Psi}$, we get
			\begin{align*}
				\Psi \multip \yoneda U \sep \paren{ \freshpsh{\yoneda U}{\Psi} \lollipsh{\yoneda U}{\Psi} \Gamma } . \freshpsh{\yoneda U}{\Psi} A \sez \apppsh{\yoneda U}{\Psi} \paren{\ftrtm{\freshpsh{\yoneda U}{\Psi}}{\mathring b}} : B \brac{\apppsh{\yoneda U}{\Psi}} \brac{  \freshpsh{\yoneda U}{\Psi} \paren{  \pi, \paren{\unmeridpsh{\yoneda U}{\Psi}}\inv(\xi)  }  }.
			\end{align*}
			Next, we apply $\wknpsh{(\in \partial U)}{\Psi \multip \yoneda U}$ and obtain something of type
			\begin{align*}
				\paren{\wknpsh{(\in \partial U)}{\Psi \multip \yoneda U} B}
				\brac{\wknpsh{(\in \partial U)}{\Psi \multip \yoneda U} \apppsh{\yoneda U}{\Psi}}
				\brac{  \wknpsh{(\in \partial U)}{\Psi \multip \yoneda U} \freshpsh{\yoneda U}{\Psi} \paren{  \pi, \paren{\unmeridpsh{\yoneda U}{\Psi}}\inv(\xi)  }  }.
			\end{align*}
			Now if we look at the context of $\wknpsh{(\in \partial U)}{\Psi \multip \yoneda U} B$, we see that the last type is the unit type by \cref{thm:poles}, so the substitution applied to $B$ is determined by its weakening. So we rewrite:
			\begin{align*}
				\ldots
				&=\paren{\wknpsh{(\in \partial U)}{\Psi \multip \yoneda U} B}
				[(\id, \_)]
				[\pi]
				\brac{\wknpsh{(\in \partial U)}{\Psi \multip \yoneda U} \apppsh{\yoneda U}{\Psi}}
				\brac{  \wknpsh{(\in \partial U)}{\Psi \multip \yoneda U} \freshpsh{\yoneda U}{\Psi} \paren{  \pi, \paren{\unmeridpsh{\yoneda U}{\Psi}}\inv(\xi)  }  } \\
				&=\paren{\wknpsh{(\in \partial U)}{\Psi \multip \yoneda U} B}
				[(\id, \_)]
				\brac{\wknpsh{(\in \partial U)}{\Psi \multip \yoneda U} \apppsh{\yoneda U}{\Psi}}
				[\pi]
				\brac{  \wknpsh{(\in \partial U)}{\Psi \multip \yoneda U} \freshpsh{\yoneda U}{\Psi} \paren{  \pi, \paren{\unmeridpsh{\yoneda U}{\Psi}}\inv(\xi)  }  } \\
				&=\paren{\wknpsh{(\in \partial U)}{\Psi \multip \yoneda U} B}
				[(\id, \_)]
				\brac{\wknpsh{(\in \partial U)}{\Psi \multip \yoneda U} \apppsh{\yoneda U}{\Psi}}
				\brac{  \wknpsh{(\in \partial U)}{\Psi \multip \yoneda U} \freshpsh{\yoneda U}{\Psi} \paren{ \pi \circ \paren{  \pi, \paren{\unmeridpsh{\yoneda U}{\Psi}}\inv(\xi)  } }  } \\
				&=\paren{\wknpsh{(\in \partial U)}{\Psi \multip \yoneda U} B}
				[(\id, \_)]
				\brac{\wknpsh{(\in \partial U)}{\Psi \multip \yoneda U} \apppsh{\yoneda U}{\Psi}}
				\brac{  \wknpsh{(\in \partial U)}{\Psi \multip \yoneda U} \freshpsh{\yoneda U}{\Psi} \pi  } \\
				&= \paren{\wknpsh{(\in \partial U)}{\Psi \multip \yoneda U} B}
				[(\id, \_)]
				\brac{\wknpsh{(\in \partial U)}{\Psi \multip \yoneda U} \paren{\apppsh{\yoneda U}{\Psi} \circ \pi}}.
			\end{align*}
			
			\item It is immediate that the substitution applied to $b_\partial$ yields the given type.
		\end{itemize}
	\end{itemize}
	
	\textbf{Soundness of the coherence criterion.} Note that, if we apply to $b$ the same reasoning that we applied to $B$ to show well-formedness of the last 3 premises, we find that the coherence criterion does hold if $b_\partial$ and $\mathring b$ arise from a common $b$.

	\textbf{Completeness of the elimination clauses.} We now show that $b$ is fully determined by the $b_\partial$ and $\mathring b$ that can be derived from it. Afterwards, we will show that the given coherence condition is sufficient to make sure that $b_\partial$ and $\mathring b$ determine some $b$.
	
	Note that $B$, being a type in a presheaf CwF, is a presheaf over the category of elements of $\Gamma . \paren{ \transppsh{\yoneda U}{\Psi} A }\brac{ \reindexpsh{\yoneda U}{\Psi} }$. Hence it acts on cells
	\begin{align*}
		\paren{
			V, \vfi^{\DSub{V}{\Psi \multip \yoneda U}}, \gamma^{\DSub{(V, \vfi)}{\Gamma}}, a^{\DSub{(V, \vfi, \gamma)}{  \paren{ \transppsh{\yoneda U}{\Psi} A }\brac{ \reindexpsh{\yoneda U}{\Psi} }  }}
		}.
	\end{align*}
	Now we divide such cells in two classes: on-boundary cells (for which $(V, \vfi)$ is on the boundary) and total cells (the others). As $\wknpsh{(\in \partial U)}{\Psi \multip \yoneda U}$ is exactly the restriction of presheaves to the on-boundary cells, it is clear that $b_\partial$ determines the action of $b$ on those.
	
	For total cells, note that the full subcategory of $\catV / (\Psi \multip \yoneda U)$ consisting of the total elements, is (by \cref{thm:pw-quotient}) equivalent to $\catW / \Psi$, with one direction given by $\freshslice{U}{\Psi}$. Restriction to total cells is then given by the central lifting of that functor, being $\lollipsh{\yoneda U}{\Psi}$. Combined with the knowledge that $\lollipsh{\yoneda U}{\Psi} \transppsh{\yoneda U}{\Psi} \cong \Id$ (\cref{thm:pw-quantification}), this reveals that $\mathring b$ determines the action of $b$ on total cells.
	
	\textbf{Completeness of the coherence criterion.} The action of a term on cells should be natural with respect to restriction. This is automatic when considered with respect to morphisms between cells that are either both total or both on-boundary. Moreover, there are no morphisms $\chi : (V, \vfi) \to (V', \vfi') : \catV / (\Psi \multip U)$ from a total cell to an on-boundary cell, since the boundary is a well-defined presheaf. So we still need to prove naturality w.r.t. morphisms from on-boundary cells to total cells.
	
	Let $\chi : (V, \vfi) \to (V', \vfi')$ be such a morphism. Then $(V', \vfi') \cong_\iota \freshslice U \Psi (W, \psi) \cong \freshslice U \Psi \sumslice U \Psi \freshslice U \Psi (W, \psi) \cong \freshslice U \Psi \sumslice U \Psi (V', \vfi')$ by an isomorphism
	\begin{align*}
		&\freshslice U \Psi \sumslice U \Psi \iota\inv \circ \freshslice U \Psi (\dropslice U \Psi)\inv \circ \iota \\
		&= \freshslice U \Psi \sumslice U \Psi \iota\inv \circ \copyslice U \Psi \circ \iota \\
		&= \copyslice U \Psi \circ \iota\inv \circ \iota = \copyslice U \Psi.
	\end{align*}
	Hence, by naturality, $\chi = (\copyslice U \Psi)\inv \circ \copyslice U \Psi \circ \chi = (\copyslice U \Psi)\inv \circ \freshslice U \Psi \sumslice U \Psi \chi \circ \copyslice U \Psi$. Thus, we have factored $\chi$ as an instance of the unit $\copyslice U \Psi$ followed by a morphism between total cells. This means it is sufficient to show naturality with respect to $\copyslice U \Psi : (V, \vfi) \to \freshslice U \Psi \sumslice U \Psi (V, \vfi)$. (The cells of $\Gamma$ and the transpension type available for $(V', \vfi')$ carry over to $\freshslice U \Psi \sumslice U \Psi (V, \vfi)$ by restriction.)
	
	Now the action of $b$ on $(V, \vfi)$ is given by the action of $b_\partial$ on $(V, \vfi)$. Meanwhile, the action of $b$ on $\freshslice U \Psi \sumslice U \Psi (V, \vfi)$ is given by the action of $\mathring b$ on $\sumslice U \Psi (V, \vfi)$, which is the action of $\ftrtm{\freshslice U \Psi}{\mathring b}$ on $(V, \vfi)$. These have to correspond via $\copyslice U \Psi : (V, \vfi) \to \freshslice U \Psi \sumslice U \Psi (V, \vfi)$, which corresponds via central lifting to the natural transformation $\apppsh{\yoneda U} \Psi$ on presheaves.
	This is exactly what happens in the coherence criterion: we use $\apppsh{\yoneda U} \Psi : \freshpsh{\yoneda U} \Psi \lollipsh{\yoneda U} \Psi \to \Id$ to bring $b_\partial$ and $\mathring b$ to the same context and type, and then equate them. Since $b_\partial$ only exists on the boundary, we also have to restrict $\mathring b$ to the boundary, but that's OK since we were interested in an on-boundary cell anyway.
\end{proof}
\begin{example}[Affine cubes]
	We instantiate \cref{thm:transp-elim} for the multiplier $\loch * \IX : \Box^k \to \Box^k$ (\cref{ex:affine-cubes}). There, $\partial \IX$ is essentially the constant presheaf with $k$ elements. So $b_\partial$ determines the images of the $k$ poles of the transpension type. The term $\mathring b$ determines the action on paths (for $k = 2$, for general $k$ perhaps `webs' is a better term), and the paths/webs of the transpension type are essentially the elements of $A$. The coherence condition says that the image of such paths/webs should always have the endpoints given by $b_\partial$.
\end{example}
\begin{example}[Clocks]
	We instantiate \cref{thm:transp-elim} for the multiplier $\loch * (i : \clocksym_k)$ (\cref{ex:clocks}), where we adapt the base category to forbid diagonals: a morphism may use every variable of its domain at most once.
	The boundary $\partial(i : \clocksym_k)$ is isomorphic to $\yoneda(i : \clocksym_{k-1})$ if $k > 0$ and to the empty presheaf $\bot$ if $k = 0$.
	So if we want to eliminate an element of the transpension type over $\yoneda(i : \clocksym_k)$, which means we have a clock and we don't care about what happens if the time exceeds $k$, then we need to handle two cases. The first case $b_\partial$ says what happens if we don't even care what happens at timestamp $k$; in which case the transpension type trivializes. Then, by giving $\mathring b$, we say what happens at timestamp $k$ and need to make sure that this is consistent with $b_\partial$. The elements of the transpension type at timestamp $k$ are essentially the elements of $A$, which are fresh for the clock.
\end{example}
\begin{example}[Embargoes]
	Recall that the multiplier $\loch \multip \embargo$ sends $W \in \catW$ to $(W, \top) \in \catW \times {\uparrow}$, the Yoneda-embedding of which represents the arrow $\yoneda W \to \yoneda W$, i.e. $\yoneda W . \embargo . \top$ under the convention that $\Psi.\embargo.\Theta$ denotes $(\Psi.\Theta \to \Psi)$.
	Its left lifting is $\loch \multip \yoneda \embargo : \widehat \catW \to \widehat{\catW \times \thearrowcat}$, and $\yoneda \embargo$ is the terminal object, so that $\widehat{\catW \times \thearrowcat} / \yoneda \embargo \cong \widehat{\catW \times \thearrowcat}$.
	We get 5 adjoint functors, of which we give here the action up to isomorphism:
	\begin{align*}
		\begin{array}{c c c c c c r}
			&& && \Psi & \mapsto & (\bot \to \Psi), \\
			&& \sumbase{\yoneda \embargo} &:& \Psi & \mapsfrom & (\Psi.\Theta \to \Psi), \\
			\loch \multip {\yoneda \embargo} & \text{or} & \freshbase{\yoneda \embargo} &:& \Psi & \mapsto & (\Psi \to \Psi), \\
			{\yoneda \embargo} \multimap \loch & \text{or} & \lollibase{\yoneda \embargo} &:& \Psi.\Theta & \mapsfrom & (\Psi.\Theta \to \Psi), \\
			{\yoneda \embargo} \amaze \loch & \text{or} & \transpbase{\yoneda \embargo} &:& \Psi & \mapsto & (\Psi \to \top).
		\end{array}
	\end{align*}
	The boundary of $\yoneda \embargo$ is $\partial \embargo \cong \yoneda(\top, \bot)$ which is isomorphic to the arrow $\bot \to \top$.
	Thus, we see:
	\begin{description}
		\item[$\sumbase{\yoneda \embargo}$] If, for some unknown embargo, we have information partly under that embargo, then we can only extract the unembargoed information,
		\item[$\freshbase{\yoneda \embargo}$] If information is fresh for an embargo, then it is unembargoed,
		\item[$\lollibase{\yoneda \embargo}$] If, for any embargo, we have information partly under that embargo, then we can extract the information,
		\item[$\transpbase{\yoneda \embargo}$] If information is transpended over an embargo, then it is completely embargoed.
	\end{description}
	Perhaps the above is more intuitive if we think of an embargo as a key or a password.
	
	So let us now instantiate \cref{thm:transp-elim}, which allows us to eliminate an element of the transpension type, i.e. essentially an element of $A \to \top$. The boundary case exists over the boundary $\bot \to \top$ and allows us to consider only the codomain of the arrow, i.e. the part of the context before the embargo, where the transpension type is trivial. The case $\mathring{b}$ then requires us to say how to act on embargoed data in a coherent way with what we already specified in $b_\partial$. The embargoed data is essentially an element of $A$, which comes from the mode where the embargo does not apply.
\end{example}

\section{Prior modalities} \label{sec:prior}
Many modalities arise as central or right liftings of functors between base categories \cite{paramdtt,reldtt,reldtt-techreport,clock-cat}. The following definition allows us to use such modalities even when part of the context is in front of a pipe.
\begin{definition} \label{def:slice-mod}
	A functor $G : \catW \to \catW'$ yields a functor $\baseslice G \Psi : \catW / \Psi \to \catW' / \lpsh G \Psi : (W, \psi) \mapsto (GW, \lpsh G \psi)$.
	This in turn yields three adjoint functors between presheaf categories:
	\begin{equation}
		\lpshpsh G \Psi \dashv \fpshpsh G \Psi \dashv \rpshpsh G \Psi.
	\end{equation}
\end{definition}
If a modality is both a right and a central lifting, then the following theorem relates the corresponding `piped' modalities:
\begin{theorem}
	If $G : \catW \to \catW'$ has a right adjoint $G \dashv S$, then we have
	\begin{equation}
		\begin{array}{c c c c c c | c c c c c c}
			&&
			\pairslice{}{\lpsh \eps} \circ \baseslice{G}{\lpsh S \Psi'}
			&\dashv&
			\baseslice{S}{\Psi'}
			&\qquad&\qquad&
			\baseslice{G}{\Psi}
			&\dashv&
			\wknslice{}{\lpsh \eta} \circ \baseslice{S}{\lpsh G \Psi}
			&&
			\\ \hline
			&&
			\pairpsh{}{\lpsh \eps} \circ \lpshpsh{G}{\lpsh S \Psi'}
			&\dashv&
			\lpshpsh{S}{\Psi'}
			&\qquad&\qquad&
			\lpshpsh{G}{\Psi}
			&\dashv&
			\wknpsh{}{\lpsh \eta} \circ \lpshpsh{S}{\lpsh G \Psi}
			&\cong&
			\fpshpsh{G}{\Psi}
			\\
			\lpshpsh{S}{\Psi'}
			&\cong&
			\fpshpsh{G}{\lpsh S \Psi'} \circ \wknpsh{}{\lpsh \eps}
			&\dashv&
			\fpshpsh{S}{\Psi'}
			&\qquad&\qquad&
			\fpshpsh{G}{\Psi}
			&\dashv&
			\fpshpsh{S}{\lpsh G \Psi} \circ \funcpsh{}{\lpsh \eta}
			&\cong&
			\rpshpsh{G}{\Psi}
			\\
			\fpshpsh{S}{\Psi'}
			&\cong&
			\funcpsh{}{\lpsh \eps} \circ \rpshpsh{G}{\lpsh S \Psi'}
			&\dashv&
			\rpshpsh{S}{\Psi'}
			&\qquad&\qquad&
			\rpshpsh{G}{\Psi}
			&\dashv&
			\cartransppsh{}{\lpsh \eta} \circ \rpshpsh{S}{\lpsh G \Psi}
			&&
		\end{array}
	\end{equation}
	assuming -- where mentioned -- that $\wknslice{}{\lpsh \eta}$ exists.
\end{theorem}
\begin{proof}
	For the left half of the table, we only prove the first line. The other adjunctions follow from the fact that $\lpsh \loch$, $\fpsh \loch$ and $\rpsh \loch$ are pseudofunctors, and the isomorphisms follow from uniqueness of the adjoint.
	We have a correspondence of diagrams
	\begin{equation}
		\xymatrix{
			W
				\ar[rr]
				\ar@{=>}[rd]_\psi
			&& SW'
				\ar@{=>}[ld]^{\lpsh S \psi'}
			&& GW
				\ar[rr]
				\ar@{=>}[rd]_{\lpsh \eps \circ \lpsh G \psi}
			&& W'
				\ar@{=>}[ld]^{\psi'}
			\\
			& \lpsh S \Psi'
			&&&&
			\Psi'
		}
	\end{equation}
	i.e. morphisms $(W, \psi) \to \baseslice{S}{\Psi'}(W', \psi') : \catW / \lpsh S \Psi'$ correspond to morphisms $\pairslice{}{\lpsh \eps} \baseslice{G}{\lpsh S \Psi'} (W, \psi) \to (W', \psi') : \catW' / \Psi'$.
	
	On the right side of the table, we similarly only need to prove the first line, and we prove it from the first line on the left side.
	The left adjoint to $\wknslice{}{\lpsh \eta} \circ \baseslice{S}{\lpsh G \Psi}$ is
	$\paren{\pairslice{}{\lpsh \eps} \circ \baseslice{G}{\lpsh S \lpsh G \Psi}} \circ \pairslice{}{\lpsh \eta}$.
	We prove that this is equal to $\baseslice G \Psi$:
	\begin{align*}
		&\pairslice{}{\lpsh \eps} \baseslice{G}{\lpsh S \lpsh G \Psi} \pairslice{}{\lpsh \eta} (W, \psi : W \to \Psi) \\
		&= \pairslice{}{\lpsh \eps} \baseslice{G}{\lpsh S \lpsh G \Psi} (W, \lpsh \eta \circ \psi : W \to \lpsh S \lpsh G \Psi) \\
		&= \pairslice{}{\lpsh \eps} (GW, \lpsh G \lpsh \eta \circ \lpsh G \psi : GW \to \lpsh G \lpsh S \lpsh G \Psi) \\
		&= (GW, \lpsh \eps \circ \lpsh G \lpsh \eta \circ \lpsh G \psi : GW \to \lpsh G \Psi)
		= (GW, \lpsh G \psi : GW \to \lpsh G \Psi). & \qedhere
	\end{align*}
%
\end{proof}

\section{Commutation rules} \label{sec:commut}

\subsection{Substitution and substitution}
See \cref{thm:commut-subst-subst}.

\subsection{Modality and substitution} \label{sec:commut-mod-subst}
\begin{theorem} \label{thm:commut-mod-subst} \label{thm:commut-mod-subst-pullback}
	Assume a functor $G : \catW \to \catW'$ and a morphism $\sigma : \Psi_1 \to \Psi_2 : \widehat \catW$. Then we have a commutative diagram
	\begin{equation}
		\xymatrix{
			\catW / \Psi_1
				\ar[r]^{\baseslice{G}{\Psi_1}}
				\ar[d]_{\pairslice{}{\sigma}}
			& \catW' / \lpsh G \Psi_1
				\ar[d]^{\pairslice{}{\lpsh G \sigma}}
			\\
			\catW / \Psi_2
				\ar[r]_{\baseslice{G}{\Psi_2}}
			& \catW' / \lpsh G \Psi_2
		}
	\end{equation}
	and hence
	\begin{align*}
		\begin{array}{c || c | c | c}
			& \lpsh G & \fpsh G & \rpsh G
			\\ \hline \hline
			\pairsym
			& \pairpsh{}{\lpsh G \sigma} \lpshpsh{G}{\Psi_1} \cong \lpshpsh{G}{\Psi_2} \pairpsh{}{\sigma}
			& \pairpsh{}{\sigma} \fpshpsh{G}{\Psi_1} \to \fpshpsh{G}{\Psi_2} \pairpsh{}{\lpsh G \sigma}
			&
			\\ \hline
			\wknsym
			& \wknpsh{}{\lpsh G \sigma} \lpshpsh{G}{\Psi_2} \leftarrow \lpshpsh{G}{\Psi_1} \wknpsh{}{\sigma}
			& \wknpsh{}{\sigma} \fpshpsh{G}{\Psi_2} = \fpshpsh{G}{\Psi_1} \wknpsh{}{\lpsh G \sigma}
			& \wknpsh{}{\lpsh G \sigma} \rpshpsh{G}{\Psi_2} \to \rpshpsh{G}{\Psi_1} \wknpsh{}{\sigma}
			\\ \hline
			\funcsym
			& 
			& \funcpsh{}{\sigma} \fpshpsh{G}{\Psi_1} \leftarrow \fpshpsh{G}{\Psi_2} \funcpsh{}{\lpsh G \sigma}
			& \funcpsh{}{\lpsh G \sigma} \rpshpsh{G}{\Psi_1} \cong \rpshpsh{G}{\Psi_2} \funcpsh{}{\sigma}
			\\ \hline
			\cartranspsym
			&
			& 
			& \cartransppsh{}{\lpsh G \sigma} \rpshpsh{G}{\Psi_2} \leftarrow \rpshpsh{G}{\Psi_1} \cartransppsh{}{\sigma}
		\end{array}
	\end{align*}
	where every statement holds if the mentioned functors exist.
\end{theorem}
\begin{proof}
	It is evident from the definitions that the given diagram commutes. Then by applying $\fpsh \loch$, we find the that $\wknpsh{}{\sigma} \fpshpsh{G}{\Psi_2} = \fpshpsh{G}{\Psi_1} \wknpsh{}{\lpsh G \sigma}$. The rest of the table then follows by \cref{thm:commut-adj}.
\end{proof}

\begin{remark} \label{rem:commut-mod-subst}
	\begin{itemize}
		\item If $\sigma = \pi : \Psi.A \to \Psi$, then this says something about weakening and the $\Sigma$- and $\Pi$-types over $A$.
		\item If $\lpsh G$ moreover happens to be a CwF morphism, then this relates weakening and the $\Sigma$- and $\Pi$-types over $A$ to those over $\lpsh G A$.
		\item If $\loch \times U$ is a cartesian multiplier and we take $\sigma = \pi_1 : \Psi \times \yoneda U \to \Psi$, then by \cref{thm:pw-quantification}, this says something about $\sumpsh{\yoneda U}{\Psi} \dashv \freshpsh{\yoneda U}{\Psi} \dashv \lollipsh{\yoneda U}{\Psi} \dashv \transppsh{\yoneda U}{\Psi}$.
		\wip{In this case, the stronger result applies if (but not only if) $G$ preserves products.}
	\end{itemize}
\end{remark}

\subsection{Multiplier and substitution} \label{sec:commut-multip-subst}
If, in \cref{sec:commut-mod-subst}, we take $G$ equal to some multiplier $\loch \multip U : \catW \to \catV$, then we have
\begin{equation}
	\baseslice G \Psi = \freshslice U \Psi, \qquad
	\lpsh G = \loch \multip \yoneda U, \qquad
	\lpshpsh G \Psi = \freshpsh{\yoneda U} \Psi, \qquad
	\fpshpsh G \Psi = \lollipsh{\yoneda U} \Psi, \qquad
	\rpshpsh G \Psi = \transppsh{\yoneda U} \Psi.
\end{equation}
This immediately yields the general case of the following theorem:
\begin{theorem} \label{thm:commut-multip-subst} \label{thm:commut-multip-subst-cartesian} \label{thm:commut-multip-subst-ff}
	Assume a multiplier $\loch \multip U : \catW \to \catV$ and a morphism $\sigma : \Psi_1 \to \Psi_2$ in $\widehat \catW$. Write $\tau = \sigma \multip \yoneda U$. Then we have:
	\begin{equation}
		\begin{array}{c || c | c | c | c}
			& \sumsym & \freshsym & \lollisym & \transpsym
			\\ \hline \hline
			\pairsym
			& \pairpsh{}{\sigma} \sumpsh{\yoneda U}{\Psi_1} \lhd^1 \sumpsh{\yoneda U}{\Psi_2} \pairpsh{}{\tau}
			& \pairpsh{}{\tau} \freshpsh{\yoneda U}{\Psi_1} \cong \freshpsh{\yoneda U}{\Psi_2} \pairpsh{}{\sigma}
			& \pairpsh{}{\sigma} \lollipsh{\yoneda U}{\Psi_1} \rhd_1 \lollipsh{\yoneda U}{\Psi_2} \pairpsh{}{\tau}
			& \pairpsh{}{\tau} \transppsh{\yoneda U}{\Psi_1} \rhd_2 \transppsh{\yoneda U}{\Psi_2} \pairpsh{}{\sigma}
			\\ \hline
			\wknsym
			& \wknpsh{}{\sigma} \sumpsh{\yoneda U}{\Psi_2} \lhd^2 \sumpsh{\yoneda U}{\Psi_1} \wknpsh{}{\tau}
			& \wknpsh{}{\tau} \freshpsh{\yoneda U}{\Psi_2} \lhd^1 \freshpsh{\yoneda U}{\Psi_1} \wknpsh{}{\sigma}
			& \wknpsh{}{\sigma} \lollipsh{\yoneda U}{\Psi_2} = \lollipsh{\yoneda U}{\Psi_1} \wknpsh{}{\tau}
			& \wknpsh{}{\tau} \transppsh{\yoneda U}{\Psi_2} \rhd_1 \transppsh{\yoneda U}{\Psi_1} \wknpsh{}{\sigma}
			\\ \hline
			\funcsym
			& \funcpsh{}{\sigma} \sumpsh{\yoneda U}{\Psi_1} \lhd^3 \sumpsh{\yoneda U}{\Psi_2} \funcpsh{}{\tau}
			& \funcpsh{}{\tau} \freshpsh{\yoneda U}{\Psi_1} \lhd^2 \freshpsh{\yoneda U}{\Psi_2} \funcpsh{}{\sigma}
			& \funcpsh{}{\sigma} \lollipsh{\yoneda U}{\Psi_1} \lhd^1 \lollipsh{\yoneda U}{\Psi_2} \funcpsh{}{\tau}
			& \funcpsh{}{\tau} \transppsh{\yoneda U}{\Psi_1} \cong \transppsh{\yoneda U}{\Psi_2} \funcpsh{}{\sigma}
			\\ \hline
			\cartranspsym
			&
			& \cartransppsh{}{\tau} \freshpsh{\yoneda U}{\Psi_2} \lhd^3 \freshpsh{\yoneda U}{\Psi_1} \cartransppsh{}{\sigma}
			& \cartransppsh{}{\sigma} \lollipsh{\yoneda U}{\Psi_2} \lhd^2 \lollipsh{\yoneda U}{\Psi_1} \cartransppsh{}{\tau}
			& \cartransppsh{}{\tau} \transppsh{\yoneda U}{\Psi_2} \lhd^1 \transppsh{\yoneda U}{\Psi_1} \cartransppsh{}{\sigma}
		\end{array}
	\end{equation}
	where every statement holds if the mentioned functors exist, and where
	\begin{enumerate}
		\item In general, $\lhd^1$ means $\leftarrow$, $\rhd_1$ means $\to$ and the other symbols mean nothing.
		\item If $\loch \multip U$ is $\top$-slice right adjoint, then $\lhd^1$ upgades to $\cong$ and $\lhd^2$ upgrades to $\leftarrow$.
		\item If $\loch \multip U$ is cartesian (hence $\top$-slice right adjoint), then $\lhd^1$ and $\lhd^2$ upgade to $\cong$ and $\lhd^3$ upgrades to $\leftarrow$.
		\item If $\loch \multip U$ is $\top$-slice fully faithful, then we have
		\begin{equation}
			\pairpsh{}{\sigma} \lollipsh{\yoneda U}{\Psi_1} \cong \lollipsh{\yoneda U}{\Psi_2} \pairpsh{}{\tau} : \widehat{\overbrace{\catV / (\Psi_1 \multip \yoneda U)}} \to \widehat{\catW / \Psi_2}
		\end{equation}
		so that $\rhd_1$ upgrades to $\cong$ and $\rhd_2$ upgrades to $\to$.
	\end{enumerate}
\end{theorem}
\begin{proof}
	\begin{enumerate}
		\item The general case is a corollary of \cref{thm:commut-mod-subst} for $G = \loch \multip U$.
		\item To prove the $\top$-slice right adjoint case, we show in the base category that $\pairslice{}{\sigma} \sumslice{U}{\Psi_1} = \sumslice{U}{\Psi_2} \pairslice{}{(\sigma \multip \yoneda U)}$.
		We use the construction of $\sumslice{U}{\Psi}$ in the proof of presheafwise right adjointness (\cref{thm:pwra}).
		On one hand, we have:
		\begin{align*}
			\pairslice{}{\sigma} \sumslice{U}{\Psi_1} (V, (\psi_1^{\DSub{W_0}{\Psi_1}} \multip \yoneda U) \circ \vfi^{\DSub{V}{W_0 \multip U}})
			&= \pairslice{}{\sigma} \pairslice{}{\psi_1} \sumslice{U}{W_0} (V, \vfi)
			= \pairslice{}{\sigma \circ \psi_1} \sumslice{U}{W_0} (V, \vfi).
		\end{align*}
		On the other hand:
		\begin{align*}
			\sumslice{U}{\Psi_2} \pairslice{}{(\sigma \multip \yoneda U)} (V, (\psi_1^{\DSub{W_0}{\Psi_1}} \multip \yoneda U) \circ \vfi^{\DSub{V}{W_0 \multip U}})
			&= \sumslice{U}{\Psi_2} (V, ((\sigma \circ \psi_1) \multip \yoneda U) \circ \vfi)
			\\
			&= \pairslice{}{\sigma \circ \psi_1} \sumslice{U}{W_0} (V, \vfi).
		\end{align*}
		\item This follows from \cref{thm:commut-subst-subst}.
		\item We show that $\pairpsh{}{\sigma} \lollipsh{\yoneda U}{\Psi_1} \cong \lollipsh{\yoneda U}{\Psi_2} \pairpsh{}{\tau}$.
		Pick a presheaf $\Gamma$ over $\catV / (\Psi_1 \multip \yoneda U)$. On the one hand, we have:
		\begin{align*}
			&\DSub{(W_2, \psi_2^{\DSub{W_2}{\Psi_2}})}{\pairpsh{}{\sigma} \lollipsh{\yoneda U}{\Psi_1} \Gamma} \\
			&= \exists (W_1, \psi_1^{\DSub{W_1}{\Psi_1}}) .
				\paren{ \theta : (W_2, \psi_2) \to \pairslice{}{\sigma}(W_1, \psi_1) } \times
				\paren{ \DSub{(W_1, \psi_1)}{\lollipsh{\yoneda U}{\Psi_1} \Gamma} } \\
			&= \exists (W_1, \psi_1^{\DSub{W_1}{\Psi_1}}) .
				\paren{ \theta : (W_2, \psi_2) \to (W_1, \sigma \circ \psi_1) } \times
				\paren{ \DSub{(W_1 \multip U, \psi_1 \multip \yoneda U)}{\Gamma} } \\
			&\cong \exists W_1, \psi_1^{\DSub{W_1}{\Psi_1}}, \theta^{W_2 \to W_1} .
				\paren{ \psi_2 = \sigma \circ \psi_1 \circ \theta } \times
				\paren{ \DSub{(W_1 \multip U, \psi_1 \multip \yoneda U)}{\Gamma} } \\
				& {}\qquad \text{We now absorb $\theta$ into $\psi_1$:} \\
			&\cong \psi_1^{\DSub{W_2}{\Psi_1}} .
				\paren{ \psi_2 = \sigma \circ \psi_1 } \times
				\paren{ \DSub{(W_2 \multip U, \psi_1 \multip \yoneda U)}{\Gamma} }.
		\end{align*}
		On the other hand, we have:
		\begin{align*}
			&\DSub{(W_2, \psi_2^{\DSub{W_2}{\Psi_2}})}{\lollipsh{\yoneda U}{\Psi_2} \pairpsh{}{\tau} \Gamma} \\
			&= \DSub{(W_2 \multip U, \psi_2 \multip \yoneda U)}{\pairpsh{}{\tau} \Gamma} \\
			&= \exists (V_1, \vfi_1^{\DSub{V_1}{\Psi_1 \multip \yoneda U}}) .
				\paren{ \omega : (W_2 \multip U, \psi_2 \multip \yoneda U) \to \pairslice{}{\tau}(V_1, \vfi_1) } \times
				\paren{ \DSub{(V_1, \vfi_1)}{\Gamma} } \\
			&= \exists (V_1, \vfi_1^{\DSub{V_1}{\Psi_1 \multip \yoneda U}}) .
				\paren{ \omega : (W_2 \multip U, \psi_2 \multip \yoneda U) \to (V_1, (\sigma \multip \yoneda U) \circ \vfi_1) } \times
				\paren{ \DSub{(V_1, \vfi_1)}{\Gamma} } \\
				& {}\qquad \text{We now deconstruct $\vfi_1 = (\psi_1 \multip \yoneda U) \circ \chi$:} \\
			&\cong \exists V_1, W_1, \chi^{V_1 \to W_1 \multip U}, \psi_1^{\DSub{W_1}{\Psi_1}} . \\
				& {}\qquad
				\paren{ \omega : (W_2 \multip U, \psi_2 \multip \yoneda U) \to (V_1, ((\sigma \circ \psi_1) \multip \yoneda U) \circ \chi) } \times
				\paren{ \DSub{(V_1, (\psi_1 \multip \yoneda U) \circ \chi)}{\Gamma} } \\
			&\cong \exists V_1, W_1, \chi^{V_1 \to W_1 \multip U}, \psi_1^{\DSub{W_1}{\Psi_1}}, \omega^{W_2 \multip U \to V_1} . \\
				& {}\qquad
				(\psi_2 \multip \yoneda U = ((\sigma \circ \psi_1) \multip \yoneda U) \circ \chi \circ \omega) \times
				\paren{ \DSub{(V_1, (\psi_1 \multip \yoneda U) \circ \chi)}{\Gamma} } \\
				& {}\qquad \text{We now absorb $\omega$ into $\chi$:} \\
			&\cong \exists W_1, \psi_1^{\DSub{W_1}{\Psi_1}}, \chi^{W_2 \multip U \to W_1 \multip U} . \\
				& {}\qquad (\psi_2 \multip \yoneda U = ((\sigma \circ \psi_1) \multip \yoneda U) \circ \chi) \times
				\paren{ \DSub{(W_2 \multip U, (\psi_1 \multip \yoneda U) \circ \chi)}{\Gamma} } \\
				& {}\qquad \text{Let $\chi = \freshslice{U}{\Psi_2} \theta : \freshslice{U}{\Psi_2}(W_2, \psi_2) \to \freshslice{U}{\Psi_2}(W_1, \sigma \circ \psi_1)$:} \\
			&\cong \exists W_1, \psi_1^{\DSub{W_1}{\Psi_1}}, \theta^{W_2 \to W_1} .
				(\psi_2 = \sigma \circ \psi_1 \circ \theta) \times
				\paren{ \DSub{(W_2 \multip U, ((\psi_1 \circ \theta) \multip \yoneda U))}{\Gamma} } \\
				& {}\qquad \text{We now absorb $\theta$ into $\psi_1$:} \\
			&\cong \psi_1^{\DSub{W_2}{\Psi_1}} .
				(\psi_2 = \sigma \circ \psi_1) \times
				\paren{ \DSub{(W_2 \multip U, (\psi_1 \multip \yoneda U))}{\Gamma} }
		\end{align*}
		This proves the isomorphism. The rest follows from \cref{thm:commut-adj}. \qedhere
	\end{enumerate}
\end{proof}

\subsection{Multiplier and modality}
\begin{theorem} \label{thm:commut-multip-mod} \label{thm:commut-multip-mod-cartesian}
	Assume a commutative diagram (up to natural isomorphism $\nu : F(\loch \multip U) \cong G\loch \multip U'$)
	\begin{equation}
		\xymatrix{
			\catW \ar[r]^G \ar[d]_{\loch \multip U}
			& \catW' \ar[d]^{\loch \multip U'}
			\\
			\catV \ar[r]_F
			& \catV'
		}
	\end{equation}
	where $\loch \multip U$ and $\loch \multip U'$ are multipliers for $U$ and $U'$.
	
	Then $\pairslice{}{\lpsh \nu}$ is a strictly invertible functor and hence we have
	\begin{equation}
		\pairpsh{}{\lpsh{\nu}} \cong
		\wknpsh{}{\lpsh{\nu}\inv} \cong
		\funcpsh{}{\lpsh{\nu}} \cong
		\transppsh{}{\lpsh{\nu}\inv}
		\qquad
		\pairpsh{}{\lpsh{\nu}\inv} \cong
		\wknpsh{}{\lpsh{\nu}} \cong
		\funcpsh{}{\lpsh{\nu}\inv} \cong
		\transppsh{}{\lpsh{\nu}},
	\end{equation}
	where $\wknpsh{}{\lpsh{\nu}\inv}$ is the strict inverse to $\wknpsh{}{\lpsh{\nu}}$.
	
	Then we have $\pairslice{}{\lpsh{\nu}\inv} \freshslice{U'}{\lpsh G \Psi} \baseslice{G}{\Psi} \cong \baseslice{F}{\Psi \multip U} \freshslice U \Psi$. This yields the following commutation table:
	\small
	\begin{align*}
		\begin{array}{c || c | c | c}
			& \lpsh F, \lpsh G & \fpsh F, \fpsh G & \rpsh F, \rpsh G
			\\ \hline \hline
			\sumsym
			& \sumpsh{\yoneda U'}{\lpsh G \Psi} \wknpsh{}{\lpsh{\nu}\inv} \lpshpsh{F}{\Psi \multip \yoneda U} \rhd_1 \lpshpsh{G}{\Psi} \sumpsh{\yoneda U} \Psi
			& \sumpsh{\yoneda U} \Psi \fpshpsh{F}{\Psi} \rhd_2 \fpshpsh G \Psi \sumpsh{\yoneda U'}{\lpsh G \Psi} \wknpsh{}{\lpsh{\nu}\inv}
			&
			\\ \hline
			\freshsym
			& \wknpsh{}{\lpsh{\nu}} \freshpsh{\yoneda U'}{\lpsh G \Psi} \lpshpsh{G}{\Psi} \cong \lpshpsh{F}{\Psi \multip \yoneda U} \freshpsh{\yoneda U} \Psi
			& \freshpsh{\yoneda U} \Psi \fpshpsh{G}{\Psi} \rhd_1 \fpshpsh{F}{\Psi \multip \yoneda U} \wknpsh{}{\lpsh{\nu}} \freshpsh{\yoneda U'}{\lpsh G \Psi}
			& \wknpsh{}{\lpsh{\nu}} \freshpsh{\yoneda U'}{\lpsh G \Psi} \rpshpsh{G}{\Psi} \rhd_2 \rpshpsh{F}{\Psi \multip \yoneda U} \freshpsh{\yoneda U} \Psi
			\\ \hline
			\lollisym
			& \lollipsh{\yoneda U'}{\lpsh G \Psi} \wknpsh{}{\lpsh{\nu}\inv} \lpshpsh{F}{\Psi \multip \yoneda U} \leftarrow \lpshpsh{G}{\Psi} \lollipsh{\yoneda U} \Psi
			& \lollipsh{\yoneda U} \Psi \fpshpsh{F}{\Psi} \cong \fpshpsh G \Psi \lollipsh{\yoneda U'}{\lpsh G \Psi} \wknpsh{}{\lpsh{\nu}\inv}
			& \lollipsh{\yoneda U'}{\lpsh G \Psi} \wknpsh{}{\lpsh{\nu}\inv} \rpshpsh{F}{\Psi \multip \yoneda U} \rhd_1 \rpshpsh{G}{\Psi} \lollipsh{\yoneda U} \Psi
			\\ \hline
			\transpsym
			&
			& \transppsh{\yoneda U} \Psi \fpshpsh{G}{\Psi} \leftarrow \fpshpsh{F}{\Psi \multip \yoneda U} \wknpsh{}{\lpsh{\nu}} \transppsh{\yoneda U'}{\lpsh G \Psi}
			& \wknpsh{}{\lpsh{\nu}} \transppsh{\yoneda U'}{\lpsh G \Psi} \rpshpsh{G}{\Psi} \cong \rpshpsh{F}{\Psi \multip \yoneda U} \transppsh{\yoneda U} \Psi
		\end{array}
	\end{align*}
	\normalsize
	where any statement holds if the mentioned functors exist, and where
	\begin{enumerate}
		\item In general, $\rhd_1$ means $\to$ and $\rhd_2$ means nothing.
		\wip{\item If $\loch \multip U$ and $\loch \multip U'$ are quantifiable and the morphism $\theta : \sumbase{U'} \circ \baseslice{F}{U} \to G \circ \sumbase U : \catV/U \to \catW'$ is invertible,\footnote{This is a slight abuse of notation, as we know that $U' \cong FU$ but not that $U' = FU$.} then $\rhd_1$ upgrades to $\cong$ and $\rhd_2$ upgrades to $\to$.
		
		This applies in particular if $\catW = \catV$, $\catW' = \catV'$, $G = F$ preserves finite products and the multipliers are cartesian.}
		\item If $\catW = \catV$, $\catW' = \catV'$, $F = G$, both multipliers are cartesian and $\nu$ respects the first projection, i.e.\ $\pi_1 \circ \nu = G\pi_1$, then $\rhd_1$ upgrades to $\cong$ and $\rhd_2$ upgrades to $\to$. Note that in this case we have $GU \cong G(\top \times U) \cong_\nu G\top \times U'$.
	\end{enumerate}
\end{theorem}
\begin{remark}
	In the above theorem, we think of $F$ and $G$ as similar functors; if we are dealing with endomultipliers, we will typically take $F = G$. The multipliers, however, will typically be different, as in general $U \not\cong FU$.
\end{remark}
\begin{proof}
	Since $\lpsh{\nu}$ is an isomorphism, $\pairslice{}{\lpsh{\nu}}$ is a strictly invertible functor with inverse $\pairslice{}{\lpsh{\nu}\inv}$. Since $\fpsh \loch$ is a 2-functor, $\wknpsh{}{\lpsh{\nu}}$ is also strictly invertible with inverse $\wknpsh{}{\lpsh{\nu}\inv}$. Because equivalences of categories are adjoint to their inverse, we get the chains of isomorphisms displayed.
	\begin{enumerate}
		\item The given commutation property in the base category follows immediately from the definitions and naturality of $\nu$ and its image under $\lpsh \loch$. The rest of the table then follows by \cref{thm:commut-adj}.

		\item We invoke \cref{thm:commut-mod-subst} with $\sigma = \pi_2 : \Psi \times \yoneda U \to \Psi$.
		This yields $\wknpsh{\yoneda U}{\Psi} \fpshpsh{G}{\Psi} = \fpshpsh{G}{\Psi \times \yoneda U} \wknpsh{}{\lpsh G \pi_2}$.
		Now $\lpsh G \pi_2 = \pi_2 \circ \lpsh \nu$ so we can rewrite this to $\wknpsh{\yoneda U}{\Psi} \fpshpsh{G}{\Psi} = \fpshpsh{G}{\Psi \times \yoneda U} \wknpsh{}{\lpsh \nu} \wknpsh{\yoneda U'}{\lpsh G \Psi}$.
		The rest of the table then follows by \cref{thm:commut-adj}. \qedhere
	\end{enumerate}
\end{proof}

\subsection{Multiplier and multiplier} \label{sec:commut-multip-multip}
\begin{figure}
	\begin{center}
	\begin{adjustbox}{angle=270}
	$
		\begin{array}{c || c | c | c | c }
			& \sumbase I, \sumbase J & \freshbase I, \freshbase J & \lollibase I, \lollibase J & \transpbase I, \transpbase J
			\\ \hline \hline
			\sumbase U
			& \sumpsh{\yoneda U}{\Psi} \sumpsh{\yoneda I}{\Psi \multip \yoneda U} \cong 
			\sumpsh{\yoneda J}{\Psi} \sumpsh{\yoneda U'}{\Psi \multip \yoneda J} \wknpsh{}{\lpsh{\nu}\inv}
			& \sumpsh{\yoneda U'}{\Psi \multip \yoneda J} \wknpsh{}{\lpsh{\nu}\inv} \freshpsh{\yoneda I}{\Psi \multip \yoneda U}
			\rhd_1 \freshpsh{\yoneda J}{\Psi} \sumpsh{\yoneda U}{\Psi}
			& \sumpsh{\yoneda U}{\Psi} \lollipsh{\yoneda I}{\Psi \multip \yoneda U} \rhd_2
			\lollipsh{\yoneda J}{\Psi} \sumpsh{\yoneda U'}{\Psi \multip \yoneda J} \wknpsh{}{\lpsh{\nu}\inv} 
			& \sumpsh{\yoneda U'}{\Psi \multip \yoneda J} \wknpsh{}{\lpsh{\nu}\inv} \transppsh{\yoneda I}{\Psi \multip \yoneda U}
			\rhd_3 \transppsh{\yoneda J}{\Psi} \sumpsh{\yoneda U}{\Psi}
			\\ \hline
			\freshbase U
			& \freshpsh{\yoneda U}{\Psi} \sumpsh{\yoneda J}{\Psi} \lhd^1
			\sumpsh{\yoneda I}{\Psi \multip \yoneda U} \wknpsh{}{\lpsh{\nu}} \freshpsh{\yoneda U'}{\Psi \multip \yoneda J}
			& \wknpsh{}{\lpsh{\nu}} \freshpsh{\yoneda U'}{\Psi \multip \yoneda J} \freshpsh{\yoneda J}{\Psi}
			\cong \freshpsh{\yoneda I}{\Psi \multip \yoneda U} \freshpsh{\yoneda U}{\Psi}
			& \freshpsh{\yoneda U}{\Psi} \lollipsh{\yoneda J}{\Psi} \rhd_1
			\lollipsh{\yoneda I}{\Psi \multip \yoneda U} \wknpsh{}{\lpsh{\nu}} \freshpsh{\yoneda U'}{\Psi \multip \yoneda J}
			& \wknpsh{}{\lpsh{\nu}} \freshpsh{\yoneda U'}{\Psi \multip \yoneda J} \transppsh{\yoneda J}{\Psi}
			\rhd_2 \transppsh{\yoneda I}{\Psi \multip \yoneda U} \freshpsh{\yoneda U}{\Psi}
			\\ \hline
			\lollibase U
			& \lollipsh{\yoneda U}{\Psi} \sumpsh{\yoneda I}{\Psi \multip \yoneda U} \lhd^2
			\sumpsh{\yoneda J}{\Psi} \lollipsh{\yoneda U'}{\Psi \multip \yoneda J} \wknpsh{}{\lpsh{\nu}\inv}
			& \lollipsh{\yoneda U'}{\Psi \multip \yoneda J} \wknpsh{}{\lpsh{\nu}\inv} \freshpsh{\yoneda I}{\Psi \multip \yoneda U}
			\lhd^1 \freshpsh{\yoneda J}{\Psi} \lollipsh{\yoneda U}{\Psi}
			& \lollipsh{\yoneda U}{\Psi} \lollipsh{\yoneda I}{\Psi \multip \yoneda U} \cong
			\lollipsh{\yoneda J}{\Psi} \lollipsh{\yoneda U'}{\Psi \multip \yoneda J} \wknpsh{}{\lpsh{\nu}\inv}
			& \lollipsh{\yoneda U'}{\Psi \multip \yoneda J} \wknpsh{}{\lpsh{\nu}\inv} \transppsh{\yoneda I}{\Psi \multip \yoneda U}
			\rhd_1 \transppsh{\yoneda J}{\Psi} \lollipsh{\yoneda U}{\Psi}
			\\ \hline
			\transpbase U
			& \transppsh{\yoneda U}{\Psi} \sumpsh{\yoneda J}{\Psi} \lhd^3
			\sumpsh{\yoneda I}{\Psi \multip \yoneda U} \wknpsh{}{\lpsh{\nu}} \transppsh{\yoneda U'}{\Psi \multip \yoneda J}
			& \wknpsh{}{\lpsh{\nu}} \transppsh{\yoneda U'}{\Psi \multip \yoneda J} \freshpsh{\yoneda J}{\Psi}
			\lhd^2 \freshpsh{\yoneda I}{\Psi \multip \yoneda U} \transppsh{\yoneda U}{\Psi}
			& \transppsh{\yoneda U}{\Psi} \lollipsh{\yoneda J}{\Psi} \lhd^1
			\lollipsh{\yoneda I}{\Psi \multip \yoneda U} \wknpsh{}{\lpsh{\nu}} \transppsh{\yoneda U'}{\Psi \multip \yoneda J}
			& \wknpsh{}{\lpsh{\nu}} \transppsh{\yoneda U'}{\Psi \multip \yoneda J} \transppsh{\yoneda J}{\Psi}
			\cong \transppsh{\yoneda I}{\Psi \multip \yoneda U} \transppsh{\yoneda U}{\Psi}
		\end{array}
	$
	\end{adjustbox}
	\end{center}
	\caption{Commutation table for 2 multipliers (\cref{thm:commut-multip-multip}).}
	\label{fig:commut-multip-multip}
\end{figure}
\begin{theorem} \label{thm:commut-multip-multip}
	Assume we have a commutative diagram (up to natural isomorphism $\nu : \loch \multip U \multip I \cong \loch \multip J \multip U'$) of multipliers
	\begin{equation}
		\xymatrix{
			\catW \ar[r]^{\loch \multip J} \ar[d]_{\loch \multip U}
			& \catW' \ar[d]^{\loch \multip U'}
			\\
			\catV \ar[r]_{\loch \multip I}
			& \catV'.
		}
	\end{equation}
	Then we have the commutation table given in \cref{fig:commut-multip-multip}
	where every statement holds if the mentioned functors exist, and where
	\begin{enumerate}
		\item In general, $\rhd_1$ means $\to$, $\lhd^1$ means $\leftarrow$ and the other symbols mean nothing.
		\item If $\catW = \catW'$, $\catV = \catV'$, $\loch \multip U = \loch \multip U'$, the multipliers $\loch \multip J$ and $\loch \multip I$ are cartesian and $(\pi_1 \multip U) \circ \nu = \pi_1 : (\loch \multip U) \times I \to \loch \multip U$, then $\lhd^1$ upgrades to $\cong$ and $\lhd^2$ upgrades to $\leftarrow$.
		\begin{enumerate}
			\item If moreover $\loch \multip U$ is $\top$-slice fully faithful, then $\lhd^2$ upgrades to $\cong$ and $\lhd^3$ upgrades to $\leftarrow$.
		\end{enumerate}
		\item The symbols $\rhd_i$ upgrade under symmetric conditions.
	\end{enumerate}
\end{theorem}
\begin{proof}
	\begin{enumerate}
		\item In the base category, it is clear that $\freshslice{U'}{\Psi \multip \yoneda J} \freshslice{J}{\Psi} \cong \pairslice{}{\lpsh\nu}\freshslice{I}{\Psi \multip \yoneda U} \freshslice{U}{\Psi}$. Applying the 2-functor $\fpsh \loch$ yields the commutation law for $\forall$ and hence, by \cref{thm:commut-adj}, the general case.
		\item We invoke \cref{thm:commut-multip-subst} with $\sigma = \pi_1 : \Psi \times \yoneda J \to \Psi$.
		This yields
		\[
			\wknpsh{\yoneda J}{\Psi} \lollipsh{\yoneda U}{\Psi} = \lollipsh{\yoneda U}{\Psi \times \yoneda J} \wknpsh{}{\pi_1 \multip \yoneda U} : \widehat{\overbrace{\catV / \Psi \multip \yoneda U}} \to \widehat{\overbrace{\catW / \Psi \times \yoneda J}}.
		\]
		Now $\pi_1 \multip \yoneda U = \pi_1 \circ \lpsh\nu\inv$ so we can rewrite this to
		\[
			\wknpsh{\yoneda J}{\Psi} \lollipsh{\yoneda U}{\Psi} = \lollipsh{\yoneda U}{\Psi \times \yoneda J} \wknpsh{}{\lpsh\nu\inv} \wknpsh{\yoneda I}{\Psi \multip \yoneda U} : \widehat{\overbrace{\catV / \Psi \multip \yoneda U}} \to \widehat{\overbrace{\catW / \Psi \times \yoneda J}}.
		\]
		The rest then follows by \cref{thm:commut-adj}.
		\begin{enumerate}
			\item Also follows from the same invocation of \cref{thm:commut-multip-subst}.
		\end{enumerate}
		\item By symmetry. \qedhere
	\end{enumerate}
\end{proof}

\section*{Acknowledgements}
We thank
	Jean-Philippe Bernardy,
	Lars Birkedal,
	Daniel Gratzer,
	Alex Kavvos,
	Magnus Baunsgaard Kristensen,
	Daniel Licata,
	Rasmus Ejlers M\o{}gelberg
	and
	Andrea Vezzosi
for relevant discussions, and to the anonymous reviewers at LMCS for their valuable feedback.
Special thanks to Dominique Devriese, whose pesky questions led to the current research.

\appendix
\section{Changelog} \label{sec:changelog}
The first version of this technical report and the associated paper appeared in 2020.
Since then, there have been significant changes, primarily terminological ones.
To help out readers coming back to these texts after having consulted earlier versions (or associated presentations), we list the most important changes here.

\subsection{\texorpdfstring{\Cref{def:pointable}}{Definition \ref{def:pointable}}}
\begin{itemize}
	\item \textbf{Unpointable} objects were formerly called \textbf{spooky},
	\item \textbf{Not objectwise pointable} categories were formerly called \textbf{spooky}.
\end{itemize}

\subsection{\texorpdfstring{\Cref{def:multiplier}}{Definition \ref{def:multiplier}}}
\begin{itemize}
	\item \textbf{Copointed} multipliers were formerly called \textbf{semicartesian},
	\item Multipliers that are \textbf{comonads} were formerly called \textbf{3/4-cartesian},
	\item \textbf{$\top$-slice faithful} multipliers were formerly called \textbf{cancellative},
	\item \textbf{$\top$-slice full} multipliers were formerly called \textbf{affine},
	\item \textbf{Not $\top$-slice objectwise pointable} multipliers were formerly called \textbf{spooky},
	\item \textbf{$\top$-slice shard-free} multipliers were formerly called \textbf{connection-free}, and \textbf{shards} were formerly called \textbf{connections},
	\item \textbf{$\top$-slice right adjoint} multipliers were formerly called \textbf{quantifiable}.
\end{itemize}

\subsection{\texorpdfstring{\Cref{def:multip-hom}}{Definition \ref{def:multip-hom}}}
\begin{itemize}
	\item A \textbf{morphism of copointed multipliers} was formerly called a \textbf{semicartesian} morphism of multipliers,
	\item A \textbf{comonad morphism of multipliers} was formerly called a \textbf{3/4-cartesian} morphism of multipliers.
\end{itemize}

\subsection{Quotient theorem}
The \textbf{quotient theorem} was formerly called \textbf{kernel theorem}.

\subsection{\texorpdfstring{\Cref{def:act-slices}}{Definition \ref{def:act-slices}}}
\textbf{Slicewise} faithful / full / shard-free / right adjoint multipliers were formerly called \textbf{strongly} cancellative / affine / connection-free / quantifiable.

A previous version of paper and technical report featured only the now obsoleted definition of \emph{indirect} shards/connections and slicewise \emph{indirect} shard-freedom / strong \emph{indirect} connection-freedom, that was based on the \emph{indirect} boundary and \emph{indirectly} dimensionally split morphisms.
These notions are obsolete and are retained solely for consistency with \cite[ch.\ 7]{nuyts-phd}.
The qualifier `indirect' was only added a posteriori to distinguish with the more appropriate \emph{direct} notions.

\subsection{\texorpdfstring{\Cref{def:act-elements}}{Definition \ref{def:act-elements}}}
\textbf{Presheafwise} faithful / full / shard-free / right adjoint multipliers were formerly called \textbf{providently} cancellative / affine / connection-free / quantifiable.

\textbf{$\top$-slice elementally} faithful / full multipliers were formerly called \textbf{elementally} cancellative / affine.

A similar note as above applies to (in)direct shards/connections, shard/connection-freedom, boundaries and dimensional splitness.


\bibliographystyle{alphaurl}
\bibliography{../intern-psh-refs.bib}

\end{document}